\documentclass[prl,twocolumn,superscriptaddress,floatfix,noshowpacs,10pt,longbibliography]{revtex4-2}

\usepackage{graphicx,bm,times}
\usepackage{amsmath}
\usepackage{amsfonts}
\usepackage{amssymb}
\usepackage{bm}
\usepackage{color}
\usepackage{times}
\usepackage[%
  colorlinks=true,
  urlcolor=blue,
  linkcolor=blue,
  citecolor=blue
  ]{hyperref}
\DeclareUnicodeCharacter{2212}{-}
\DeclareUnicodeCharacter{2009}{\,}

\begin{document}

\title{Unveiling the quasiparticle behaviour
in the pressure-induced high-$E_c$  phase \\
of an iron-chalcogenide superconductor}

\author{Z. Zajicek}
\thanks{These two authors contributed equally}
\affiliation{Clarendon Laboratory, Department of Physics,
	University of Oxford, Parks Road, Oxford OX1 3PU, UK}

\author{P. Reiss}
\thanks{These two authors contributed equally}
\affiliation{Clarendon Laboratory, Department of Physics,
University of Oxford, Parks Road, Oxford OX1 3PU, UK}
\affiliation{Max Planck Institute for Solid State Research, Heisenbergstr. 1, 70569 Stuttgart, Germany}

\author{D. Graf}
\affiliation{National High Magnetic Field Laboratory and Department of Physics, Florida State University, Tallahassee, Florida 32306, USA}

\author{J. C. A. Prentice}
\affiliation{Department of Materials, University of Oxford, Parks Road, Oxford OX1 3PH, United Kingdom}

\author{Y. Sadki}
\affiliation{Clarendon Laboratory, Department of Physics, University of Oxford, Parks Road, Oxford OX1 3PU, UK}

\author{A. A. Haghighirad}
\affiliation{Clarendon Laboratory, Department of Physics,
University of Oxford, Parks Road, Oxford OX1 3PU, UK}
\affiliation{Institute for Quantum Materials and Technologies, Karlsruhe Institute of Technology, 76021 Karlsruhe, Germany}

\author{A. I. Coldea}
\email[corresponding author:]{amalia.coldea@physics.ox.ac.uk}
\affiliation{Clarendon Laboratory, Department of Physics, University of Oxford, Parks Road, Oxford OX1 3PU, UK}

\begin{abstract}

Superconductivity of iron chalocogenides is strongly enhanced  under applied pressure
yet its underlying pairing mechanism remains elusive.
Here, we present a quantum oscillations study up to 45~T
in the high-$T_{\rm c}$ phase of tetragonal FeSe$_{0.82}$S$_{0.18}$ up to 22\,kbar.
Under applied pressure, the
quasi-two dimensional multi-band Fermi surface expands and the effective masses remain large,
whereas the superconductivity displays a three-fold enhancement.
Comparing with chemical pressure tuning of FeSe$_{1-x}$S$_{x}$,
the Fermi surface enlarges in a similar manner
but the effective masses and  $T_{\rm c}$ are suppressed.
These differences may be attributed to the
changes in the density of states influenced by the chalcogen height, which could promote stronger spin fluctuations pairing under pressure.
Furthermore, our study also reveals unusual scattering
and broadening of superconducting transitions in the high-pressure phase,
indicating the presence of a complex pairing mechanism.

\end{abstract}
\date{\today}
\maketitle

{\bf Introduction}

One direct route to enhance conventional superconductivity towards room temperature
is to apply extreme pressures to light elements to strengthen the electron-phonon coupling  \cite{Drozdov2019}.
In the case of iron-based superconductors, applying  pressure to bulk FeSe leads to a fourfold enhancement
in superconductivity towards 37\,K  under $\sim$40\,kbar \cite{Medvedev2009,Kothapalli2016,Sun2016pressure}.
Due to the small Fermi surface of FeSe,
various competing electronic nematic, magnetic, and superconducting orders can emerge on similar energy scales
which could be stabilized in different pressure regimes \cite{Chubukov2016}.
At low pressures, FeSe has  a small Fermi energy
and exhibits a nematic electronic phase driven by orbital degrees of freedom and strong electronic correlations,
leading to highly anisotropic electronic and superconducting behavior
 \cite{Sprau2016,Coldea2019,Kreisel2020}.
On the other hand, an increase in the Fermi energy may
favour  the stability of a magnetic phase  \cite{Chubukov2016}.
Experimentally, both superconducting and magnetic phases could coexist  under pressure \cite{Bendele2012},
along with additional structural effects  \cite{Bohmer2016}
 posing challenges in understanding the high-$T_{\rm c}$ high-pressure phase.

In a similar manner to applied pressure, chemical pressure induced by the isovalent substitution of selenium for sulphur can suppress nematicity \cite{Reiss2017}.
Surprisingly, within  the tetragonal phase the superconductivity is not enhanced,
no magnetic order is detected, and electronic correlations are significantly weakened
\cite{Reiss2017,Coldea2021,Hosoi2016,Bristow2020}.
Across the nematic end point of FeSe$_{1-x}$S$_x$,
changes in the superconducting gap structure  \cite{Hanaguri2018,Sato2018}, and
a topological transition into an ultranodal phase with Bogoliubov Fermi surface phases has been proposed \cite{Setty2020}.
Notably, by combining both applied and chemical pressure,
the anomalies associated with the magnetic order are shifted to higher pressures
with increased sulphur substitution in FeSe$_{1-x}$S$_x$ \cite{Matsuura2017}.
This decoupling of overlapping nematic and magnetic  phases allows for a deeper understanding of their individual
contribution to superconductivity and provides an opportunity to explore the region of a quantum nematic phase transition
 \cite{Reiss2020,Rana2020,Xiang2017,Bristow2020}.
 Despite the robustness of the high-$T_{\rm c}$  phase under applied pressure,
 the presence of the magnetic order is highly sensitive to the isoelectronic substitution,
 disorder, and uniaxial effects \cite{Matsuura2017,Xie2021,Zajicek2022Cupressure,Reiss2022}.
 This raises the fundamental question whether the magnetically mediated pairing is responsible for superconductivity,
 which needs to be clarified by having direct access to the
 Fermi surface topology and electronic correlations in the high-pressure phase.

Quantum oscillations provide a direct measurement of Fermi surfaces and the properties of quasiparticles involved in the superconducting pairing mechanism.
 Unlike other spectroscopic techniques like ARPES and STM, which probe the electronic structure, quantum oscillations offer access to the high-$T_{\rm c}$
  high-pressure phase of FeSe$_{1-x}$S$_{x}$.
Quantum oscillations have been observed at ambient pressure in
FeSe \cite{Terashima2014,Watson2015a,Audouard2015} and FeSe$_{1-x}$S$_{x}$ \cite{Coldea2019,Terashima2019},
 revealing multiple Fermi surface pockets and relatively large effective masses which are strongly suppressed by chemical pressure
 \cite{Coldea2021}.  Fermi surfaces expand via chemical pressure towards FeS
and a Lifshitz transition has been identified at the boundary of the nematic phase \cite{Coldea2019,Reiss2020,Terashima2019}.
 In contrast, quantum oscillations in FeSe under high pressure primarily detect low frequencies,
 suggesting a Fermi surface reconstruction in the presence of the magnetic order~\cite{Terashima2016}.
These observations highlight the importance of understanding
Fermi surfaces and quasiparticles in different regimes
to identify the relevant features for superconductivity.

In this study, we experimentally investigate the electronic behaviour of
the tetragonal high-$T_{\rm c}$ phase of FeSe$_{0.82}$S$_{0.18}$
under high magnetic fields up to 45\,T and applied pressures up to 22\,kbar.
The observed quantum oscillations probe directly the evolution of the Fermi surfaces  and
the quasiparticle behaviour with increasing pressure.
Our results reveal the expansion of the Fermi surface and
large and weakly varying cyclotron effective masses, in particular for the outer hole pockets.
In comparison, the superconducting critical temperature displays a gentle decrease
at low pressures  (around 7~K at $\sim 11$\,kbar),  followed by a significant threefold  increase   ($\sim 19$\,K  at $\sim 21$\,kbar).
Additionally,  the high-pressure phase harbours broad superconducting transition widths and larger residual resistivity
and we detect unusual disparity between the small and large angle scattering.
These findings reveal a complex normal behaviour
and suggest the involvement of additional pairing channels to stabilize
the high-$T_{\rm c}$ phase under pressure.

\begin{figure}[htbp]
	\centering
	\includegraphics[trim={0cm 0cm 0cm 0cm}, width=0.8\linewidth,clip=true]{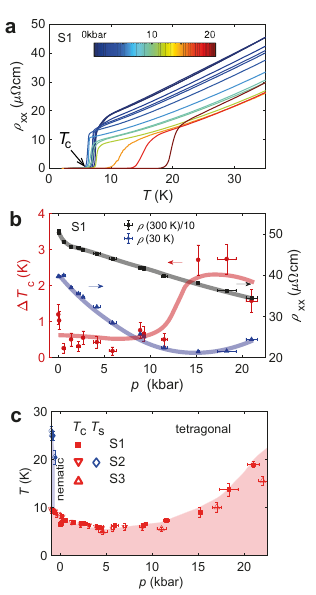}
\caption{{\bf Transport behaviour of  FeSe$_{0.82}$S$_{0.18}$ under pressure.}
(a) The temperature dependence of resistivity for the sample S1 for different applied pressures up to 21\,kbar.
$T_{\rm c}$ is defined as the  offset superconducting transition temperature (extrapolated towards the zero resistivity state).
		(b) The superconducting transition width, $\Delta T_{\rm c} = T_{\rm on} - T_{\rm off}$ (red circles),
and resistivity at 300\,K (black squares, scaled by a factor of 0.1), and at 30\,K (blue triangles) as a function of pressure. Solid lines are guide to the eye.
(c) The temperature-pressure phase diagram of FeSe$_{0.82}$S$_{0.18}$ for different single crystals.
Sample S1 (solid squares) and  S3 (open up triangles) are in the tetragonal phase at ambient pressure.
Sample S2 (open down triangles for $T_{\rm c}$ and open diamonds for the nematic transition, $T_{\rm s}$) is just inside the nematic phase
corresponding to a negative pressure of  -1\,kbar, as shown in Supplementary Fig.~1 
in the SM.
}
	\label{fig1}
\end{figure}

{\bf Results}

{\bf Temperature dependence of resistivity with applied pressure.}
Fig.~\ref{fig1}(a) shows the temperature dependence of the longitudinal resistivity, $\rho_{\rm xx}$, of a
 tetragonal sample S1 under various applied pressures.
At ambient pressure,  sample S1 displays a sharp superconducting transition.
In contrast, sample S2 shows an additional weak anomaly at $T_{\rm s} \sim 26$~K, which is
 quickly suppressed at low pressures ($\sim 1$\,kbar)
    (see the first derivative in  Supplementary Fig.~1  
  in the Supplemental Materials (SM))
  \cite{Coldea2021}.
Different samples have similar values of $T_{\rm c}\sim 7(1)$~K in the tetragonal phase,
which remains relatively unchanged up to 11\,kbar, as shown in Figs.~\ref{fig1}(c).
Interestingly, at high pressures the superconductivity is enhanced by up to a factor of 3
towards 19\,K at 21\,kbar for sample S1
 (see also Supplementary Fig.~5 
 and Supplementary Fig.~7 in the SM).  
 Furthermore, the width of the superconducting transition broadens at higher pressures, both in temperature and magnetic field,
 despite the absence of any competing phases in this regime
 (see Fig.~\ref{fig1}(a,b), Supplementary Fig.~5(b), and Supplementary Fig.~12 in the SM).  
 Notably, comparable $T_{\rm c} \sim 20$~K  is also detected in FeSe under a similar applied pressure,
but in the vicinity of the magnetic phase \cite{Terashima2016}, as well as in Cu$_x$Fe$_{1-x}$Se,
in which the signatures of magnetism are washed away
by impurity scattering \cite{Zajicek2022Cupressure}.
Normally, applied pressure is expected to increase the bandwidth and decrease the resistivity,
as shown in Fig.~\ref{fig1}(b)
 and Supplementary Fig.~3 
 in the SM.
 However, in the low temperature regime $\sim$30\,K, the resistivity initially decreases with pressure, followed by
  a slight increase above 15\,kbar  (Fig.~\ref{fig1}(b) and Supplementary Fig.~2 in the SM). 
 These findings suggest that
  the high-$T_{\rm c}$ high-pressure phase harbours
  additional  scattering mechanisms that slightly enhance resistivity and broaden the superconducting-to-normal transition.

\begin{figure*}[htbp]
	\includegraphics[trim={0cm 0cm 0cm 0cm}, width=0.93\linewidth,clip=true]{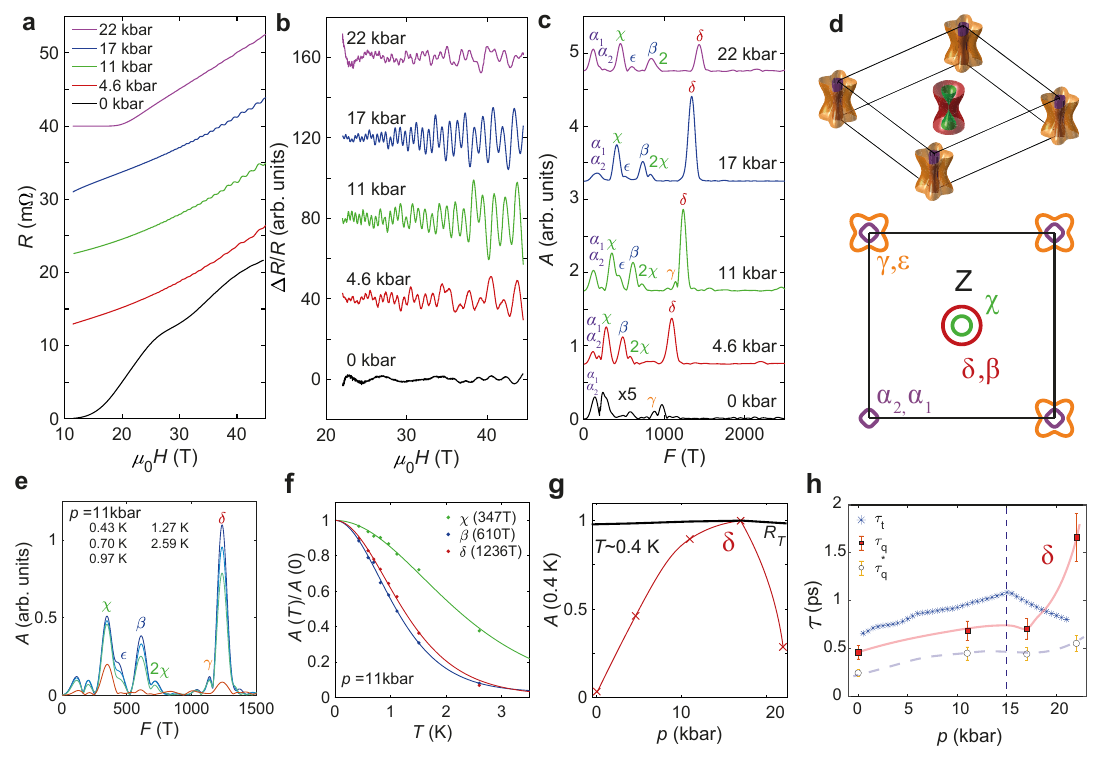}
	\caption{{\bf Evolution of quantum oscillations with pressure in FeSe$_{0.82}$S$_{0.18}$. }
		(a) Longitudinal resistivity measurements in magnetic fields up to 45\,T
at 0.4\,K for a range of pressures.
		(b) Shubnikov-de Haas oscillations obtained from extracting the background magnetoresistance
described by a 7$^{th}$ order polynomial over a window $ \mu_0 \Delta H= 22-45$\,T.
		(c) Fast Fourier transforms (FFT)  of the signal in (b) using a Hanning windowing function
having the different peaks and the corresponding harmonics labelled.
The ambient signal is amplified five times above 200\,T and curves in panels (a-c) are shifted vertically for clarity.
	(d) The proposed quasi-3D Fermi surface of FeSe$_{0.82}$S$_{0.18}$ and the slice at the top of the Brillouin zone, $k_{z} = \pi/c$.
		(e) FFT amplitude spectra at different temperatures for $p$ = 11\,kbar. Amplitude is fitted to the Lifshitz-Kosevich
formula \cite{Shoenberg1984} to determine the quasi-particle effective masses of each orbit in (f).
			(g) The pressure dependence of the FFT amplitudes of the main peaks at base temperature of $\sim 0.4$~K.
The solid line corresponds to the values of $R_{T}$ term normalized at the 17\,kbar value.
(h) The changes in the quantum, $\tau_{\rm q}$ and classical scattering time, $\tau_{\rm t}$, as a function of pressure.
 $\tau_{\rm q}$ is estimated from Dingle plots  at $\sim 0.4~$K by filtering the $\delta$ orbit
and  $\tau_{\rm q}^*$ is obtained from the subtracted signal $\Delta R$ (see Supplementary Fig.~11 in the SM). 
The classical time is estimated from the resistivity data in Fig.~ S2.
The solid and dashed lines are guides to the eye
and the vertical dashed line indicates the position at which the classical time reaches its largest value.
}
	\label{fig2}
\end{figure*}

{\bf The evolution of the Fermi surface with pressure.}
Fig.~\ref{fig2}a illustrates the longitudinal magnetoresistance up to $45$\,T for sample S3 at 0.4~K for pressures up to $22$\,kbar.
The high quality of the crystals and the relatively low upper critical fields allow for the observation of
 quantum oscillations arising from Landau quantization, which are superimposed on the magnetoresistance background.
The frequencies of quantum oscillations, determined by the Onsager relation $F_{k}=\frac{\hbar}{2\pi e}A_{k}$,
directly correspond to the extremal cross-sectional areas, $A_{k}$, of a Fermi surface \cite{Shoenberg1984}.
At ambient pressure, the magnetotransport data is primarily characterized by a low-frequency oscillation,
which has been previously linked to a potential Lifshitz transition at the boundary of the nematic phase   \cite{Coldea2019}.
As the pressure increases, the low frequency disappears and the spectrum is dominated by high frequency oscillations.
These frequencies, visible in the raw data (see Figs.~\ref{fig2}a and b), are attributed to larger Fermi surface sheets,
resembling those found in the tetragonal FeSe$_{1-x}$S$_{x}$ system
tuned by chemical pressure \cite{Coldea2019}.

Figures~\ref{fig2}b and c show the extracted oscillatory signal and the fast Fourier transform spectra which correspond
to the cross-sectional areas of the minimum and maximum orbits on different Fermi surface sheets.
With increasing pressure, all the frequencies increase in size linearly,
surpassing the expected enlargement of the Brillouin zone size
(see Supplementary Fig.~14 
in the SM), as seen in Fig.~\ref{fig3}d.
At ambient pressure,
the Fermi surface of FeSe$_{0.82}$S$_{0.18}$ is anticipated to contain two electron and two hole pockets,
 based on ARPES and quantum oscillations studies \cite{Coldea2021,Reiss2017}.
  However, DFT calculations tend to overestimate the size of the Fermi surface, requiring band shifts and
  renormalization to align with the experimental observations,
 as detailed in Supplementary Fig.~13  in the SM. 
 Based on the comparison between the data and different simulations,
 the  frequencies, $\beta$ and $\delta$,  are assigned to
the minimum and maximum of the outer hole band orbits,
whereas $\epsilon$ and $\gamma$
 are associated with the outer electron pocket, as shown in  Fig.~\ref{fig2}(d).
Additionally, the $\chi$ frequency corresponds to the maximal orbit of the 3D inner hole,
while the two lowest frequencies, $\alpha_1$ and $\alpha_2$
 are assigned to the small inner electron pocket \cite{Reiss2017}.
Interestingly, the FFT spectra obtained outside the nematic phase exhibit striking similarities between
FeSe$_{0.82}$S$_{0.18}$  at 4.6\,kbar and FeSe$_{0.81}$S$_{0.19}$ at ambient pressure
(see Supplementary Fig.~10 
 in the SM) \cite{Coldea2019},
implying a similar evolution of the Fermi surface in the tetragonal phase.

In order to gain insights into the intricate alterations in the Fermi surface topography of FeSe$_{0.82}$S$_{0.18}$ under pressure,
 we employ a tight-binding-like decomposition
 of the Fermi vector in cylindrical coordinates.
 This approach takes into account the allowed symmetries \cite{Bergemann2000},
while ensuring charge compensation, as detailed in Supplementary Fig.~8
 in the SM.
Since  the interlayer compressibility of FeSe under pressure
 increases by a factor of 2.5 compared to the in-plane ($ab$) plane value
 \cite{Margadonna2009},
 it implies the presence
of soft Se-Se interlayer interactions.
This, in turn, influences the interplane distortions of the Fermi surface, which can be quantified
 by the $k_{10}\sim 2\pi/c$
term (see Supplementary Table~1 and Supplementary Fig.~8(d) in the SM).
Through a comparison between experimental data and these simulations,
we observe that the Fermi surface pockets expand, and the outer hole pocket becomes increasingly two-dimensional
as pressure increases.
Additionally, the pockets that are most sensitive to interplane distortion are the small inner pockets
 (Figs.~S8(c) and (d) in the SM).  
 When comparing with the chemical pressure tuning from FeSe to FeS,
the hole cylinders  increase in size and become
less warped compared to the quasi-two-dimensional electron pockets
\cite{Terashima2019}.
Thus, both chemical and applied pressure in FeSe$_{1-x}$S$_{x}$
mainly result in an expansion of the Fermi surface  \cite{Terashima2019,Coldea2019,Coldea2021,Reiss2020}.
However, the enhancement of superconductivity is only achieved through the application of physical pressure.

{\bf Quasiparticle effective masses.}
The damping effects on the quantum oscillations amplitude,
caused by temperature and magnetic field, provide direct information about the cyclotron effective mass
(Fig.~\ref{fig2}(e) and (f)) and scattering times of
quasiparticles  (Fig.~\ref{fig2}(h) and Supplementary Fig.~11 in the SM). 
 Fig.~\ref{fig2}(f) shows the temperature dependence of the amplitude for each orbit at $p$ = 11\,kbar
 (other pressures are in Supplementary Fig.~9
 in the SM), from which the effective masses are extracted using the thermal damping term of the Lifshitz-Kosevich
formula  \cite{Shoenberg1984}.
The temperature dependence of the amplitudes of the $\beta$ and $\delta$ hole orbits decrease more rapidly,
 indicating a larger effective mass ($\sim 4$~$m_{\rm e}$), compared to the $\chi$ hole orbit which has a lighter mass  ($\sim 2$~$m_{\rm e}$).
 These results are in good agreement with the values of the inner and outer hole pockets obtained from ARPES studies
  \cite{Reiss2017}.
 The effective masses of the hole pockets show relatively weak variations
 with applied pressure, despite the significant changes
  in $T_{\rm c}$ up to 22\,kbar, as shown in Fig.~\ref{fig3}(f).
The effective mass of the $\chi$ orbit slightly decreases, while that of the
 $\beta$ orbit increases.
 However, the effective mass associated with the $\epsilon$
 electron pocket has a higher level of uncertainty due to its weak signal.
Overall, the effective masses in  FeSe$_{0.82}$S$_{0.18}$ under pressure are much larger than those detected in FeS,
where the largest effective mass is  $\sim 2.4$~$m_{\rm e}$  \cite{Terashima2019}.
The effective mass can be enhanced by both electron-electron correlations and
electron-phonon coupling as $m^*$ = $m_b (1+\lambda_{\rm el-ph}) (1+\lambda_{\rm e-e}$),
where $m_b$ is the band mass \cite{Shoenberg1984}.
The electron-phonon coupling of FeSe is predicted to slightly increase with applied pressure
(from a maximum value $\lambda_{\rm el-ph} = 0.98$ at 0\,kbar to 1.159 at 26\,kbar) \cite{Mandal2014}
and it could lead to small increase in the effective mass  of 0.4-0.7 $m_{\rm e}$,
falling within the range of values measured for the hole orbits
(Fig.~\ref{fig3}f).

\begin{figure}[htbp]
	\centering
\includegraphics[trim={0cm 0cm 0cm 0cm},width=1\linewidth,clip=true]{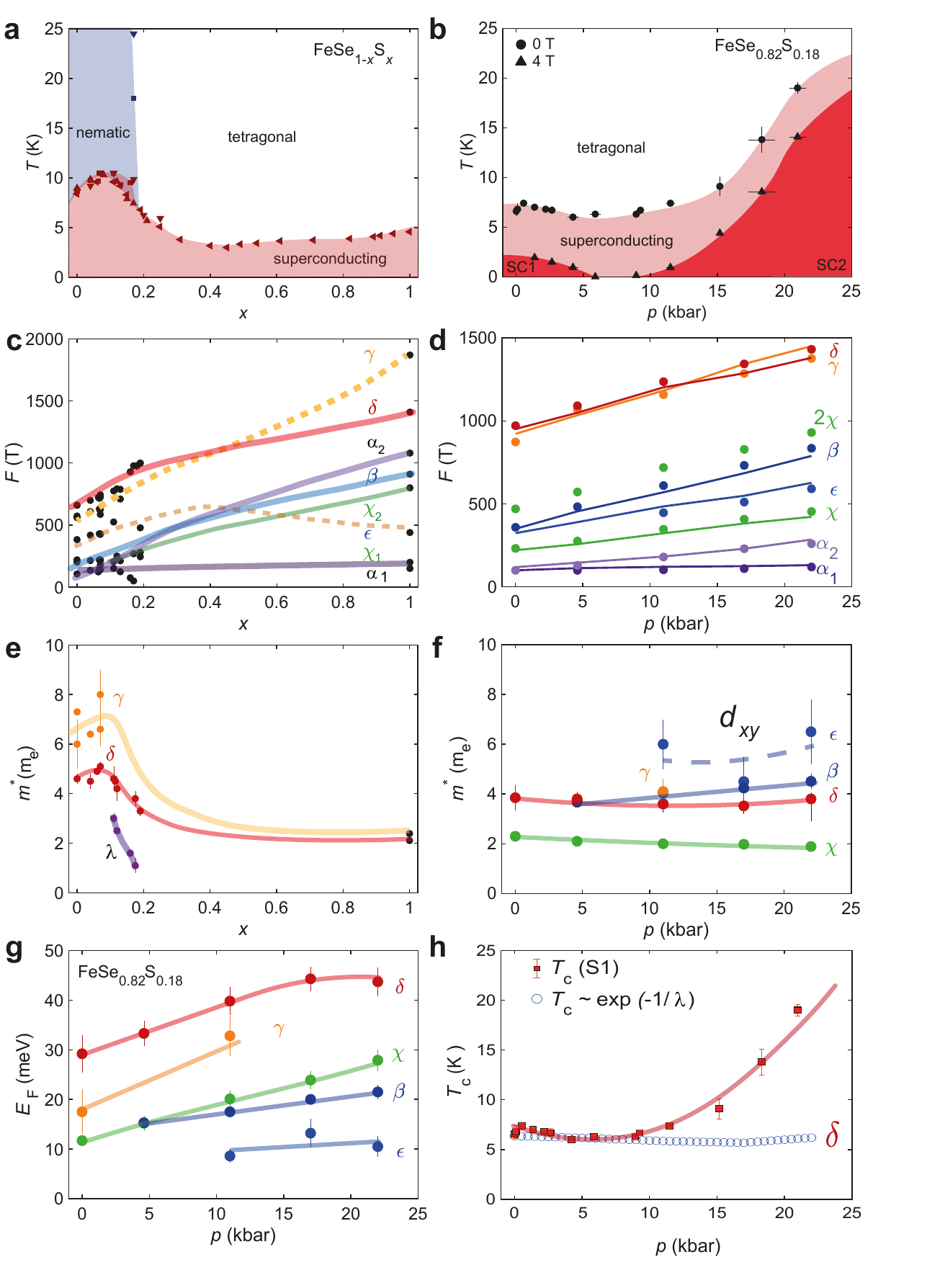}
	\caption{{\bf Comparison between the physical and chemical pressure effects.}
		(a) The phase diagram of FeSe$_{1-x}$S$_{x}$ tuned by the sulphur substitution $x$ (after Ref.~\cite{Coldea2019})
and (b)  the $p-T$ phase diagram of FeSe$_{0.82}$S$_{0.18}$  up to 22\,kbar, in 0\,T (as in Fig.~\ref{fig1}(g)) and in 4\,T field (darker areas),
which separates two regimes of superconductivity (SC1 and SC2).
 Frequencies extracted from quantum oscillations for  (c) FeSe$_{1-x}$S$_{x}$  versus $x$ (after Ref.~\cite{Coldea2019}) and \cite{Terashima2019})
 and (d) FeSe$_{0.82}$S$_{0.18}$ versus pressure.
 The corresponding quasiparticle effective masses, $m^*$,  are in  (e) and (f), respectively.
 Solid and dashed lines are guide to the eye.
 (g) The calculated Fermi energy, $E_{\rm F}$ for FeSe$_{0.82}$S$_{0.18}$
 assuming parabolic band dispersion using the values in (d) and (f).
(h) The pressure dependence of the $T_{\rm c}$ for S1 (solid square) and a rough estimate
assuming a BCS dependence which depends only on the density of states of a quasi-2D pocket
for the $\delta$ pocket
(open circles).
}	
	\label{fig3}
\end{figure}

{\bf Scattering under pressure.}
Scattering mechanisms
 under pressure can be assessed from our measurements
  at the lowest temperature, as shown in Fig.~\ref{fig2}(h).
  The first mechanism is the transport classical scattering time, $\tau_{\rm t}$,
  which primarily accounts for large-angle scattering events resulting in significant momentum changes.
  This can be estimated from the zero temperature resistivity, $\rho_0$,
and considering that  the  carrier density,  $n$,
 increases with pressure (Supplementary Fig.~8(f) in the SM). 
 Secondly, the quantum
lifetime, $\tau_{\rm q}$, which encompasses all scattering events, can be extracted from Dingle plots.
These reflect the damping of the quantum oscillation amplitude
of the $\delta$ pocket as a function of the
 inverse magnetic field at constant temperature (see Supplementary Fig.~11 in the SM). 
We find that $\tau_{\rm q}$ varies between 0.5-0.7~ps  up to  17\,kbar
(which corresponds to a mean free path close to $\sim $450~\AA),
similar to values found in the tetragonal phase of
 FeSe$_{0.89}$S$_{0.11}$ \cite{Reiss2020} or FeSe  \cite{Watson2015b}.
  It is worth noting that the values of
$\tau_{\rm q}$ is highly sensitive
 to the background magnetoresistance
 and any potential interference from the $\gamma$ pocket at 4.6~kbar,
 as shown in Supplementary Fig.~11 in the SM. 
 Moreover, $\tau_{\rm t}$ is larger than $\tau_{\rm q}$, similar to FeSe \cite{Farrar2022},
and they both increase  with applied pressure up to 15\,kbar (Fig.~\ref{fig2}(h)).
At the highest pressures $\sim $22~kbar,
a significant discrepancy arises between the two scattering times,
as the $\tau_{\rm t}$ decreases whereas $\tau_{\rm q}$ increases, as shown in Fig.~\ref{fig2}(h).
Since the original impurity concentration remains constant under pressure,
any additional changes in the scattering times indicate modifications in the electronic phase
induced at high pressure.

{\bf Discussion.}
The evolution of the electronic structure and effective masses of FeSe$_{0.82}$S$_{0.18}$ under applied hydrostatic pressure
can be compared  to the effects of chemical substitution
in FeSe$_{1-x}$S$_{x}$ \cite{Coldea2019}, as illustrated in Fig.~\ref{fig3}(c) and (d).
The observed frequencies exhibit a linear expansion with increasing pressure,
 similar to trends observed in the tetragonal phases of FeSe$_{1-x}$S$_x$ \cite{Coldea2019,Reiss2020}.
 This behaviour rules out the reconstruction of the Fermi surface at high pressures,  in contrast
to FeSe under applied pressure where only a very small pocket was detected
 \cite{Terashima2016}.
 The expansion of the Fermi surface leads to an increase in
the carrier density,  $n$,  (according to the Luttinger's theorem \cite{Luttinger1960}),
which doubles in value (Supplementary Fig.~8(f) in the SM).
Moreover, the extracted values of the Fermi liquid
coefficient, $A$, decrease with pressure, before flattening
off above 15~kbar (see Supplementary Fig. 6 in the SM),
which also reflects the
expansion of the Fermi surface.
Meanwhile, the value of $T_{\rm c}$ increases by a factor of 3 over
the same pressure range  (Fig.~\ref{fig3}g).
It is worth noting that FeS has an even larger carrier density,
yet its $T_{\rm c}$ remains relatively small around 5\,K \cite{Terashima2016FeS}.
Thus, having a large Fermi surface alone is not sufficient to
explain the changes in superconductivity.

The effective masses of the different orbits of the Fermi surface of FeSe$_{0.82}$S$_{0.18}$ remain almost unchanged under pressure,
in particular, the value for the $\delta$ orbit is only slightly lighter in the tetragonal phase as compared to the nematic phase \cite{Coldea2019}.
The effective mass for a quasi-two dimensional system
is related to the density of states at the Fermi level, $g$($E_{\rm F})  \sim m^*/(\pi \hbar^2)$.
Thus,   it is expected that higher $T_{\rm c}$  in FeSe$_{0.82}$S$_{0.18}$ could be associated with
the larger effective masses, in comparison with FeS, which has lower $T_{\rm c}$ and lighter masses.
These experimental observations are further supported by estimated
density of states based on the adjusted Fermi surfaces based on DFT calculations,
as shown in Supplementary Fig.~14(f) in the SM. 
 One parameter which influences density of states is the chalcogen height above the Fe planes, $h$,
(see Supplementary Fig.~14 in the SM)
which increases under applied pressure
 ($\sim 1.45 $~\AA~ at 18.2~kbar in FeSe$_{0.80}$S$_{0.20}$
 \cite{Tomita2015}, but
 decreases for
FeS at ambient pressure (1.269~\AA~) \cite{Terashima2016}). 
However, the observed variation in effective masses of the hole pockets 
is minimal under applied pressure (see Fig.~\ref{fig3}(f)).
Therefore,  the significant enhancement of $T_{\rm c}$ at high pressure requires
an additional ingredient to boost the superconducting pairing (see Fig.~\ref{fig3}(h)).

In order to enhance the superconducting transition temperature $T_{\rm c}$ by a factor of three, either
the contribution to $g$($E_{\rm F}$) from other pockets increases, or an additional pairing interaction
becomes operative under pressure (SC2 regime in Fig.~\ref{fig3}(b)).
One potential mechanism involves shifting the
hole pocket with $d_{xy}$ orbital character
close to the Fermi level,  as found in  FeSe$_{1-x}$Te$_{x}$  \cite{Morfoot2022}.
The presence of a third hole pocket is predicted to enhance the spin fluctuations
in the $d_{xy}$ channel increasing the pairing interaction \cite{Yamakawa2017}.
However, our quantum oscillations do not detect a third hole pocket.
 Alternatively, the heavy electron pockets,
whose orbits $\epsilon, \gamma$ have regions with $d_{xz/yz}$ and  $d_{xy}$ orbital character  (Fig.~\ref{fig2}(c))
could become
even more correlated under pressure
 and potentially contribute to the missing density of states.
 Interestingly,  systems containing only electron pockets, like electron-doped intercalated FeSe systems  \cite{Xiao2022}
 or the FeSe monolayer  grown on  SrTiO$_3$
 have a higher $T_{\rm c}$ exceeding  40~K.
 Thus, pairing channels involving  electron pockets could be potentially enhanced
under high pressures \cite{Lee2018,Kreisel2020}.

The increase  in the
density of states  at the Fermi level, $g$($E_{\rm F}$) under pressure,
can contribute to the stabilization of magnetically ordered states based on the Stoner criteria
 ($I \cdot g$($E_{\rm F})>1$,  where $I$ is the Stoner coefficient).
The density of states can be enhanced with increasing chalcogen height,
under applied pressure   \cite{Okabe2010}, or  by the isoelectronic substitution with Te \cite{Kumar2012},
which create conditions to stabilize magnetically ordered phases.
 The nearest-neighbour Coulomb repulsion and electronic correlations in FeSe
would generally decrease with pressure and lead to enhanced spin fluctuations
\cite{Scherer2017}.
  The nesting between large and isotropic pockets with large Fermi energy (50~meV  at 22~kbar as shown in Fig.~\ref{fig3}(g))
 in the presence of a larger density of states could induce a spin-density wave  \cite{Chubukov2016}.
   While in  our system, the density of states increases under pressure,
  we do not detect any clear anomalies in resistivity associated to a spin-density wave up to 22~kbar,
  which was proposed  to occur around $50$~kbar \cite{Matsuura2017}.
  On the other hand, FeSe shows a magnetic phase under pressure
  and quantum oscillations only detect a very small pocket with light effective mass,
 potentially reflecting the Fermi surface reconstruction \cite{Terashima2016}.           
   Unusually, NMR studies in FeSe$_{1-x}$S$_x$
   find weak and potentially short-range spin fluctuations in the tetragonal phase at high pressures  \cite{Rana2023},
   different from  strong stripe-type AFM spin fluctuations were found inside the nematic phase \cite{Wiecki2018}.

The high-pressure high-$T_{\rm c}$ phase exhibits distinct electronic signatures
 compared to the nematic phase of FeSe$_{1-x}$S$_x$ systems.
Firstly, there is no correlation between the strong enhancement of $T_{\rm c}$ and  the effective mass
of the $\delta$ hole orbit (Figs.~\ref{fig3}(b, f)), contrary to the trends
observed inside the nematic phase (Fig.~\ref{fig3}(a, e))
\cite{Bristow2020,Reiss2020,Coldea2021}.
 Secondly, there are significant differences in spin fluctuations
which are mainly detected only  inside the nematic phase \cite{Wiecki2018},
but could transform into short-range correlations in the tetragonal phase at high pressures
\cite{Kuwayama2021,Rana2023}.
Thirdly, there is a notable difference between the quantum and classical scattering times, and the amplitude of the $\delta$ hole orbit is significantly suppressed
(Fig.~\ref{fig2}(a) and (g), and Supplementary Fig.~4 in the SM). 
Lastly, the superconducting transition broadens both in temperature and field,
accompanied by a reduction in the oscillatory signal relative to the background magnetoresistance (Supplementary Fig.~5 in the SM). 

Interestingly,  superconductivity is enhanced under pressure for various FeSe$_{1-x}$S$_x$ systems,
 even in the presence of Cu impurities \cite{Zajicek2022Cupressure}.
Thus, the high-$T_{\rm c}$ phase in FeSe$_{1-x}$S$_x$ demonstrates remarkable robustness
to impurities and substitutions, indicating a potential non-sign changing pairing mechanism
and reduced sensitivity to any long-range magnetic order.
At very high pressures (50~kbar) in FeSe$_{0.89}$S$_{0.11}$ both the superconductivity and metallicity are lost
  \cite{Reiss2022},  and the diamagnetic signal is suppressed \cite{Yip2017},
triggered by the development of uniaxial pressure effects.
 These effects may reflect a complex high-pressure phase
having  different conductivity channels at high pressures \cite{Ayres2022},
or be dominated by strong magnetic or superconducting fluctuations \cite{Gati2019,Okabe2010}.
Alternatively, the distribution of chalcogen ions outside the conducting
planes could create conditions to develop
electronic inhomogeneities and quantum Griffiths phases
\cite{Reiss2021} or potential short-range stripe patterns of
 incipient charge order correlations
\cite{Walker2023}.

The electron-phonon coupling could also play a role in stabilizing higher superconducting
phase in the high pressure in FeSe$_{1-x}$S$_x$,
similar to other high-$T_{\rm c}$ iron-based superconductors.
Raman studies in the electron-doped intercalated  (Li,Fe)OHFeSe
suggest that lattice-induced orbital fluctuations are responsible for pairing
\cite{Xiao2022}, whereas in the monolayer FeSe on
interfacial coupling with the oxygen optical phonons in SrTiO3 could play an important role \cite{Lee2018}.
The lattice itself is sensitive to pressure-induced changes,
 and the chalcogen height  influences the out-of-plane phonon mode, $A_{1g}$  and the
 electron-phonon coupling \cite{Mandal2014}.
 In such cases, local deformation potentials can result in strong coupling between electrons and the lattice,
 forming polarons \cite{Mandal2014}
which could condense as pairs in bipolarons \cite{Zhao1997}.
In the presence of electronic correlations,
the electron-phonon coupling could be further enhanced in FeSe
  \cite{Gerber2017,Ding2022}.
Thus, the route to induce a high-$T_{\rm c}$ phase under applied pressure phase could be promoted by
the formation of a novel electronic phase,  with potentially new quasiparticles, in which lattice effects can influence the density of states,
the strength of electronic correlations and it can lead to unconventional scattering.

{\bf Methods}

{\it Single crystals.}
Single crystals of FeSe$_{0.82}$S$_{0.18}$ were grown by the KCl/AlCl$_3$ chemical vapor transport method, as reported previously \cite{Bohmer2013,Bohmer2016g}.
More than 10 crystals were screened at ambient pressure and were found to have large residual resistivity ratios (RRR) up to 25
between room temperature and the onset of superconductivity. Crystals from
the same batch were previously used in quantum oscillations and ARPES studies \cite{Reiss2017,Coldea2019}.
All samples measured were from the same batch and EDX measurements measured the composition to contain a sulphur content of
$x$ = 0.18(1), which places the batch in the vicinity of the nematic end point \cite{Coldea2021,Reiss2017}.

{\it High-pressure studies in magnetic fields.}
High magnetic field measurements up to 45\,T at ambient pressure and under hydrostatic pressure were performed using
the hybrid magnet dc facility at the NHMFL in Tallahassee, FL, USA on sample S3.
Pressures up to 22\,kbar were generated using a MP35N piston cylinder cell,
using Daphne Oil 7575 as pressure medium. The pressure inside the cell was determined by means of ruby fluorescence at low temperatures where quantum oscillations were observed.
Magnetotransport and Hall effect measurements under pressure using a 5-contact configuration were carried out on samples S1 and S2 in low fields up to 16\,T in Oxford
using the Quantum Design PPMS and an ElectroLab High Pressure Cell, using Daphne Oil 7373 which ensures hydrostatic conditions up to about 21\,kbar.
The pressure inside this cell was determined via the superconducting transition temperature of Sn after cancelling the remnant field in the magnet.
The magnetic field was applied along the crystallographic $c$ axis for all samples. A maximum current of up to 2\,mA flowing in the conducting tetragonal $ab$ plane was used.
The cooling rate was kept 0.5\,K/min below 100\,K to reduce
the thermal lag between the sample and the pressure cell.

{\it Quantum oscillations}
The amplitude of the quantum oscillations is affected by different damping terms
given by the Lifshitz-Kosevich equation \cite{Shoenberg1984}.
The thermal damping term, $R_T =\frac{X}{\mathrm{sinh}(X)} $, where $X = \frac{2 \pi^2 k_{\rm B} T \, p m^*}{e \hbar B}$,
enables to extract the quasiparticle effective masses $m^*$, and
the Dingle term, $R_{D}= \exp(-\frac{\pi p m^*}{eB \tau_{\rm q}})$,
 allows to estimate the quantum scattering rate $\tau_{\rm q}$
(or the mean free path $\ell$ using the equivalent expression
$R_D =\exp(-1140 \frac{\sqrt{F}}{\ell B})$ \cite{Carrington2011}.
Other damping factors which have exponential forms could be caused
by small random sample inhomogeneities, magnetic field inhomogeneities
and additional damping within the vortex regime \cite{Rourke2010}.

\vspace{0.5cm}

{\bf Acknowledgements}
We thank Roemer Hinlopen for the development of the software to model the Fermi surface, and Matthew Bristow  for computational support.
We thank Steve Simon and Sean Hartnoll for useful discussions.
This work was mainly supported by EPSRC (EP/I004475/1, EP/I017836/1).
A.A.H. acknowledges the financial support of the Oxford Quantum Materials Platform Grant (EP/M020517/1).
The research was partly funded by the Oxford Centre for Applied Superconductivity at Oxford University.
We also acknowledge financial support from the John Fell Fund of Oxford University.
A portion of this work was performed at the National High Magnetic Field Laboratory,
which is supported by National Science Foundation Cooperative Agreement No. DMR-1157490 and the State of Florida.
 Z.Z. acknowledges financial support from EPSRC Studentships
EP/N509711/1 and EP/R513295/1.
J.C.A.P. acknowledges the support of St. Edmund Hall, University of Oxford,
through the Cooksey Early Career Teaching and Research
Fellowship. The DFT calculations were performed on the
University of Oxford Advanced Research Computing Service.
AIC
is grateful to KITP for hospitality and this research was supported
in part by the National Science Foundation under
Grants No. NSF PHY-1748958 and PHY-2309135.
 A.I.C. acknowledges EPSRC Career
Acceleration Fellowship EP/I004475/1.

{\bf Data availability.}
All data are available in the manuscript or the supplementary materials.
The data that support the findings of this study will be available through the
 open access data archive at the University of Oxford (ORA) (https://doi.org/xxx).
Additional information about the data will be made available
from the corresponding author upon reasonable request.

{\bf Author contributions}
P.R., Z.Z., D.G. and A.I.C. performed transport experiments under pressure.
Z.Z., P.R., A.I.C. analysed the transport and quantum oscillation data.
A.A. synthesized single crystals. J.C.A.P
performed DFT calculations.
Y.S. performed the Dingle analysis.
A.I.C. and Z.Z. wrote the paper with inputs from all authors.
A.I.C. supervised the projects.

{\bf Additional information}
Supplementary Information accompanies this paper at doi:xxx

{\bf Competing interests}
The authors declare no competing financial interests.

\bibliography{FeSeS_pressure_bib_april23}

\begin{thebibliography}{64}%
\makeatletter
\providecommand \@ifxundefined [1]{%
 \@ifx{#1\undefined}
}%
\providecommand \@ifnum [1]{%
 \ifnum #1\expandafter \@firstoftwo
 \else \expandafter \@secondoftwo
 \fi
}%
\providecommand \@ifx [1]{%
 \ifx #1\expandafter \@firstoftwo
 \else \expandafter \@secondoftwo
 \fi
}%
\providecommand \natexlab [1]{#1}%
\providecommand \enquote  [1]{``#1''}%
\providecommand \bibnamefont  [1]{#1}%
\providecommand \bibfnamefont [1]{#1}%
\providecommand \citenamefont [1]{#1}%
\providecommand \href@noop [0]{\@secondoftwo}%
\providecommand \href [0]{\begingroup \@sanitize@url \@href}%
\providecommand \@href[1]{\@@startlink{#1}\@@href}%
\providecommand \@@href[1]{\endgroup#1\@@endlink}%
\providecommand \@sanitize@url [0]{\catcode `\\12\catcode `\$12\catcode
  `\&12\catcode `\#12\catcode `\^12\catcode `\_12\catcode `\%12\relax}%
\providecommand \@@startlink[1]{}%
\providecommand \@@endlink[0]{}%
\providecommand \url  [0]{\begingroup\@sanitize@url \@url }%
\providecommand \@url [1]{\endgroup\@href {#1}{\urlprefix }}%
\providecommand \urlprefix  [0]{URL }%
\providecommand \Eprint [0]{\href }%
\providecommand \doibase [0]{https://doi.org/}%
\providecommand \selectlanguage [0]{\@gobble}%
\providecommand \bibinfo  [0]{\@secondoftwo}%
\providecommand \bibfield  [0]{\@secondoftwo}%
\providecommand \translation [1]{[#1]}%
\providecommand \BibitemOpen [0]{}%
\providecommand \bibitemStop [0]{}%
\providecommand \bibitemNoStop [0]{.\EOS\space}%
\providecommand \EOS [0]{\spacefactor3000\relax}%
\providecommand \BibitemShut  [1]{\csname bibitem#1\endcsname}%
\let\auto@bib@innerbib\@empty
\bibitem [{\citenamefont {Drozdov}\ \emph {et~al.}(2019)\citenamefont
  {Drozdov}, \citenamefont {Kong}, \citenamefont {Minkov}, \citenamefont
  {Besedin}, \citenamefont {Kuzovnikov}, \citenamefont {Mozaffari},
  \citenamefont {Balicas}, \citenamefont {Balakirev}, \citenamefont {Graf},
  \citenamefont {Prakapenka}, \citenamefont {Greenberg}, \citenamefont
  {Knyazev}, \citenamefont {Tkacz},\ and\ \citenamefont
  {Eremets}}]{Drozdov2019}%
  \BibitemOpen
  \bibfield  {author} {\bibinfo {author} {\bibfnamefont {A.~P.}\ \bibnamefont
  {Drozdov}}, \bibinfo {author} {\bibfnamefont {P.~P.}\ \bibnamefont {Kong}},
  \bibinfo {author} {\bibfnamefont {V.~S.}\ \bibnamefont {Minkov}}, \bibinfo
  {author} {\bibfnamefont {S.~P.}\ \bibnamefont {Besedin}}, \bibinfo {author}
  {\bibfnamefont {M.~A.}\ \bibnamefont {Kuzovnikov}}, \bibinfo {author}
  {\bibfnamefont {S.}~\bibnamefont {Mozaffari}}, \bibinfo {author}
  {\bibfnamefont {L.}~\bibnamefont {Balicas}}, \bibinfo {author} {\bibfnamefont
  {F.~F.}\ \bibnamefont {Balakirev}}, \bibinfo {author} {\bibfnamefont {D.~E.}\
  \bibnamefont {Graf}}, \bibinfo {author} {\bibfnamefont {V.~B.}\ \bibnamefont
  {Prakapenka}}, \bibinfo {author} {\bibfnamefont {E.}~\bibnamefont
  {Greenberg}}, \bibinfo {author} {\bibfnamefont {D.~A.}\ \bibnamefont
  {Knyazev}}, \bibinfo {author} {\bibfnamefont {M.}~\bibnamefont {Tkacz}},\
  and\ \bibinfo {author} {\bibfnamefont {M.~I.}\ \bibnamefont {Eremets}},\
  }\bibfield  {title} {\bibinfo {title} {{Superconductivity at 250~K in
  lanthanum hydride under high pressures}},\ }\href
  {https://doi.org/10.1038/s41586-019-1201-8} {\bibfield  {journal} {\bibinfo
  {journal} {Nature}\ }\textbf {\bibinfo {volume} {569}},\ \bibinfo {pages}
  {528} (\bibinfo {year} {2019})}\BibitemShut {NoStop}%
\bibitem [{\citenamefont {Medvedev}\ \emph {et~al.}(2009)\citenamefont
  {Medvedev}, \citenamefont {McQueen}, \citenamefont {Troyan}, \citenamefont
  {Palasyuk}, \citenamefont {Eremets}, \citenamefont {Cava}, \citenamefont
  {Naghavi}, \citenamefont {Casper}, \citenamefont {Ksenofontov}, \citenamefont
  {Wortmann},\ and\ \citenamefont {Felser}}]{Medvedev2009}%
  \BibitemOpen
  \bibfield  {author} {\bibinfo {author} {\bibfnamefont {S.}~\bibnamefont
  {Medvedev}}, \bibinfo {author} {\bibfnamefont {T.~M.}\ \bibnamefont
  {McQueen}}, \bibinfo {author} {\bibfnamefont {I.~A.}\ \bibnamefont {Troyan}},
  \bibinfo {author} {\bibfnamefont {T.}~\bibnamefont {Palasyuk}}, \bibinfo
  {author} {\bibfnamefont {M.~I.}\ \bibnamefont {Eremets}}, \bibinfo {author}
  {\bibfnamefont {R.~J.}\ \bibnamefont {Cava}}, \bibinfo {author}
  {\bibfnamefont {S.}~\bibnamefont {Naghavi}}, \bibinfo {author} {\bibfnamefont
  {F.}~\bibnamefont {Casper}}, \bibinfo {author} {\bibfnamefont
  {V.}~\bibnamefont {Ksenofontov}}, \bibinfo {author} {\bibfnamefont
  {G.}~\bibnamefont {Wortmann}},\ and\ \bibinfo {author} {\bibfnamefont
  {C.}~\bibnamefont {Felser}},\ }\bibfield  {title} {\bibinfo {title}
  {{Electronic and magnetic phase diagram of $\beta$-Fe$_{1.01}$Se with
  superconductivity at 36.7 K under pressure}},\ }\href
  {https://doi.org/10.1038/nmat2491} {\bibfield  {journal} {\bibinfo  {journal}
  {Nat. Mater.}\ }\textbf {\bibinfo {volume} {8}},\ \bibinfo {pages} {630}
  (\bibinfo {year} {2009})}\BibitemShut {NoStop}%
\bibitem [{\citenamefont {Kothapalli}\ \emph {et~al.}(2016)\citenamefont
  {Kothapalli}, \citenamefont {B{\"{o}}hmer}, \citenamefont {Jayasekara},
  \citenamefont {Ueland}, \citenamefont {Das}, \citenamefont {Sapkota},
  \citenamefont {Taufour}, \citenamefont {Xiao}, \citenamefont {Alp},
  \citenamefont {Bud'ko}, \citenamefont {Canfield}, \citenamefont {Kreyssig},\
  and\ \citenamefont {Goldman}}]{Kothapalli2016}%
  \BibitemOpen
  \bibfield  {author} {\bibinfo {author} {\bibfnamefont {K.}~\bibnamefont
  {Kothapalli}}, \bibinfo {author} {\bibfnamefont {A.~E.}\ \bibnamefont
  {B{\"{o}}hmer}}, \bibinfo {author} {\bibfnamefont {W.~T.}\ \bibnamefont
  {Jayasekara}}, \bibinfo {author} {\bibfnamefont {B.~G.}\ \bibnamefont
  {Ueland}}, \bibinfo {author} {\bibfnamefont {P.}~\bibnamefont {Das}},
  \bibinfo {author} {\bibfnamefont {A.}~\bibnamefont {Sapkota}}, \bibinfo
  {author} {\bibfnamefont {V.}~\bibnamefont {Taufour}}, \bibinfo {author}
  {\bibfnamefont {Y.}~\bibnamefont {Xiao}}, \bibinfo {author} {\bibfnamefont
  {E.}~\bibnamefont {Alp}}, \bibinfo {author} {\bibfnamefont {S.~L.}\
  \bibnamefont {Bud'ko}}, \bibinfo {author} {\bibfnamefont {P.~C.}\
  \bibnamefont {Canfield}}, \bibinfo {author} {\bibfnamefont {A.}~\bibnamefont
  {Kreyssig}},\ and\ \bibinfo {author} {\bibfnamefont {A.~I.}\ \bibnamefont
  {Goldman}},\ }\bibfield  {title} {\bibinfo {title} {{Strong cooperative
  coupling of pressure-induced magnetic order and nematicity in FeSe}},\
  }\href@noop {} {\bibfield  {journal} {\bibinfo  {journal} {Nat. Comm.}\
  }\textbf {\bibinfo {volume} {7}},\ \bibinfo {pages} {12728} (\bibinfo {year}
  {2016})}\BibitemShut {NoStop}%
\bibitem [{\citenamefont {Sun}\ \emph {et~al.}(2016)\citenamefont {Sun},
  \citenamefont {Matsuura}, \citenamefont {Ye}, \citenamefont {Mizukami},
  \citenamefont {Shimozawa}, \citenamefont {Matsubayashi}, \citenamefont
  {Yamashita}, \citenamefont {Watashige}, \citenamefont {Kasahara},
  \citenamefont {Matsuda}, \citenamefont {Yan}, \citenamefont {Sales},
  \citenamefont {Uwatoko}, \citenamefont {Cheng},\ and\ \citenamefont
  {Shibauchi}}]{Sun2016pressure}%
  \BibitemOpen
  \bibfield  {author} {\bibinfo {author} {\bibfnamefont {J.~P.}\ \bibnamefont
  {Sun}}, \bibinfo {author} {\bibfnamefont {K.}~\bibnamefont {Matsuura}},
  \bibinfo {author} {\bibfnamefont {G.~Z.}\ \bibnamefont {Ye}}, \bibinfo
  {author} {\bibfnamefont {Y.}~\bibnamefont {Mizukami}}, \bibinfo {author}
  {\bibfnamefont {M.}~\bibnamefont {Shimozawa}}, \bibinfo {author}
  {\bibfnamefont {K.}~\bibnamefont {Matsubayashi}}, \bibinfo {author}
  {\bibfnamefont {M.}~\bibnamefont {Yamashita}}, \bibinfo {author}
  {\bibfnamefont {T.}~\bibnamefont {Watashige}}, \bibinfo {author}
  {\bibfnamefont {S.}~\bibnamefont {Kasahara}}, \bibinfo {author}
  {\bibfnamefont {Y.}~\bibnamefont {Matsuda}}, \bibinfo {author} {\bibfnamefont
  {J.-Q.}\ \bibnamefont {Yan}}, \bibinfo {author} {\bibfnamefont {B.~C.}\
  \bibnamefont {Sales}}, \bibinfo {author} {\bibfnamefont {Y.}~\bibnamefont
  {Uwatoko}}, \bibinfo {author} {\bibfnamefont {J.-G.}\ \bibnamefont {Cheng}},\
  and\ \bibinfo {author} {\bibfnamefont {T.}~\bibnamefont {Shibauchi}},\
  }\bibfield  {title} {\bibinfo {title} {{Dome-shaped magnetic order competing
  with high-temperature superconductivity at high pressures in FeSe}},\ }\href
  {https://doi.org/10.1038/ncomms12146} {\bibfield  {journal} {\bibinfo
  {journal} {Nat. Commun.}\ }\textbf {\bibinfo {volume} {7}},\ \bibinfo {pages}
  {12146} (\bibinfo {year} {2016})}\BibitemShut {NoStop}%
\bibitem [{\citenamefont {Chubukov}\ \emph {et~al.}(2016)\citenamefont
  {Chubukov}, \citenamefont {Khodas},\ and\ \citenamefont
  {Fernandes}}]{Chubukov2016}%
  \BibitemOpen
  \bibfield  {author} {\bibinfo {author} {\bibfnamefont {A.~V.}\ \bibnamefont
  {Chubukov}}, \bibinfo {author} {\bibfnamefont {M.}~\bibnamefont {Khodas}},\
  and\ \bibinfo {author} {\bibfnamefont {R.~M.}\ \bibnamefont {Fernandes}},\
  }\bibfield  {title} {\bibinfo {title} {{Magnetism, Superconductivity, and
  Spontaneous Orbital Order in Iron-Based Superconductors: Which Comes First
  and Why?}},\ }\href {https://doi.org/10.1103/PhysRevX.6.041045} {\bibfield
  {journal} {\bibinfo  {journal} {Phys. Rev. X}\ }\textbf {\bibinfo {volume}
  {6}},\ \bibinfo {pages} {041045} (\bibinfo {year} {2016})}\BibitemShut
  {NoStop}%
\bibitem [{\citenamefont {{Sprau}}\ \emph {et~al.}(2016)\citenamefont
  {{Sprau}}, \citenamefont {{Kostin}}, \citenamefont {{Kreisel}}, \citenamefont
  {{B{\"o}hmer}}, \citenamefont {{Taufour}}, \citenamefont {{Canfield}},
  \citenamefont {{Mukherjee}}, \citenamefont {{Hirschfeld}}, \citenamefont
  {{Andersen}},\ and\ \citenamefont {{S{\'e}amus Davis}}}]{Sprau2016}%
  \BibitemOpen
  \bibfield  {author} {\bibinfo {author} {\bibfnamefont {P.~O.}\ \bibnamefont
  {{Sprau}}}, \bibinfo {author} {\bibfnamefont {A.}~\bibnamefont {{Kostin}}},
  \bibinfo {author} {\bibfnamefont {A.}~\bibnamefont {{Kreisel}}}, \bibinfo
  {author} {\bibfnamefont {A.~E.}\ \bibnamefont {{B{\"o}hmer}}}, \bibinfo
  {author} {\bibfnamefont {V.}~\bibnamefont {{Taufour}}}, \bibinfo {author}
  {\bibfnamefont {P.~C.}\ \bibnamefont {{Canfield}}}, \bibinfo {author}
  {\bibfnamefont {S.}~\bibnamefont {{Mukherjee}}}, \bibinfo {author}
  {\bibfnamefont {P.~J.}\ \bibnamefont {{Hirschfeld}}}, \bibinfo {author}
  {\bibfnamefont {B.~M.}\ \bibnamefont {{Andersen}}},\ and\ \bibinfo {author}
  {\bibfnamefont {J.~C.}\ \bibnamefont {{S{\'e}amus Davis}}},\ }\bibfield
  {title} {\bibinfo {title} {{Discovery of Orbital-Selective Cooper Pairing in
  FeSe}},\ }\href@noop {} {\bibfield  {journal} {\bibinfo  {journal} {Science}\
  }\textbf {\bibinfo {volume} {357}},\ \bibinfo {pages} {75} (\bibinfo {year}
  {2016})}\BibitemShut {NoStop}%
\bibitem [{\citenamefont {Coldea}\ \emph {et~al.}(2019)\citenamefont {Coldea},
  \citenamefont {Blake}, \citenamefont {Kasahara}, \citenamefont {Haghighirad},
  \citenamefont {Watson}, \citenamefont {Knafo}, \citenamefont {Choi},
  \citenamefont {McCollam}, \citenamefont {Reiss}, \citenamefont {Yamashita},
  \citenamefont {Bruma}, \citenamefont {Speller}, \citenamefont {Matsuda},
  \citenamefont {Wolf}, \citenamefont {Shibauchi},\ and\ \citenamefont
  {Schofield}}]{Coldea2019}%
  \BibitemOpen
  \bibfield  {author} {\bibinfo {author} {\bibfnamefont {A.~I.}\ \bibnamefont
  {Coldea}}, \bibinfo {author} {\bibfnamefont {S.~F.}\ \bibnamefont {Blake}},
  \bibinfo {author} {\bibfnamefont {S.}~\bibnamefont {Kasahara}}, \bibinfo
  {author} {\bibfnamefont {A.~A.}\ \bibnamefont {Haghighirad}}, \bibinfo
  {author} {\bibfnamefont {M.~D.}\ \bibnamefont {Watson}}, \bibinfo {author}
  {\bibfnamefont {W.}~\bibnamefont {Knafo}}, \bibinfo {author} {\bibfnamefont
  {E.~S.}\ \bibnamefont {Choi}}, \bibinfo {author} {\bibfnamefont
  {A.}~\bibnamefont {McCollam}}, \bibinfo {author} {\bibfnamefont
  {P.}~\bibnamefont {Reiss}}, \bibinfo {author} {\bibfnamefont
  {T.}~\bibnamefont {Yamashita}}, \bibinfo {author} {\bibfnamefont
  {M.}~\bibnamefont {Bruma}}, \bibinfo {author} {\bibfnamefont
  {S.}~\bibnamefont {Speller}}, \bibinfo {author} {\bibfnamefont
  {Y.}~\bibnamefont {Matsuda}}, \bibinfo {author} {\bibfnamefont
  {T.}~\bibnamefont {Wolf}}, \bibinfo {author} {\bibfnamefont {T.}~\bibnamefont
  {Shibauchi}},\ and\ \bibinfo {author} {\bibfnamefont {A.~J.}\ \bibnamefont
  {Schofield}},\ }\bibfield  {title} {\bibinfo {title} {{Evolution of the
  low-temperature Fermi surface of superconducting FeSe$_{1-x}$S$_x$ across a
  nematic phase transition}},\ }\href
  {https://doi.org/10.1038/s41535-018-0141-0} {\bibfield  {journal} {\bibinfo
  {journal} {npj Quantum Mater.}\ }\textbf {\bibinfo {volume} {4}},\ \bibinfo
  {pages} {2} (\bibinfo {year} {2019})}\BibitemShut {NoStop}%
\bibitem [{\citenamefont {Kreisel}\ \emph {et~al.}(2020)\citenamefont
  {Kreisel}, \citenamefont {Hirschfeld},\ and\ \citenamefont
  {Andersen}}]{Kreisel2020}%
  \BibitemOpen
  \bibfield  {author} {\bibinfo {author} {\bibfnamefont {A.}~\bibnamefont
  {Kreisel}}, \bibinfo {author} {\bibfnamefont {P.~J.}\ \bibnamefont
  {Hirschfeld}},\ and\ \bibinfo {author} {\bibfnamefont {B.~M.}\ \bibnamefont
  {Andersen}},\ }\bibfield  {title} {\bibinfo {title} {{On the Remarkable
  Superconductivity of FeSe and Its Close Cousins}},\ }\href
  {https://doi.org/10.3390/sym12091402} {\bibfield  {journal} {\bibinfo
  {journal} {Symmetry}\ }\textbf {\bibinfo {volume} {12}},\ \bibinfo {pages}
  {1402} (\bibinfo {year} {2020})}\BibitemShut {NoStop}%
\bibitem [{\citenamefont {Bendele}\ \emph {et~al.}(2012)\citenamefont
  {Bendele}, \citenamefont {Ichsanow}, \citenamefont {Pashkevich},
  \citenamefont {Keller}, \citenamefont {Str\"assle}, \citenamefont {Gusev},
  \citenamefont {Pomjakushina}, \citenamefont {Conder}, \citenamefont
  {Khasanov},\ and\ \citenamefont {Keller}}]{Bendele2012}%
  \BibitemOpen
  \bibfield  {author} {\bibinfo {author} {\bibfnamefont {M.}~\bibnamefont
  {Bendele}}, \bibinfo {author} {\bibfnamefont {A.}~\bibnamefont {Ichsanow}},
  \bibinfo {author} {\bibfnamefont {Y.}~\bibnamefont {Pashkevich}}, \bibinfo
  {author} {\bibfnamefont {L.}~\bibnamefont {Keller}}, \bibinfo {author}
  {\bibfnamefont {T.}~\bibnamefont {Str\"assle}}, \bibinfo {author}
  {\bibfnamefont {A.}~\bibnamefont {Gusev}}, \bibinfo {author} {\bibfnamefont
  {E.}~\bibnamefont {Pomjakushina}}, \bibinfo {author} {\bibfnamefont
  {K.}~\bibnamefont {Conder}}, \bibinfo {author} {\bibfnamefont
  {R.}~\bibnamefont {Khasanov}},\ and\ \bibinfo {author} {\bibfnamefont
  {H.}~\bibnamefont {Keller}},\ }\bibfield  {title} {\bibinfo {title}
  {{Coexistence of superconductivity and magnetism in
  FeSe${}_{1\ensuremath{-}x}$ under pressure}},\ }\href
  {https://doi.org/10.1103/PhysRevB.85.064517} {\bibfield  {journal} {\bibinfo
  {journal} {Phys. Rev. B}\ }\textbf {\bibinfo {volume} {85}},\ \bibinfo
  {pages} {064517} (\bibinfo {year} {2012})}\BibitemShut {NoStop}%
\bibitem [{\citenamefont {B\"ohmer}\ \emph
  {et~al.}(2016{\natexlab{a}})\citenamefont {B\"ohmer}, \citenamefont
  {Taufour}, \citenamefont {Straszheim}, \citenamefont {Wolf},\ and\
  \citenamefont {Canfield}}]{Bohmer2016}%
  \BibitemOpen
  \bibfield  {author} {\bibinfo {author} {\bibfnamefont {A.~E.}\ \bibnamefont
  {B\"ohmer}}, \bibinfo {author} {\bibfnamefont {V.}~\bibnamefont {Taufour}},
  \bibinfo {author} {\bibfnamefont {W.~E.}\ \bibnamefont {Straszheim}},
  \bibinfo {author} {\bibfnamefont {T.}~\bibnamefont {Wolf}},\ and\ \bibinfo
  {author} {\bibfnamefont {P.~C.}\ \bibnamefont {Canfield}},\ }\bibfield
  {title} {\bibinfo {title} {{Variation of transition temperatures and residual
  resistivity ratio in vapor-grown FeSe}},\ }\href
  {https://doi.org/10.1103/PhysRevB.94.024526} {\bibfield  {journal} {\bibinfo
  {journal} {Phys. Rev. B}\ }\textbf {\bibinfo {volume} {94}},\ \bibinfo
  {pages} {024526} (\bibinfo {year} {2016}{\natexlab{a}})}\BibitemShut
  {NoStop}%
\bibitem [{\citenamefont {Reiss}\ \emph {et~al.}(2017)\citenamefont {Reiss},
  \citenamefont {Watson}, \citenamefont {Kim}, \citenamefont {Haghighirad},
  \citenamefont {Woodruff}, \citenamefont {Bruma}, \citenamefont {Clarke},\
  and\ \citenamefont {Coldea}}]{Reiss2017}%
  \BibitemOpen
  \bibfield  {author} {\bibinfo {author} {\bibfnamefont {P.}~\bibnamefont
  {Reiss}}, \bibinfo {author} {\bibfnamefont {M.~D.}\ \bibnamefont {Watson}},
  \bibinfo {author} {\bibfnamefont {T.~K.}\ \bibnamefont {Kim}}, \bibinfo
  {author} {\bibfnamefont {A.~A.}\ \bibnamefont {Haghighirad}}, \bibinfo
  {author} {\bibfnamefont {D.~N.}\ \bibnamefont {Woodruff}}, \bibinfo {author}
  {\bibfnamefont {M.}~\bibnamefont {Bruma}}, \bibinfo {author} {\bibfnamefont
  {S.~J.}\ \bibnamefont {Clarke}},\ and\ \bibinfo {author} {\bibfnamefont
  {A.~I.}\ \bibnamefont {Coldea}},\ }\bibfield  {title} {\bibinfo {title}
  {{Suppression of electronic correlations by chemical pressure from FeSe to
  FeS}},\ }\href {https://doi.org/10.1103/PhysRevB.96.121103} {\bibfield
  {journal} {\bibinfo  {journal} {Phys. Rev. B}\ }\textbf {\bibinfo {volume}
  {96}},\ \bibinfo {pages} {121103} (\bibinfo {year} {2017})}\BibitemShut
  {NoStop}%
\bibitem [{\citenamefont {Coldea}(2021)}]{Coldea2021}%
  \BibitemOpen
  \bibfield  {author} {\bibinfo {author} {\bibfnamefont {A.~I.}\ \bibnamefont
  {Coldea}},\ }\bibfield  {title} {\bibinfo {title} {{Electronic Nematic States
  Tuned by Isoelectronic Substitution in Bulk FeSe$_{1−x}$S$_x$}},\ }\href
  {https://doi.org/10.3389/fphy.2020.594500} {\bibfield  {journal} {\bibinfo
  {journal} {Frontiers in Physics}\ }\textbf {\bibinfo {volume} {8}},\ \bibinfo
  {pages} {528} (\bibinfo {year} {2021})}\BibitemShut {NoStop}%
\bibitem [{\citenamefont {Hosoi}\ \emph {et~al.}(2016)\citenamefont {Hosoi},
  \citenamefont {Matsuura}, \citenamefont {Ishida}, \citenamefont {Wang},
  \citenamefont {Mizukami}, \citenamefont {Watashige}, \citenamefont
  {Kasahara}, \citenamefont {Matsuda},\ and\ \citenamefont
  {Shibauchi}}]{Hosoi2016}%
  \BibitemOpen
  \bibfield  {author} {\bibinfo {author} {\bibfnamefont {S.}~\bibnamefont
  {Hosoi}}, \bibinfo {author} {\bibfnamefont {K.}~\bibnamefont {Matsuura}},
  \bibinfo {author} {\bibfnamefont {K.}~\bibnamefont {Ishida}}, \bibinfo
  {author} {\bibfnamefont {H.}~\bibnamefont {Wang}}, \bibinfo {author}
  {\bibfnamefont {Y.}~\bibnamefont {Mizukami}}, \bibinfo {author}
  {\bibfnamefont {T.}~\bibnamefont {Watashige}}, \bibinfo {author}
  {\bibfnamefont {S.}~\bibnamefont {Kasahara}}, \bibinfo {author}
  {\bibfnamefont {Y.}~\bibnamefont {Matsuda}},\ and\ \bibinfo {author}
  {\bibfnamefont {T.}~\bibnamefont {Shibauchi}},\ }\bibfield  {title} {\bibinfo
  {title} {{Nematic quantum critical point without magnetism in
  FeSe$_{1-x}$S$_x$ superconductors}},\ }\href
  {https://doi.org/doi:10.1073/pnas.1605806113} {\bibfield  {journal} {\bibinfo
   {journal} {PNAS}\ }\textbf {\bibinfo {volume} {113}},\ \bibinfo {pages}
  {8139} (\bibinfo {year} {2016})}\BibitemShut {NoStop}%
\bibitem [{\citenamefont {Bristow}\ \emph {et~al.}(2020)\citenamefont
  {Bristow}, \citenamefont {Reiss}, \citenamefont {Haghighirad}, \citenamefont
  {Zajicek}, \citenamefont {Singh}, \citenamefont {Wolf}, \citenamefont {Graf},
  \citenamefont {Knafo}, \citenamefont {McCollam},\ and\ \citenamefont
  {Coldea}}]{Bristow2020}%
  \BibitemOpen
  \bibfield  {author} {\bibinfo {author} {\bibfnamefont {M.}~\bibnamefont
  {Bristow}}, \bibinfo {author} {\bibfnamefont {P.}~\bibnamefont {Reiss}},
  \bibinfo {author} {\bibfnamefont {A.~A.}\ \bibnamefont {Haghighirad}},
  \bibinfo {author} {\bibfnamefont {Z.}~\bibnamefont {Zajicek}}, \bibinfo
  {author} {\bibfnamefont {S.~J.}\ \bibnamefont {Singh}}, \bibinfo {author}
  {\bibfnamefont {T.}~\bibnamefont {Wolf}}, \bibinfo {author} {\bibfnamefont
  {D.}~\bibnamefont {Graf}}, \bibinfo {author} {\bibfnamefont {W.}~\bibnamefont
  {Knafo}}, \bibinfo {author} {\bibfnamefont {A.}~\bibnamefont {McCollam}},\
  and\ \bibinfo {author} {\bibfnamefont {A.~I.}\ \bibnamefont {Coldea}},\
  }\bibfield  {title} {\bibinfo {title} {{Anomalous high-magnetic field
  electronic state of the nematic superconductors
  ${\mathrm{FeSe}}_{1\ensuremath{-}x}{\mathrm{S}}_{x}$}},\ }\href
  {https://doi.org/10.1103/PhysRevResearch.2.013309} {\bibfield  {journal}
  {\bibinfo  {journal} {Phys. Rev. Research}\ }\textbf {\bibinfo {volume}
  {2}},\ \bibinfo {pages} {013309} (\bibinfo {year} {2020})}\BibitemShut
  {NoStop}%
\bibitem [{\citenamefont {Hanaguri}\ \emph {et~al.}(2018)\citenamefont
  {Hanaguri}, \citenamefont {Iwaya}, \citenamefont {Kohsaka}, \citenamefont
  {Machida}, \citenamefont {Watashige}, \citenamefont {Kasahara}, \citenamefont
  {Shibauchi},\ and\ \citenamefont {Matsuda}}]{Hanaguri2018}%
  \BibitemOpen
  \bibfield  {author} {\bibinfo {author} {\bibfnamefont {T.}~\bibnamefont
  {Hanaguri}}, \bibinfo {author} {\bibfnamefont {V.}~\bibnamefont {Iwaya}},
  \bibinfo {author} {\bibfnamefont {Y.}~\bibnamefont {Kohsaka}}, \bibinfo
  {author} {\bibfnamefont {T.}~\bibnamefont {Machida}}, \bibinfo {author}
  {\bibfnamefont {T.}~\bibnamefont {Watashige}}, \bibinfo {author}
  {\bibfnamefont {S.}~\bibnamefont {Kasahara}}, \bibinfo {author}
  {\bibfnamefont {T.}~\bibnamefont {Shibauchi}},\ and\ \bibinfo {author}
  {\bibfnamefont {Y.}~\bibnamefont {Matsuda}},\ }\bibfield  {title} {\bibinfo
  {title} {{Two distinct superconducting pairing states divided by the nematic
  end point in FeSe$_{1-x}$S$_x$}},\ }\href
  {https://doi.org/10.1126/sciadv.aar6419} {\bibfield  {journal} {\bibinfo
  {journal} {Sci. Adv.}\ }\textbf {\bibinfo {volume} {4}},\ \bibinfo {pages}
  {eaar6419} (\bibinfo {year} {2018})}\BibitemShut {NoStop}%
\bibitem [{\citenamefont {Sato}\ \emph {et~al.}(2018)\citenamefont {Sato},
  \citenamefont {Kasahara}, \citenamefont {Taniguchi}, \citenamefont {Xing},
  \citenamefont {Kasahara}, \citenamefont {Tokiwa}, \citenamefont {Yamakawa},
  \citenamefont {Kontani}, \citenamefont {Shibauchi},\ and\ \citenamefont
  {Matsuda}}]{Sato2018}%
  \BibitemOpen
  \bibfield  {author} {\bibinfo {author} {\bibfnamefont {Y.}~\bibnamefont
  {Sato}}, \bibinfo {author} {\bibfnamefont {S.}~\bibnamefont {Kasahara}},
  \bibinfo {author} {\bibfnamefont {T.}~\bibnamefont {Taniguchi}}, \bibinfo
  {author} {\bibfnamefont {X.}~\bibnamefont {Xing}}, \bibinfo {author}
  {\bibfnamefont {Y.}~\bibnamefont {Kasahara}}, \bibinfo {author}
  {\bibfnamefont {Y.}~\bibnamefont {Tokiwa}}, \bibinfo {author} {\bibfnamefont
  {Y.}~\bibnamefont {Yamakawa}}, \bibinfo {author} {\bibfnamefont
  {H.}~\bibnamefont {Kontani}}, \bibinfo {author} {\bibfnamefont
  {T.}~\bibnamefont {Shibauchi}},\ and\ \bibinfo {author} {\bibfnamefont
  {Y.}~\bibnamefont {Matsuda}},\ }\bibfield  {title} {\bibinfo {title} {{Abrupt
  change of the superconducting gap structure at the nematic critical point in
  in FeSe$_{1-x}$S$_x$}},\ }\href {https://doi.org/10.1073/pnas.1717331115}
  {\bibfield  {journal} {\bibinfo  {journal} {Proceedings of the National
  Academy of Sciences}\ }\textbf {\bibinfo {volume} {115}},\ \bibinfo {pages}
  {1227} (\bibinfo {year} {2018})}\BibitemShut {NoStop}%
\bibitem [{\citenamefont {Setty}\ \emph {et~al.}(2020)\citenamefont {Setty},
  \citenamefont {Bhattacharyya}, \citenamefont {Cao}, \citenamefont {Kreisel},\
  and\ \citenamefont {Hirschfeld}}]{Setty2020}%
  \BibitemOpen
  \bibfield  {author} {\bibinfo {author} {\bibfnamefont {C.}~\bibnamefont
  {Setty}}, \bibinfo {author} {\bibfnamefont {S.}~\bibnamefont
  {Bhattacharyya}}, \bibinfo {author} {\bibfnamefont {Y.}~\bibnamefont {Cao}},
  \bibinfo {author} {\bibfnamefont {A.}~\bibnamefont {Kreisel}},\ and\ \bibinfo
  {author} {\bibfnamefont {P.~J.}\ \bibnamefont {Hirschfeld}},\ }\bibfield
  {title} {\bibinfo {title} {{Topological ultranodal pair states in iron-based
  superconductors}},\ }\href {https://doi.org/10.1038/s41467-020-14357-2}
  {\bibfield  {journal} {\bibinfo  {journal} {Nature Communications}\ }\textbf
  {\bibinfo {volume} {11}},\ \bibinfo {pages} {523} (\bibinfo {year}
  {2020})}\BibitemShut {NoStop}%
\bibitem [{\citenamefont {Matsuura}\ \emph {et~al.}(2017)\citenamefont
  {Matsuura}, \citenamefont {Mizukami}, \citenamefont {Arai}, \citenamefont
  {Sugimura}, \citenamefont {Maejima}, \citenamefont {Machida}, \citenamefont
  {Watanuki}, \citenamefont {Fukuda}, \citenamefont {Yajima}, \citenamefont
  {Hiroi}, \citenamefont {Yip}, \citenamefont {Chan}, \citenamefont {Niu},
  \citenamefont {Hosoi}, \citenamefont {Ishida}, \citenamefont {Mukasa},
  \citenamefont {Kasahara}, \citenamefont {Cheng}, \citenamefont {Goh},
  \citenamefont {Matsuda}, \citenamefont {Uwatoko},\ and\ \citenamefont
  {Shibauchi}}]{Matsuura2017}%
  \BibitemOpen
  \bibfield  {author} {\bibinfo {author} {\bibfnamefont {K.}~\bibnamefont
  {Matsuura}}, \bibinfo {author} {\bibfnamefont {Y.}~\bibnamefont {Mizukami}},
  \bibinfo {author} {\bibfnamefont {Y.}~\bibnamefont {Arai}}, \bibinfo {author}
  {\bibfnamefont {Y.}~\bibnamefont {Sugimura}}, \bibinfo {author}
  {\bibfnamefont {N.}~\bibnamefont {Maejima}}, \bibinfo {author} {\bibfnamefont
  {A.}~\bibnamefont {Machida}}, \bibinfo {author} {\bibfnamefont
  {T.}~\bibnamefont {Watanuki}}, \bibinfo {author} {\bibfnamefont
  {T.}~\bibnamefont {Fukuda}}, \bibinfo {author} {\bibfnamefont
  {T.}~\bibnamefont {Yajima}}, \bibinfo {author} {\bibfnamefont
  {Z.}~\bibnamefont {Hiroi}}, \bibinfo {author} {\bibfnamefont {K.~Y.}\
  \bibnamefont {Yip}}, \bibinfo {author} {\bibfnamefont {Y.~C.}\ \bibnamefont
  {Chan}}, \bibinfo {author} {\bibfnamefont {Q.}~\bibnamefont {Niu}}, \bibinfo
  {author} {\bibfnamefont {S.}~\bibnamefont {Hosoi}}, \bibinfo {author}
  {\bibfnamefont {K.}~\bibnamefont {Ishida}}, \bibinfo {author} {\bibfnamefont
  {K.}~\bibnamefont {Mukasa}}, \bibinfo {author} {\bibfnamefont
  {S.}~\bibnamefont {Kasahara}}, \bibinfo {author} {\bibfnamefont {J.-G.}\
  \bibnamefont {Cheng}}, \bibinfo {author} {\bibfnamefont {S.~K.}\ \bibnamefont
  {Goh}}, \bibinfo {author} {\bibfnamefont {Y.}~\bibnamefont {Matsuda}},
  \bibinfo {author} {\bibfnamefont {Y.}~\bibnamefont {Uwatoko}},\ and\ \bibinfo
  {author} {\bibfnamefont {T.}~\bibnamefont {Shibauchi}},\ }\bibfield  {title}
  {\bibinfo {title} {{Maximizing $T_{\rm c}$ by tuning nematicity and magnetism
  in FeSe$_{1-x}$S$_ x$ superconductors}},\ }\href
  {https://doi.org/10.1038/s41467-017-01277-x} {\bibfield  {journal} {\bibinfo
  {journal} {Nat. Comm.}\ }\textbf {\bibinfo {volume} {8}},\ \bibinfo {pages}
  {1143} (\bibinfo {year} {2017})}\BibitemShut {NoStop}%
\bibitem [{\citenamefont {Reiss}\ \emph {et~al.}(2020)\citenamefont {Reiss},
  \citenamefont {Graf}, \citenamefont {Haghighirad}, \citenamefont {Knafo},
  \citenamefont {Drigo}, \citenamefont {Bristow}, \citenamefont {Schofield},\
  and\ \citenamefont {Coldea}}]{Reiss2020}%
  \BibitemOpen
  \bibfield  {author} {\bibinfo {author} {\bibfnamefont {P.}~\bibnamefont
  {Reiss}}, \bibinfo {author} {\bibfnamefont {D.}~\bibnamefont {Graf}},
  \bibinfo {author} {\bibfnamefont {A.~A.}\ \bibnamefont {Haghighirad}},
  \bibinfo {author} {\bibfnamefont {W.}~\bibnamefont {Knafo}}, \bibinfo
  {author} {\bibfnamefont {L.}~\bibnamefont {Drigo}}, \bibinfo {author}
  {\bibfnamefont {M.}~\bibnamefont {Bristow}}, \bibinfo {author} {\bibfnamefont
  {A.~J.}\ \bibnamefont {Schofield}},\ and\ \bibinfo {author} {\bibfnamefont
  {A.~I.}\ \bibnamefont {Coldea}},\ }\bibfield  {title} {\bibinfo {title}
  {{Quenched nematic criticality and two superconducting domes in an iron-based
  superconductor}},\ }\href {https://doi.org/10.1038/s41567-019-0694-2}
  {\bibfield  {journal} {\bibinfo  {journal} {Nature Physics}\ }\textbf
  {\bibinfo {volume} {16}},\ \bibinfo {pages} {89} (\bibinfo {year}
  {2020})}\BibitemShut {NoStop}%
\bibitem [{\citenamefont {Rana}\ \emph {et~al.}(2020)\citenamefont {Rana},
  \citenamefont {Xiang}, \citenamefont {Wiecki}, \citenamefont {Ribeiro},
  \citenamefont {Lesseux}, \citenamefont {B\"ohmer}, \citenamefont {Bud'ko},
  \citenamefont {Canfield},\ and\ \citenamefont {Furukawa}}]{Rana2020}%
  \BibitemOpen
  \bibfield  {author} {\bibinfo {author} {\bibfnamefont {K.}~\bibnamefont
  {Rana}}, \bibinfo {author} {\bibfnamefont {L.}~\bibnamefont {Xiang}},
  \bibinfo {author} {\bibfnamefont {P.}~\bibnamefont {Wiecki}}, \bibinfo
  {author} {\bibfnamefont {R.~A.}\ \bibnamefont {Ribeiro}}, \bibinfo {author}
  {\bibfnamefont {G.~G.}\ \bibnamefont {Lesseux}}, \bibinfo {author}
  {\bibfnamefont {A.~E.}\ \bibnamefont {B\"ohmer}}, \bibinfo {author}
  {\bibfnamefont {S.~L.}\ \bibnamefont {Bud'ko}}, \bibinfo {author}
  {\bibfnamefont {P.~C.}\ \bibnamefont {Canfield}},\ and\ \bibinfo {author}
  {\bibfnamefont {Y.}~\bibnamefont {Furukawa}},\ }\bibfield  {title} {\bibinfo
  {title} {{Impact of nematicity on the relationship between antiferromagnetic
  fluctuations and superconductivity in FeSe$_{0.91}$S$_{0.09}$ under
  pressure}},\ }\href {https://doi.org/10.1103/PhysRevB.101.180503} {\bibfield
  {journal} {\bibinfo  {journal} {Phys. Rev. B}\ }\textbf {\bibinfo {volume}
  {101}},\ \bibinfo {pages} {180503} (\bibinfo {year} {2020})}\BibitemShut
  {NoStop}%
\bibitem [{\citenamefont {Xiang}\ \emph {et~al.}(2017)\citenamefont {Xiang},
  \citenamefont {Kaluarachchi}, \citenamefont {B\"ohmer}, \citenamefont
  {Taufour}, \citenamefont {Tanatar}, \citenamefont {Prozorov}, \citenamefont
  {Bud'ko},\ and\ \citenamefont {Canfield}}]{Xiang2017}%
  \BibitemOpen
  \bibfield  {author} {\bibinfo {author} {\bibfnamefont {L.}~\bibnamefont
  {Xiang}}, \bibinfo {author} {\bibfnamefont {U.~S.}\ \bibnamefont
  {Kaluarachchi}}, \bibinfo {author} {\bibfnamefont {A.~E.}\ \bibnamefont
  {B\"ohmer}}, \bibinfo {author} {\bibfnamefont {V.}~\bibnamefont {Taufour}},
  \bibinfo {author} {\bibfnamefont {M.~A.}\ \bibnamefont {Tanatar}}, \bibinfo
  {author} {\bibfnamefont {R.}~\bibnamefont {Prozorov}}, \bibinfo {author}
  {\bibfnamefont {S.~L.}\ \bibnamefont {Bud'ko}},\ and\ \bibinfo {author}
  {\bibfnamefont {P.~C.}\ \bibnamefont {Canfield}},\ }\bibfield  {title}
  {\bibinfo {title} {{Dome of magnetic order inside the nematic phase of
  sulfur-substituted FeSe under pressure}},\ }\href
  {https://doi.org/10.1103/PhysRevB.96.024511} {\bibfield  {journal} {\bibinfo
  {journal} {Phys. Rev. B}\ }\textbf {\bibinfo {volume} {96}},\ \bibinfo
  {pages} {024511} (\bibinfo {year} {2017})}\BibitemShut {NoStop}%
\bibitem [{\citenamefont {Xie}\ \emph {et~al.}(2021)\citenamefont {Xie},
  \citenamefont {Liu}, \citenamefont {Zhang}, \citenamefont {Wong},
  \citenamefont {Zhou}, \citenamefont {Zhao}, \citenamefont {Wang},
  \citenamefont {Lai},\ and\ \citenamefont {Goh}}]{Xie2021}%
  \BibitemOpen
  \bibfield  {author} {\bibinfo {author} {\bibfnamefont {J.}~\bibnamefont
  {Xie}}, \bibinfo {author} {\bibfnamefont {X.}~\bibnamefont {Liu}}, \bibinfo
  {author} {\bibfnamefont {W.}~\bibnamefont {Zhang}}, \bibinfo {author}
  {\bibfnamefont {S.~M.}\ \bibnamefont {Wong}}, \bibinfo {author}
  {\bibfnamefont {X.}~\bibnamefont {Zhou}}, \bibinfo {author} {\bibfnamefont
  {Y.}~\bibnamefont {Zhao}}, \bibinfo {author} {\bibfnamefont {S.}~\bibnamefont
  {Wang}}, \bibinfo {author} {\bibfnamefont {K.~T.}\ \bibnamefont {Lai}},\ and\
  \bibinfo {author} {\bibfnamefont {S.~K.}\ \bibnamefont {Goh}},\ }\bibfield
  {title} {\bibinfo {title} {{Fragile Pressure-Induced Magnetism in FeSe
  Superconductors with a Thickness Reduction}},\ }\href
  {https://doi.org/10.1021/acs.nanolett.1c03508} {\bibfield  {journal}
  {\bibinfo  {journal} {Nano Letters}\ }\textbf {\bibinfo {volume} {21}},\
  \bibinfo {pages} {9310} (\bibinfo {year} {2021})}\BibitemShut {NoStop}%
\bibitem [{\citenamefont {Zajicek}\ \emph {et~al.}(2022)\citenamefont
  {Zajicek}, \citenamefont {Singh},\ and\ \citenamefont
  {Coldea}}]{Zajicek2022Cupressure}%
  \BibitemOpen
  \bibfield  {author} {\bibinfo {author} {\bibfnamefont {Z.}~\bibnamefont
  {Zajicek}}, \bibinfo {author} {\bibfnamefont {S.~J.}\ \bibnamefont {Singh}},\
  and\ \bibinfo {author} {\bibfnamefont {A.~I.}\ \bibnamefont {Coldea}},\
  }\bibfield  {title} {\bibinfo {title} {{Robust superconductivity and fragile
  magnetism induced by the strong Cu impurity scattering in the high-pressure
  phase of FeSe}},\ }\href {https://doi.org/10.1103/PhysRevResearch.4.043123}
  {\bibfield  {journal} {\bibinfo  {journal} {Phys. Rev. Res.}\ }\textbf
  {\bibinfo {volume} {4}},\ \bibinfo {pages} {043123} (\bibinfo {year}
  {2022})}\BibitemShut {NoStop}%
\bibitem [{\citenamefont {Reiss}\ \emph {et~al.}(2022)\citenamefont {Reiss},
  \citenamefont {McCollam}, \citenamefont {Zajicek}, \citenamefont
  {Haghighirad},\ and\ \citenamefont {Coldea}}]{Reiss2022}%
  \BibitemOpen
  \bibfield  {author} {\bibinfo {author} {\bibfnamefont {P.}~\bibnamefont
  {Reiss}}, \bibinfo {author} {\bibfnamefont {A.}~\bibnamefont {McCollam}},
  \bibinfo {author} {\bibfnamefont {Z.}~\bibnamefont {Zajicek}}, \bibinfo
  {author} {\bibfnamefont {A.~A.}\ \bibnamefont {Haghighirad}},\ and\ \bibinfo
  {author} {\bibfnamefont {A.~I.}\ \bibnamefont {Coldea}},\ }\bibfield  {title}
  {\bibinfo {title} {{Collapse of Metallicity and High-$T_c$ Superconductivity
  in the High-Pressure phase of FeSe$_{0.89}$S$_{0.11}$}},\ }\bibfield
  {journal} {\bibinfo  {journal} {\rm Preprint at
  https://arxiv.org/abs/2212.06824}\ }\href
  {https://doi.org/https://arxiv.org/abs/2212.06824}
  {https://arxiv.org/abs/2212.06824} (\bibinfo {year} {2022})\BibitemShut
  {NoStop}%
\bibitem [{\citenamefont {Terashima}\ \emph {et~al.}(2014)\citenamefont
  {Terashima}, \citenamefont {Kikugawa}, \citenamefont {Kiswandhi},
  \citenamefont {Choi}, \citenamefont {Brooks}, \citenamefont {Kasahara},
  \citenamefont {Watashige}, \citenamefont {Ikeda}, \citenamefont {Shibauchi},
  \citenamefont {Matsuda}, \citenamefont {Wolf}, \citenamefont {B\"ohmer},
  \citenamefont {Hardy}, \citenamefont {Meingast}, \citenamefont {L\"ohneysen},
  \citenamefont {Suzuki}, \citenamefont {Arita},\ and\ \citenamefont
  {Uji}}]{Terashima2014}%
  \BibitemOpen
  \bibfield  {author} {\bibinfo {author} {\bibfnamefont {T.}~\bibnamefont
  {Terashima}}, \bibinfo {author} {\bibfnamefont {N.}~\bibnamefont {Kikugawa}},
  \bibinfo {author} {\bibfnamefont {A.}~\bibnamefont {Kiswandhi}}, \bibinfo
  {author} {\bibfnamefont {E.-S.}\ \bibnamefont {Choi}}, \bibinfo {author}
  {\bibfnamefont {J.~S.}\ \bibnamefont {Brooks}}, \bibinfo {author}
  {\bibfnamefont {S.}~\bibnamefont {Kasahara}}, \bibinfo {author}
  {\bibfnamefont {T.}~\bibnamefont {Watashige}}, \bibinfo {author}
  {\bibfnamefont {H.}~\bibnamefont {Ikeda}}, \bibinfo {author} {\bibfnamefont
  {T.}~\bibnamefont {Shibauchi}}, \bibinfo {author} {\bibfnamefont
  {Y.}~\bibnamefont {Matsuda}}, \bibinfo {author} {\bibfnamefont
  {T.}~\bibnamefont {Wolf}}, \bibinfo {author} {\bibfnamefont {A.~E.}\
  \bibnamefont {B\"ohmer}}, \bibinfo {author} {\bibfnamefont {F.}~\bibnamefont
  {Hardy}}, \bibinfo {author} {\bibfnamefont {C.}~\bibnamefont {Meingast}},
  \bibinfo {author} {\bibfnamefont {H.~v.}\ \bibnamefont {L\"ohneysen}},
  \bibinfo {author} {\bibfnamefont {M.-T.}\ \bibnamefont {Suzuki}}, \bibinfo
  {author} {\bibfnamefont {R.}~\bibnamefont {Arita}},\ and\ \bibinfo {author}
  {\bibfnamefont {S.}~\bibnamefont {Uji}},\ }\bibfield  {title} {\bibinfo
  {title} {{Anomalous Fermi surface in FeSe seen by Shubnikov-de Haas
  oscillation measurements}},\ }\href
  {https://doi.org/10.1103/PhysRevB.90.144517} {\bibfield  {journal} {\bibinfo
  {journal} {Phys. Rev. B}\ }\textbf {\bibinfo {volume} {90}},\ \bibinfo
  {pages} {144517} (\bibinfo {year} {2014})}\BibitemShut {NoStop}%
\bibitem [{\citenamefont {Watson}\ \emph
  {et~al.}(2015{\natexlab{a}})\citenamefont {Watson}, \citenamefont {Kim},
  \citenamefont {Haghighirad}, \citenamefont {Davies}, \citenamefont
  {McCollam}, \citenamefont {Narayanan}, \citenamefont {Blake}, \citenamefont
  {Chen}, \citenamefont {Ghannadzadeh}, \citenamefont {Schofield},
  \citenamefont {Hoesch}, \citenamefont {Meingast}, \citenamefont {Wolf},\ and\
  \citenamefont {Coldea}}]{Watson2015a}%
  \BibitemOpen
  \bibfield  {author} {\bibinfo {author} {\bibfnamefont {M.~D.}\ \bibnamefont
  {Watson}}, \bibinfo {author} {\bibfnamefont {T.~K.}\ \bibnamefont {Kim}},
  \bibinfo {author} {\bibfnamefont {A.~A.}\ \bibnamefont {Haghighirad}},
  \bibinfo {author} {\bibfnamefont {N.~R.}\ \bibnamefont {Davies}}, \bibinfo
  {author} {\bibfnamefont {A.}~\bibnamefont {McCollam}}, \bibinfo {author}
  {\bibfnamefont {A.}~\bibnamefont {Narayanan}}, \bibinfo {author}
  {\bibfnamefont {S.~F.}\ \bibnamefont {Blake}}, \bibinfo {author}
  {\bibfnamefont {Y.~L.}\ \bibnamefont {Chen}}, \bibinfo {author}
  {\bibfnamefont {S.}~\bibnamefont {Ghannadzadeh}}, \bibinfo {author}
  {\bibfnamefont {A.~J.}\ \bibnamefont {Schofield}}, \bibinfo {author}
  {\bibfnamefont {M.}~\bibnamefont {Hoesch}}, \bibinfo {author} {\bibfnamefont
  {C.}~\bibnamefont {Meingast}}, \bibinfo {author} {\bibfnamefont
  {T.}~\bibnamefont {Wolf}},\ and\ \bibinfo {author} {\bibfnamefont {A.~I.}\
  \bibnamefont {Coldea}},\ }\bibfield  {title} {\bibinfo {title} {{Emergence of
  the nematic electronic state in FeSe}},\ }\href
  {https://doi.org/10.1103/PhysRevB.91.155106} {\bibfield  {journal} {\bibinfo
  {journal} {Phys. Rev. B}\ }\textbf {\bibinfo {volume} {91}},\ \bibinfo
  {pages} {155106} (\bibinfo {year} {2015}{\natexlab{a}})}\BibitemShut
  {NoStop}%
\bibitem [{\citenamefont {Audouard}\ \emph {et~al.}(2015)\citenamefont
  {Audouard}, \citenamefont {Duc}, \citenamefont {Drigo}, \citenamefont
  {Toulemonde}, \citenamefont {Karlsson}, \citenamefont {Strobel},\ and\
  \citenamefont {Sulpice}}]{Audouard2015}%
  \BibitemOpen
  \bibfield  {author} {\bibinfo {author} {\bibfnamefont {A.}~\bibnamefont
  {Audouard}}, \bibinfo {author} {\bibfnamefont {F.}~\bibnamefont {Duc}},
  \bibinfo {author} {\bibfnamefont {L.}~\bibnamefont {Drigo}}, \bibinfo
  {author} {\bibfnamefont {P.}~\bibnamefont {Toulemonde}}, \bibinfo {author}
  {\bibfnamefont {S.}~\bibnamefont {Karlsson}}, \bibinfo {author}
  {\bibfnamefont {P.}~\bibnamefont {Strobel}},\ and\ \bibinfo {author}
  {\bibfnamefont {A.}~\bibnamefont {Sulpice}},\ }\bibfield  {title} {\bibinfo
  {title} {{Quantum oscillations and upper critical magnetic field of the
  iron-based superconductor FeSe}},\ }\href
  {https://doi.org/10.1209/0295-5075/109/27003} {\bibfield  {journal} {\bibinfo
   {journal} {Europhys. Lett.}\ }\textbf {\bibinfo {volume} {109}},\ \bibinfo
  {pages} {27003} (\bibinfo {year} {2015})}\BibitemShut {NoStop}%
\bibitem [{\citenamefont {Terashima}\ \emph {et~al.}(2019)\citenamefont
  {Terashima}, \citenamefont {Kikugawa}, \citenamefont {Graf}, \citenamefont
  {Hirose}, \citenamefont {Uji}, \citenamefont {Matsushita}, \citenamefont
  {Lin}, \citenamefont {Zhu}, \citenamefont {Wen}, \citenamefont {Nomoto},
  \citenamefont {Suzuki},\ and\ \citenamefont {Ikeda}}]{Terashima2019}%
  \BibitemOpen
  \bibfield  {author} {\bibinfo {author} {\bibfnamefont {T.}~\bibnamefont
  {Terashima}}, \bibinfo {author} {\bibfnamefont {N.}~\bibnamefont {Kikugawa}},
  \bibinfo {author} {\bibfnamefont {D.}~\bibnamefont {Graf}}, \bibinfo {author}
  {\bibfnamefont {H.~T.}\ \bibnamefont {Hirose}}, \bibinfo {author}
  {\bibfnamefont {S.}~\bibnamefont {Uji}}, \bibinfo {author} {\bibfnamefont
  {Y.}~\bibnamefont {Matsushita}}, \bibinfo {author} {\bibfnamefont
  {H.}~\bibnamefont {Lin}}, \bibinfo {author} {\bibfnamefont {X.}~\bibnamefont
  {Zhu}}, \bibinfo {author} {\bibfnamefont {H.-H.}\ \bibnamefont {Wen}},
  \bibinfo {author} {\bibfnamefont {T.}~\bibnamefont {Nomoto}}, \bibinfo
  {author} {\bibfnamefont {K.}~\bibnamefont {Suzuki}},\ and\ \bibinfo {author}
  {\bibfnamefont {H.}~\bibnamefont {Ikeda}},\ }\bibfield  {title} {\bibinfo
  {title} {{Accurate determination of the Fermi surface of tetragonal FeS via
  quantum oscillation measurements and quasiparticle self-consistent GW
  calculations}},\ }\href {https://doi.org/10.1103/PhysRevB.99.134501}
  {\bibfield  {journal} {\bibinfo  {journal} {Phys. Rev. B}\ }\textbf {\bibinfo
  {volume} {99}},\ \bibinfo {pages} {134501} (\bibinfo {year}
  {2019})}\BibitemShut {NoStop}%
\bibitem [{\citenamefont {Terashima}\ \emph
  {et~al.}(2016{\natexlab{a}})\citenamefont {Terashima}, \citenamefont
  {Kikugawa}, \citenamefont {Kiswandhi}, \citenamefont {Graf}, \citenamefont
  {Choi}, \citenamefont {Brooks}, \citenamefont {Kasahara}, \citenamefont
  {Watashige}, \citenamefont {Matsuda}, \citenamefont {Shibauchi},
  \citenamefont {Wolf}, \citenamefont {B\"ohmer}, \citenamefont {Hardy},
  \citenamefont {Meingast}, \citenamefont {L\"ohneysen},\ and\ \citenamefont
  {Uji}}]{Terashima2016}%
  \BibitemOpen
  \bibfield  {author} {\bibinfo {author} {\bibfnamefont {T.}~\bibnamefont
  {Terashima}}, \bibinfo {author} {\bibfnamefont {N.}~\bibnamefont {Kikugawa}},
  \bibinfo {author} {\bibfnamefont {A.}~\bibnamefont {Kiswandhi}}, \bibinfo
  {author} {\bibfnamefont {D.}~\bibnamefont {Graf}}, \bibinfo {author}
  {\bibfnamefont {E.-S.}\ \bibnamefont {Choi}}, \bibinfo {author}
  {\bibfnamefont {J.~S.}\ \bibnamefont {Brooks}}, \bibinfo {author}
  {\bibfnamefont {S.}~\bibnamefont {Kasahara}}, \bibinfo {author}
  {\bibfnamefont {T.}~\bibnamefont {Watashige}}, \bibinfo {author}
  {\bibfnamefont {Y.}~\bibnamefont {Matsuda}}, \bibinfo {author} {\bibfnamefont
  {T.}~\bibnamefont {Shibauchi}}, \bibinfo {author} {\bibfnamefont
  {T.}~\bibnamefont {Wolf}}, \bibinfo {author} {\bibfnamefont {A.~E.}\
  \bibnamefont {B\"ohmer}}, \bibinfo {author} {\bibfnamefont {F.}~\bibnamefont
  {Hardy}}, \bibinfo {author} {\bibfnamefont {C.}~\bibnamefont {Meingast}},
  \bibinfo {author} {\bibfnamefont {H.~v.}\ \bibnamefont {L\"ohneysen}},\ and\
  \bibinfo {author} {\bibfnamefont {S.}~\bibnamefont {Uji}},\ }\bibfield
  {title} {\bibinfo {title} {{Fermi surface reconstruction in FeSe under high
  pressure}},\ }\href {https://doi.org/10.1103/PhysRevB.93.094505} {\bibfield
  {journal} {\bibinfo  {journal} {Phys. Rev. B}\ }\textbf {\bibinfo {volume}
  {93}},\ \bibinfo {pages} {094505} (\bibinfo {year}
  {2016}{\natexlab{a}})}\BibitemShut {NoStop}%
\bibitem [{\citenamefont {Shoenberg}(1984)}]{Shoenberg1984}%
  \BibitemOpen
  \bibfield  {author} {\bibinfo {author} {\bibfnamefont {D.}~\bibnamefont
  {Shoenberg}},\ }\href@noop {} {\emph {\bibinfo {title} {Magnetic Oscillations
  in Metals}}}\ (\bibinfo  {publisher} {Cambridge University Press},\ \bibinfo
  {address} {Cambridge},\ \bibinfo {year} {1984})\BibitemShut {NoStop}%
\bibitem [{\citenamefont {Bergemann}\ \emph {et~al.}(2000)\citenamefont
  {Bergemann}, \citenamefont {Julian}, \citenamefont {Mackenzie}, \citenamefont
  {NishiZaki},\ and\ \citenamefont {Maeno}}]{Bergemann2000}%
  \BibitemOpen
  \bibfield  {author} {\bibinfo {author} {\bibfnamefont {C.}~\bibnamefont
  {Bergemann}}, \bibinfo {author} {\bibfnamefont {S.~R.}\ \bibnamefont
  {Julian}}, \bibinfo {author} {\bibfnamefont {A.~P.}\ \bibnamefont
  {Mackenzie}}, \bibinfo {author} {\bibfnamefont {S.}~\bibnamefont
  {NishiZaki}},\ and\ \bibinfo {author} {\bibfnamefont {Y.}~\bibnamefont
  {Maeno}},\ }\bibfield  {title} {\bibinfo {title} {{Detailed Topography of the
  Fermi Surface of ${\mathrm{Sr}}_{2}{\mathrm{RuO}}_{4}$}},\ }\href
  {https://doi.org/10.1103/PhysRevLett.84.2662} {\bibfield  {journal} {\bibinfo
   {journal} {Phys. Rev. Lett.}\ }\textbf {\bibinfo {volume} {84}},\ \bibinfo
  {pages} {2662} (\bibinfo {year} {2000})}\BibitemShut {NoStop}%
\bibitem [{\citenamefont {Margadonna}\ \emph {et~al.}(2009)\citenamefont
  {Margadonna}, \citenamefont {Takabayashi}, \citenamefont {Ohishi},
  \citenamefont {Mizuguchi}, \citenamefont {Takano}, \citenamefont {Kagayama},
  \citenamefont {Nakagawa}, \citenamefont {Takata},\ and\ \citenamefont
  {Prassides}}]{Margadonna2009}%
  \BibitemOpen
  \bibfield  {author} {\bibinfo {author} {\bibfnamefont {S.}~\bibnamefont
  {Margadonna}}, \bibinfo {author} {\bibfnamefont {Y.}~\bibnamefont
  {Takabayashi}}, \bibinfo {author} {\bibfnamefont {Y.}~\bibnamefont {Ohishi}},
  \bibinfo {author} {\bibfnamefont {Y.}~\bibnamefont {Mizuguchi}}, \bibinfo
  {author} {\bibfnamefont {Y.}~\bibnamefont {Takano}}, \bibinfo {author}
  {\bibfnamefont {T.}~\bibnamefont {Kagayama}}, \bibinfo {author}
  {\bibfnamefont {T.}~\bibnamefont {Nakagawa}}, \bibinfo {author}
  {\bibfnamefont {M.}~\bibnamefont {Takata}},\ and\ \bibinfo {author}
  {\bibfnamefont {K.}~\bibnamefont {Prassides}},\ }\bibfield  {title} {\bibinfo
  {title} {{Pressure evolution of the low-temperature crystal structure and
  bonding of the superconductor FeSe $({T}_{\rm c}=37\text{ }\text{K})$}},\
  }\href {https://doi.org/10.1103/PhysRevB.80.064506} {\bibfield  {journal}
  {\bibinfo  {journal} {Phys. Rev. B}\ }\textbf {\bibinfo {volume} {80}},\
  \bibinfo {pages} {064506} (\bibinfo {year} {2009})}\BibitemShut {NoStop}%
\bibitem [{\citenamefont {Mandal}\ \emph {et~al.}(2014)\citenamefont {Mandal},
  \citenamefont {Cohen},\ and\ \citenamefont {Haule}}]{Mandal2014}%
  \BibitemOpen
  \bibfield  {author} {\bibinfo {author} {\bibfnamefont {S.}~\bibnamefont
  {Mandal}}, \bibinfo {author} {\bibfnamefont {R.~E.}\ \bibnamefont {Cohen}},\
  and\ \bibinfo {author} {\bibfnamefont {K.}~\bibnamefont {Haule}},\ }\bibfield
   {title} {\bibinfo {title} {{Strong pressure-dependent electron-phonon
  coupling in FeSe}},\ }\href {https://doi.org/10.1103/PhysRevB.89.220502}
  {\bibfield  {journal} {\bibinfo  {journal} {Phys. Rev. B}\ }\textbf {\bibinfo
  {volume} {89}},\ \bibinfo {pages} {220502} (\bibinfo {year}
  {2014})}\BibitemShut {NoStop}%
\bibitem [{\citenamefont {Watson}\ \emph
  {et~al.}(2015{\natexlab{b}})\citenamefont {Watson}, \citenamefont
  {Yamashita}, \citenamefont {Kasahara}, \citenamefont {Knafo}, \citenamefont
  {Nardone}, \citenamefont {B{\'{e}}ard}, \citenamefont {Hardy}, \citenamefont
  {McCollam}, \citenamefont {Narayanan}, \citenamefont {Blake}, \citenamefont
  {Wolf}, \citenamefont {Haghighirad}, \citenamefont {Meingast}, \citenamefont
  {Schofield}, \citenamefont {v.~L{\"{o}}hneysen}, \citenamefont {Matsuda},
  \citenamefont {Coldea},\ and\ \citenamefont {Shibauchi}}]{Watson2015b}%
  \BibitemOpen
  \bibfield  {author} {\bibinfo {author} {\bibfnamefont {M.~D.}\ \bibnamefont
  {Watson}}, \bibinfo {author} {\bibfnamefont {T.}~\bibnamefont {Yamashita}},
  \bibinfo {author} {\bibfnamefont {S.}~\bibnamefont {Kasahara}}, \bibinfo
  {author} {\bibfnamefont {W.}~\bibnamefont {Knafo}}, \bibinfo {author}
  {\bibfnamefont {M.}~\bibnamefont {Nardone}}, \bibinfo {author} {\bibfnamefont
  {J.}~\bibnamefont {B{\'{e}}ard}}, \bibinfo {author} {\bibfnamefont
  {F.}~\bibnamefont {Hardy}}, \bibinfo {author} {\bibfnamefont
  {A.}~\bibnamefont {McCollam}}, \bibinfo {author} {\bibfnamefont
  {A.}~\bibnamefont {Narayanan}}, \bibinfo {author} {\bibfnamefont {S.~F.}\
  \bibnamefont {Blake}}, \bibinfo {author} {\bibfnamefont {T.}~\bibnamefont
  {Wolf}}, \bibinfo {author} {\bibfnamefont {A.~A.}\ \bibnamefont
  {Haghighirad}}, \bibinfo {author} {\bibfnamefont {C.}~\bibnamefont
  {Meingast}}, \bibinfo {author} {\bibfnamefont {A.~J.}\ \bibnamefont
  {Schofield}}, \bibinfo {author} {\bibfnamefont {H.}~\bibnamefont
  {v.~L{\"{o}}hneysen}}, \bibinfo {author} {\bibfnamefont {Y.}~\bibnamefont
  {Matsuda}}, \bibinfo {author} {\bibfnamefont {A.~I.}\ \bibnamefont
  {Coldea}},\ and\ \bibinfo {author} {\bibfnamefont {T.}~\bibnamefont
  {Shibauchi}},\ }\bibfield  {title} {\bibinfo {title} {{Dichotomy between the
  Hole and Electron Behavior in Multiband Superconductor FeSe Probed by
  Ultrahigh Magnetic Fields}},\ }\href
  {https://doi.org/10.1103/PhysRevLett.115.027006} {\bibfield  {journal}
  {\bibinfo  {journal} {Phys. Rev. Lett.}\ }\textbf {\bibinfo {volume} {115}},\
  \bibinfo {pages} {027006} (\bibinfo {year} {2015}{\natexlab{b}})}\BibitemShut
  {NoStop}%
\bibitem [{\citenamefont {Farrar}\ \emph {et~al.}(2022)\citenamefont {Farrar},
  \citenamefont {Zajicek}, \citenamefont {Morfoot}, \citenamefont {Bristow},
  \citenamefont {Humphries}, \citenamefont {Haghighirad}, \citenamefont
  {McCollam}, \citenamefont {Bending},\ and\ \citenamefont
  {Coldea}}]{Farrar2022}%
  \BibitemOpen
  \bibfield  {author} {\bibinfo {author} {\bibfnamefont {L.~S.}\ \bibnamefont
  {Farrar}}, \bibinfo {author} {\bibfnamefont {Z.}~\bibnamefont {Zajicek}},
  \bibinfo {author} {\bibfnamefont {A.~B.}\ \bibnamefont {Morfoot}}, \bibinfo
  {author} {\bibfnamefont {M.}~\bibnamefont {Bristow}}, \bibinfo {author}
  {\bibfnamefont {O.~S.}\ \bibnamefont {Humphries}}, \bibinfo {author}
  {\bibfnamefont {A.~A.}\ \bibnamefont {Haghighirad}}, \bibinfo {author}
  {\bibfnamefont {A.}~\bibnamefont {McCollam}}, \bibinfo {author}
  {\bibfnamefont {S.~J.}\ \bibnamefont {Bending}},\ and\ \bibinfo {author}
  {\bibfnamefont {A.~I.}\ \bibnamefont {Coldea}},\ }\bibfield  {title}
  {\bibinfo {title} {Unconventional localization of electrons inside of a
  nematic electronic phase},\ }\href {https://doi.org/10.1073/pnas.2200405119}
  {\bibfield  {journal} {\bibinfo  {journal} {Proceedings of the National
  Academy of Sciences}\ }\textbf {\bibinfo {volume} {119}},\ \bibinfo {pages}
  {e2200405119} (\bibinfo {year} {2022})}\BibitemShut {NoStop}%
\bibitem [{\citenamefont {Luttinger}\ and\ \citenamefont
  {Ward}(1960)}]{Luttinger1960}%
  \BibitemOpen
  \bibfield  {author} {\bibinfo {author} {\bibfnamefont {J.~M.}\ \bibnamefont
  {Luttinger}}\ and\ \bibinfo {author} {\bibfnamefont {J.~C.}\ \bibnamefont
  {Ward}},\ }\bibfield  {title} {\bibinfo {title} {Ground-state energy of a
  many-fermion system. ii},\ }\href {https://doi.org/10.1103/PhysRev.118.1417}
  {\bibfield  {journal} {\bibinfo  {journal} {Phys. Rev.}\ }\textbf {\bibinfo
  {volume} {118}},\ \bibinfo {pages} {1417} (\bibinfo {year}
  {1960})}\BibitemShut {NoStop}%
\bibitem [{\citenamefont {Terashima}\ \emph
  {et~al.}(2016{\natexlab{b}})\citenamefont {Terashima}, \citenamefont
  {Kikugawa}, \citenamefont {Lin}, \citenamefont {Zhu}, \citenamefont {Wen},
  \citenamefont {Nomoto}, \citenamefont {Suzuki}, \citenamefont {Ikeda},\ and\
  \citenamefont {Uji}}]{Terashima2016FeS}%
  \BibitemOpen
  \bibfield  {author} {\bibinfo {author} {\bibfnamefont {T.}~\bibnamefont
  {Terashima}}, \bibinfo {author} {\bibfnamefont {N.}~\bibnamefont {Kikugawa}},
  \bibinfo {author} {\bibfnamefont {H.}~\bibnamefont {Lin}}, \bibinfo {author}
  {\bibfnamefont {X.}~\bibnamefont {Zhu}}, \bibinfo {author} {\bibfnamefont
  {H.-H.}\ \bibnamefont {Wen}}, \bibinfo {author} {\bibfnamefont
  {T.}~\bibnamefont {Nomoto}}, \bibinfo {author} {\bibfnamefont
  {K.}~\bibnamefont {Suzuki}}, \bibinfo {author} {\bibfnamefont
  {H.}~\bibnamefont {Ikeda}},\ and\ \bibinfo {author} {\bibfnamefont
  {S.}~\bibnamefont {Uji}},\ }\bibfield  {title} {\bibinfo {title} {{Upper
  critical field and quantum oscillations in tetragonal superconducting FeS}},\
  }\href {https://doi.org/10.1103/PhysRevB.94.100503} {\bibfield  {journal}
  {\bibinfo  {journal} {Phys. Rev. B}\ }\textbf {\bibinfo {volume} {94}},\
  \bibinfo {pages} {100503} (\bibinfo {year} {2016}{\natexlab{b}})}\BibitemShut
  {NoStop}%
\bibitem [{\citenamefont {Tomita}\ \emph {et~al.}(2015)\citenamefont {Tomita},
  \citenamefont {Takahashi}, \citenamefont {Takahashi}, \citenamefont {Okada},
  \citenamefont {Mizuguchi}, \citenamefont {Takano}, \citenamefont {Nakano},
  \citenamefont {Matsubayashi},\ and\ \citenamefont {Uwatoko}}]{Tomita2015}%
  \BibitemOpen
  \bibfield  {author} {\bibinfo {author} {\bibfnamefont {T.}~\bibnamefont
  {Tomita}}, \bibinfo {author} {\bibfnamefont {H.}~\bibnamefont {Takahashi}},
  \bibinfo {author} {\bibfnamefont {H.}~\bibnamefont {Takahashi}}, \bibinfo
  {author} {\bibfnamefont {H.}~\bibnamefont {Okada}}, \bibinfo {author}
  {\bibfnamefont {Y.}~\bibnamefont {Mizuguchi}}, \bibinfo {author}
  {\bibfnamefont {Y.}~\bibnamefont {Takano}}, \bibinfo {author} {\bibfnamefont
  {S.}~\bibnamefont {Nakano}}, \bibinfo {author} {\bibfnamefont
  {K.}~\bibnamefont {Matsubayashi}},\ and\ \bibinfo {author} {\bibfnamefont
  {Y.}~\bibnamefont {Uwatoko}},\ }\bibfield  {title} {\bibinfo {title}
  {{Correlation between $T_{\rm c}$ and Crystal Structure in S-Doped FeSe
  Superconductors under Pressure: Studied by X-ray Diffraction of
  FeSe$_{0.8}$S$_{0.2}$ at Low Temperatures}},\ }\href
  {https://doi.org/10.7566/JPSJ.84.024713} {\bibfield  {journal} {\bibinfo
  {journal} {Journal of the Physical Society of Japan}\ }\textbf {\bibinfo
  {volume} {84}},\ \bibinfo {pages} {024713} (\bibinfo {year}
  {2015})}\BibitemShut {NoStop}%
\bibitem [{\citenamefont {Morfoot}\ \emph {et~al.}(2022)\citenamefont
  {Morfoot}, \citenamefont {Kim}, \citenamefont {Watson}, \citenamefont
  {Haghighirad}, \citenamefont {Singh}, \citenamefont {Okuma},\ and\
  \citenamefont {Coldea}}]{Morfoot2022}%
  \BibitemOpen
  \bibfield  {author} {\bibinfo {author} {\bibfnamefont {A.}~\bibnamefont
  {Morfoot}}, \bibinfo {author} {\bibfnamefont {T.~K.}\ \bibnamefont {Kim}},
  \bibinfo {author} {\bibfnamefont {M.~D.}\ \bibnamefont {Watson}}, \bibinfo
  {author} {\bibfnamefont {A.~A.}\ \bibnamefont {Haghighirad}}, \bibinfo
  {author} {\bibfnamefont {S.~J.}\ \bibnamefont {Singh}}, \bibinfo {author}
  {\bibfnamefont {R.}~\bibnamefont {Okuma}},\ and\ \bibinfo {author}
  {\bibfnamefont {A.~I.}\ \bibnamefont {Coldea}},\ }\bibfield  {title}
  {\bibinfo {title} {{The evolution of the electronic structure and electronic
  correlations across the nematic electronic phase of FeSe$_{1−x}$Te$_x$}},\
  }\href@noop {} {\bibfield  {journal} {\bibinfo  {journal} {in preparation}\ }
  (\bibinfo {year} {2022})}\BibitemShut {NoStop}%
\bibitem [{\citenamefont {Yamakawa}\ and\ \citenamefont
  {Kontani}(2017)}]{Yamakawa2017}%
  \BibitemOpen
  \bibfield  {author} {\bibinfo {author} {\bibfnamefont {Y.}~\bibnamefont
  {Yamakawa}}\ and\ \bibinfo {author} {\bibfnamefont {H.}~\bibnamefont
  {Kontani}},\ }\bibfield  {title} {\bibinfo {title} {{Nematicity, magnetism,
  and superconductivity in FeSe under pressure: Unified explanation based on
  the self-consistent vertex correction theory}},\ }\href
  {https://doi.org/10.1103/PhysRevB.96.144509} {\bibfield  {journal} {\bibinfo
  {journal} {Phys. Rev. B}\ }\textbf {\bibinfo {volume} {96}},\ \bibinfo
  {pages} {144509} (\bibinfo {year} {2017})}\BibitemShut {NoStop}%
\bibitem [{\citenamefont {Xiao}\ \emph {et~al.}(2022)\citenamefont {Xiao},
  \citenamefont {Zhang}, \citenamefont {Asmara}, \citenamefont {Li},
  \citenamefont {Li}, \citenamefont {Zhang}, \citenamefont {Tseng},
  \citenamefont {Dong}, \citenamefont {Wang}, \citenamefont {Chen},
  \citenamefont {Schmitt},\ and\ \citenamefont {Peng}}]{Xiao2022}%
  \BibitemOpen
  \bibfield  {author} {\bibinfo {author} {\bibfnamefont {Q.}~\bibnamefont
  {Xiao}}, \bibinfo {author} {\bibfnamefont {W.}~\bibnamefont {Zhang}},
  \bibinfo {author} {\bibfnamefont {T.~C.}\ \bibnamefont {Asmara}}, \bibinfo
  {author} {\bibfnamefont {D.}~\bibnamefont {Li}}, \bibinfo {author}
  {\bibfnamefont {Q.}~\bibnamefont {Li}}, \bibinfo {author} {\bibfnamefont
  {S.}~\bibnamefont {Zhang}}, \bibinfo {author} {\bibfnamefont
  {Y.}~\bibnamefont {Tseng}}, \bibinfo {author} {\bibfnamefont
  {X.}~\bibnamefont {Dong}}, \bibinfo {author} {\bibfnamefont {Y.}~\bibnamefont
  {Wang}}, \bibinfo {author} {\bibfnamefont {C.-C.}\ \bibnamefont {Chen}},
  \bibinfo {author} {\bibfnamefont {T.}~\bibnamefont {Schmitt}},\ and\ \bibinfo
  {author} {\bibfnamefont {Y.}~\bibnamefont {Peng}},\ }\bibfield  {title}
  {\bibinfo {title} {{Dispersionless orbital excitations in (Li,Fe)OHFeSe
  superconductors}},\ }\href {https://doi.org/10.1038/s41535-022-00492-0}
  {\bibfield  {journal} {\bibinfo  {journal} {npj Quantum Mater.}\ }\textbf
  {\bibinfo {volume} {7}},\ \bibinfo {pages} {80} (\bibinfo {year}
  {2022})}\BibitemShut {NoStop}%
\bibitem [{\citenamefont {Lee}(2018)}]{Lee2018}%
  \BibitemOpen
  \bibfield  {author} {\bibinfo {author} {\bibfnamefont {D.-H.}\ \bibnamefont
  {Lee}},\ }\bibfield  {title} {\bibinfo {title} {{Routes to High-Temperature
  Superconductivity: A Lesson from FeSe/SrTiO$_3$}},\ }\href
  {https://doi.org/10.1146/annurev-conmatphys-033117-053942} {\bibfield
  {journal} {\bibinfo  {journal} {Annual Review of Condensed Matter Physics}\
  }\textbf {\bibinfo {volume} {9}},\ \bibinfo {pages} {261} (\bibinfo {year}
  {2018})}\BibitemShut {NoStop}%
\bibitem [{\citenamefont {Okabe}\ \emph {et~al.}(2010)\citenamefont {Okabe},
  \citenamefont {Takeshita}, \citenamefont {Horigane}, \citenamefont
  {Muranaka},\ and\ \citenamefont {Akimitsu}}]{Okabe2010}%
  \BibitemOpen
  \bibfield  {author} {\bibinfo {author} {\bibfnamefont {H.}~\bibnamefont
  {Okabe}}, \bibinfo {author} {\bibfnamefont {N.}~\bibnamefont {Takeshita}},
  \bibinfo {author} {\bibfnamefont {K.}~\bibnamefont {Horigane}}, \bibinfo
  {author} {\bibfnamefont {T.}~\bibnamefont {Muranaka}},\ and\ \bibinfo
  {author} {\bibfnamefont {J.}~\bibnamefont {Akimitsu}},\ }\bibfield  {title}
  {\bibinfo {title} {{Pressure-induced high-${T}_{c}$ superconducting phase in
  FeSe: Correlation between anion height and ${T}_{c}$}},\ }\href
  {https://doi.org/10.1103/PhysRevB.81.205119} {\bibfield  {journal} {\bibinfo
  {journal} {Phys. Rev. B}\ }\textbf {\bibinfo {volume} {81}},\ \bibinfo
  {pages} {205119} (\bibinfo {year} {2010})}\BibitemShut {NoStop}%
\bibitem [{\citenamefont {Kumar}\ \emph {et~al.}(2012)\citenamefont {Kumar},
  \citenamefont {Auluck}, \citenamefont {Ahluwalia},\ and\ \citenamefont
  {Awana}}]{Kumar2012}%
  \BibitemOpen
  \bibfield  {author} {\bibinfo {author} {\bibfnamefont {J.}~\bibnamefont
  {Kumar}}, \bibinfo {author} {\bibfnamefont {S.}~\bibnamefont {Auluck}},
  \bibinfo {author} {\bibfnamefont {P.~K.}\ \bibnamefont {Ahluwalia}},\ and\
  \bibinfo {author} {\bibfnamefont {V.~P.~S.}\ \bibnamefont {Awana}},\
  }\bibfield  {title} {\bibinfo {title} {{Chalcogen height dependence of
  magnetism and Fermiology in FeTe$_x$Se$_{1-x}$}},\ }\href
  {https://doi.org/10.1088/0953-2048/25/9/095002} {\bibfield  {journal}
  {\bibinfo  {journal} {Superconductor Science and Technology}\ }\textbf
  {\bibinfo {volume} {25}},\ \bibinfo {pages} {095002} (\bibinfo {year}
  {2012})}\BibitemShut {NoStop}%
\bibitem [{\citenamefont {Scherer}\ \emph {et~al.}(2017)\citenamefont
  {Scherer}, \citenamefont {Jacko}, \citenamefont {Friedrich}, \citenamefont
  {\ifmmode \mbox{\c{S}}\else \c{S}\fi{}a\ifmmode \mbox{\c{s}}\else
  \c{s}\fi{}\ifmmode \imath \else \i \fi{}o\ifmmode~\breve{g}\else
  \u{g}\fi{}lu}, \citenamefont {Bl\"ugel}, \citenamefont {Valent\'{\i}},\ and\
  \citenamefont {Andersen}}]{Scherer2017}%
  \BibitemOpen
  \bibfield  {author} {\bibinfo {author} {\bibfnamefont {D.~D.}\ \bibnamefont
  {Scherer}}, \bibinfo {author} {\bibfnamefont {A.~C.}\ \bibnamefont {Jacko}},
  \bibinfo {author} {\bibfnamefont {C.}~\bibnamefont {Friedrich}}, \bibinfo
  {author} {\bibfnamefont {E.}~\bibnamefont {\ifmmode \mbox{\c{S}}\else
  \c{S}\fi{}a\ifmmode \mbox{\c{s}}\else \c{s}\fi{}\ifmmode \imath \else \i
  \fi{}o\ifmmode~\breve{g}\else \u{g}\fi{}lu}}, \bibinfo {author}
  {\bibfnamefont {S.}~\bibnamefont {Bl\"ugel}}, \bibinfo {author}
  {\bibfnamefont {R.}~\bibnamefont {Valent\'{\i}}},\ and\ \bibinfo {author}
  {\bibfnamefont {B.~M.}\ \bibnamefont {Andersen}},\ }\bibfield  {title}
  {\bibinfo {title} {{Interplay of nematic and magnetic orders in FeSe under
  pressure}},\ }\href {https://doi.org/10.1103/PhysRevB.95.094504} {\bibfield
  {journal} {\bibinfo  {journal} {Phys. Rev. B}\ }\textbf {\bibinfo {volume}
  {95}},\ \bibinfo {pages} {094504} (\bibinfo {year} {2017})}\BibitemShut
  {NoStop}%
\bibitem [{\citenamefont {Rana}\ \emph {et~al.}(2023)\citenamefont {Rana},
  \citenamefont {Ambika}, \citenamefont {Bud'ko}, \citenamefont {B\"ohmer},
  \citenamefont {Canfield},\ and\ \citenamefont {Furukawa}}]{Rana2023}%
  \BibitemOpen
  \bibfield  {author} {\bibinfo {author} {\bibfnamefont {K.}~\bibnamefont
  {Rana}}, \bibinfo {author} {\bibfnamefont {D.~V.}\ \bibnamefont {Ambika}},
  \bibinfo {author} {\bibfnamefont {S.~L.}\ \bibnamefont {Bud'ko}}, \bibinfo
  {author} {\bibfnamefont {A.~E.}\ \bibnamefont {B\"ohmer}}, \bibinfo {author}
  {\bibfnamefont {P.~C.}\ \bibnamefont {Canfield}},\ and\ \bibinfo {author}
  {\bibfnamefont {Y.}~\bibnamefont {Furukawa}},\ }\bibfield  {title} {\bibinfo
  {title} {{Interrelationships between nematicity, antiferromagnetic spin
  fluctuations, and superconductivity: Role of hotspots in
  ${\mathrm{FeSe}}_{1\ensuremath{-}x}{\mathrm{S}}_{x}$ revealed by high
  pressure $^{77}\mathrm{Se}$ NMR study}},\ }\href
  {https://doi.org/10.1103/PhysRevB.107.134507} {\bibfield  {journal} {\bibinfo
   {journal} {Phys. Rev. B}\ }\textbf {\bibinfo {volume} {107}},\ \bibinfo
  {pages} {134507} (\bibinfo {year} {2023})}\BibitemShut {NoStop}%
\bibitem [{\citenamefont {Wiecki}\ \emph {et~al.}(2018)\citenamefont {Wiecki},
  \citenamefont {Rana}, \citenamefont {B\"ohmer}, \citenamefont {Lee},
  \citenamefont {Bud'ko}, \citenamefont {Canfield},\ and\ \citenamefont
  {Furukawa}}]{Wiecki2018}%
  \BibitemOpen
  \bibfield  {author} {\bibinfo {author} {\bibfnamefont {P.}~\bibnamefont
  {Wiecki}}, \bibinfo {author} {\bibfnamefont {K.}~\bibnamefont {Rana}},
  \bibinfo {author} {\bibfnamefont {A.~E.}\ \bibnamefont {B\"ohmer}}, \bibinfo
  {author} {\bibfnamefont {Y.}~\bibnamefont {Lee}}, \bibinfo {author}
  {\bibfnamefont {S.~L.}\ \bibnamefont {Bud'ko}}, \bibinfo {author}
  {\bibfnamefont {P.~C.}\ \bibnamefont {Canfield}},\ and\ \bibinfo {author}
  {\bibfnamefont {Y.}~\bibnamefont {Furukawa}},\ }\bibfield  {title} {\bibinfo
  {title} {{Persistent correlation between superconductivity and
  antiferromagnetic fluctuations near a nematic quantum critical point in
  ${\mathrm{FeSe}}_{1\ensuremath{-}x}{\mathrm{S}}_{x}$}},\ }\href
  {https://doi.org/10.1103/PhysRevB.98.020507} {\bibfield  {journal} {\bibinfo
  {journal} {Phys. Rev. B}\ }\textbf {\bibinfo {volume} {98}},\ \bibinfo
  {pages} {020507} (\bibinfo {year} {2018})}\BibitemShut {NoStop}%
\bibitem [{\citenamefont {Kuwayama}\ \emph {et~al.}(2021)\citenamefont
  {Kuwayama}, \citenamefont {Matsuura}, \citenamefont {Gouchi}, \citenamefont
  {Yamakawa}, \citenamefont {Mizukami}, \citenamefont {Kasahara}, \citenamefont
  {Matsuda}, \citenamefont {Shibauchi}, \citenamefont {Kontani}, \citenamefont
  {Uwatoko},\ and\ \citenamefont {Fujiwara}}]{Kuwayama2021}%
  \BibitemOpen
  \bibfield  {author} {\bibinfo {author} {\bibfnamefont {T.}~\bibnamefont
  {Kuwayama}}, \bibinfo {author} {\bibfnamefont {K.}~\bibnamefont {Matsuura}},
  \bibinfo {author} {\bibfnamefont {J.}~\bibnamefont {Gouchi}}, \bibinfo
  {author} {\bibfnamefont {Y.}~\bibnamefont {Yamakawa}}, \bibinfo {author}
  {\bibfnamefont {Y.}~\bibnamefont {Mizukami}}, \bibinfo {author}
  {\bibfnamefont {S.}~\bibnamefont {Kasahara}}, \bibinfo {author}
  {\bibfnamefont {Y.}~\bibnamefont {Matsuda}}, \bibinfo {author} {\bibfnamefont
  {T.}~\bibnamefont {Shibauchi}}, \bibinfo {author} {\bibfnamefont
  {H.}~\bibnamefont {Kontani}}, \bibinfo {author} {\bibfnamefont
  {Y.}~\bibnamefont {Uwatoko}},\ and\ \bibinfo {author} {\bibfnamefont
  {N.}~\bibnamefont {Fujiwara}},\ }\bibfield  {title} {\bibinfo {title}
  {{Pressure-induced reconstitution of Fermi surfaces and spin fluctuations in
  S-substituted FeSe}},\ }\href {https://doi.org/10.1038/s41598-021-96277-9}
  {\bibfield  {journal} {\bibinfo  {journal} {Scientific Reports}\ }\textbf
  {\bibinfo {volume} {11}},\ \bibinfo {pages} {17265} (\bibinfo {year}
  {2021})}\BibitemShut {NoStop}%
\bibitem [{\citenamefont {Yip}\ \emph {et~al.}(2017)\citenamefont {Yip},
  \citenamefont {Chan}, \citenamefont {Niu}, \citenamefont {Matsuura},
  \citenamefont {Mizukami}, \citenamefont {Kasahara}, \citenamefont {Matsuda},
  \citenamefont {Shibauchi},\ and\ \citenamefont {Goh}}]{Yip2017}%
  \BibitemOpen
  \bibfield  {author} {\bibinfo {author} {\bibfnamefont {K.~Y.}\ \bibnamefont
  {Yip}}, \bibinfo {author} {\bibfnamefont {Y.~C.}\ \bibnamefont {Chan}},
  \bibinfo {author} {\bibfnamefont {Q.}~\bibnamefont {Niu}}, \bibinfo {author}
  {\bibfnamefont {K.}~\bibnamefont {Matsuura}}, \bibinfo {author}
  {\bibfnamefont {Y.}~\bibnamefont {Mizukami}}, \bibinfo {author}
  {\bibfnamefont {S.}~\bibnamefont {Kasahara}}, \bibinfo {author}
  {\bibfnamefont {Y.}~\bibnamefont {Matsuda}}, \bibinfo {author} {\bibfnamefont
  {T.}~\bibnamefont {Shibauchi}},\ and\ \bibinfo {author} {\bibfnamefont
  {S.~K.}\ \bibnamefont {Goh}},\ }\bibfield  {title} {\bibinfo {title}
  {{Weakening of the diamagnetic shielding in FeSe$_{1-x}$S$_x$ at high
  pressures}},\ }\href {https://doi.org/10.1103/PhysRevB.96.020502} {\bibfield
  {journal} {\bibinfo  {journal} {Phys. Rev. B}\ }\textbf {\bibinfo {volume}
  {96}},\ \bibinfo {pages} {020502} (\bibinfo {year} {2017})}\BibitemShut
  {NoStop}%
\bibitem [{\citenamefont {Ayres}\ \emph {et~al.}(2022)\citenamefont {Ayres},
  \citenamefont {{\v{C}}ulo}, \citenamefont {Buhot}, \citenamefont
  {Bern{\'{a}}th}, \citenamefont {Kasahara}, \citenamefont {Matsuda},
  \citenamefont {Shibauchi}, \citenamefont {Carrington}, \citenamefont
  {Friedemann},\ and\ \citenamefont {Hussey}}]{Ayres2022}%
  \BibitemOpen
  \bibfield  {author} {\bibinfo {author} {\bibfnamefont {J.}~\bibnamefont
  {Ayres}}, \bibinfo {author} {\bibfnamefont {M.}~\bibnamefont {{\v{C}}ulo}},
  \bibinfo {author} {\bibfnamefont {J.}~\bibnamefont {Buhot}}, \bibinfo
  {author} {\bibfnamefont {B.}~\bibnamefont {Bern{\'{a}}th}}, \bibinfo {author}
  {\bibfnamefont {S.}~\bibnamefont {Kasahara}}, \bibinfo {author}
  {\bibfnamefont {Y.}~\bibnamefont {Matsuda}}, \bibinfo {author} {\bibfnamefont
  {T.}~\bibnamefont {Shibauchi}}, \bibinfo {author} {\bibfnamefont
  {A.}~\bibnamefont {Carrington}}, \bibinfo {author} {\bibfnamefont
  {S.}~\bibnamefont {Friedemann}},\ and\ \bibinfo {author} {\bibfnamefont
  {N.~E.}\ \bibnamefont {Hussey}},\ }\bibfield  {title} {\bibinfo {title}
  {{Transport evidence for decoupled nematic and magnetic criticality in iron
  chalcogenides}},\ }\href {https://doi.org/10.1038/s42005-022-00873-8}
  {\bibfield  {journal} {\bibinfo  {journal} {Communications Physics}\ }\textbf
  {\bibinfo {volume} {5}},\ \bibinfo {pages} {100} (\bibinfo {year}
  {2022})}\BibitemShut {NoStop}%
\bibitem [{\citenamefont {Gati}\ \emph {et~al.}(2019)\citenamefont {Gati},
  \citenamefont {B\"ohmer}, \citenamefont {Bud'ko},\ and\ \citenamefont
  {Canfield}}]{Gati2019}%
  \BibitemOpen
  \bibfield  {author} {\bibinfo {author} {\bibfnamefont {E.}~\bibnamefont
  {Gati}}, \bibinfo {author} {\bibfnamefont {A.~E.}\ \bibnamefont {B\"ohmer}},
  \bibinfo {author} {\bibfnamefont {S.~L.}\ \bibnamefont {Bud'ko}},\ and\
  \bibinfo {author} {\bibfnamefont {P.~C.}\ \bibnamefont {Canfield}},\
  }\bibfield  {title} {\bibinfo {title} {{Bulk Superconductivity and Role of
  Fluctuations in the Iron-Based Superconductor FeSe at High Pressures}},\
  }\href {https://doi.org/10.1103/PhysRevLett.123.167002} {\bibfield  {journal}
  {\bibinfo  {journal} {Phys. Rev. Lett.}\ }\textbf {\bibinfo {volume} {123}},\
  \bibinfo {pages} {167002} (\bibinfo {year} {2019})}\BibitemShut {NoStop}%
\bibitem [{\citenamefont {Reiss}\ \emph {et~al.}(2021)\citenamefont {Reiss},
  \citenamefont {Graf}, \citenamefont {Haghighirad}, \citenamefont {Vojta},\
  and\ \citenamefont {Coldea}}]{Reiss2021}%
  \BibitemOpen
  \bibfield  {author} {\bibinfo {author} {\bibfnamefont {P.}~\bibnamefont
  {Reiss}}, \bibinfo {author} {\bibfnamefont {D.}~\bibnamefont {Graf}},
  \bibinfo {author} {\bibfnamefont {A.~A.}\ \bibnamefont {Haghighirad}},
  \bibinfo {author} {\bibfnamefont {T.}~\bibnamefont {Vojta}},\ and\ \bibinfo
  {author} {\bibfnamefont {A.~I.}\ \bibnamefont {Coldea}},\ }\bibfield  {title}
  {\bibinfo {title} {{Signatures of a Quantum Griffiths Phase Close to an
  Electronic Nematic Quantum Phase Transition}},\ }\href
  {https://doi.org/10.1103/PhysRevLett.127.246402} {\bibfield  {journal}
  {\bibinfo  {journal} {Phys. Rev. Lett.}\ }\textbf {\bibinfo {volume} {127}},\
  \bibinfo {pages} {246402} (\bibinfo {year} {2021})}\BibitemShut {NoStop}%
\bibitem [{\citenamefont {Walker}\ \emph {et~al.}(2023)\citenamefont {Walker},
  \citenamefont {Scott}, \citenamefont {Boyle}, \citenamefont {Byland},
  \citenamefont {B{\"{o}}tzel}, \citenamefont {Zhao}, \citenamefont {Day},
  \citenamefont {Zhdanovich}, \citenamefont {Gorovikov}, \citenamefont
  {Pedersen}, \citenamefont {Klavins}, \citenamefont {Damascelli},
  \citenamefont {Eremin}, \citenamefont {Gozar}, \citenamefont {Taufour},\ and\
  \citenamefont {{da Silva Neto}}}]{Walker2023}%
  \BibitemOpen
  \bibfield  {author} {\bibinfo {author} {\bibfnamefont {M.}~\bibnamefont
  {Walker}}, \bibinfo {author} {\bibfnamefont {K.}~\bibnamefont {Scott}},
  \bibinfo {author} {\bibfnamefont {T.~J.}\ \bibnamefont {Boyle}}, \bibinfo
  {author} {\bibfnamefont {J.~K.}\ \bibnamefont {Byland}}, \bibinfo {author}
  {\bibfnamefont {S.}~\bibnamefont {B{\"{o}}tzel}}, \bibinfo {author}
  {\bibfnamefont {Z.}~\bibnamefont {Zhao}}, \bibinfo {author} {\bibfnamefont
  {R.~P.}\ \bibnamefont {Day}}, \bibinfo {author} {\bibfnamefont
  {S.}~\bibnamefont {Zhdanovich}}, \bibinfo {author} {\bibfnamefont
  {S.}~\bibnamefont {Gorovikov}}, \bibinfo {author} {\bibfnamefont {T.~M.}\
  \bibnamefont {Pedersen}}, \bibinfo {author} {\bibfnamefont {P.}~\bibnamefont
  {Klavins}}, \bibinfo {author} {\bibfnamefont {A.}~\bibnamefont {Damascelli}},
  \bibinfo {author} {\bibfnamefont {I.~M.}\ \bibnamefont {Eremin}}, \bibinfo
  {author} {\bibfnamefont {A.}~\bibnamefont {Gozar}}, \bibinfo {author}
  {\bibfnamefont {V.}~\bibnamefont {Taufour}},\ and\ \bibinfo {author}
  {\bibfnamefont {E.~H.}\ \bibnamefont {{da Silva Neto}}},\ }\bibfield  {title}
  {\bibinfo {title} {{Electronic stripe patterns near the Fermi level of
  tetragonal Fe(Se,S)}},\ }\href {https://doi.org/10.1038/s41535-023-00592-5}
  {\bibfield  {journal} {\bibinfo  {journal} {npj Quantum Mater.}\ }\textbf
  {\bibinfo {volume} {8}},\ \bibinfo {pages} {60} (\bibinfo {year}
  {2023})}\BibitemShut {NoStop}%
\bibitem [{\citenamefont {Zhao}\ \emph {et~al.}(1997)\citenamefont {Zhao},
  \citenamefont {Hunt}, \citenamefont {Keller},\ and\ \citenamefont
  {M{\"{u}}ller}}]{Zhao1997}%
  \BibitemOpen
  \bibfield  {author} {\bibinfo {author} {\bibfnamefont {G.-M.}\ \bibnamefont
  {Zhao}}, \bibinfo {author} {\bibfnamefont {M.~B.}\ \bibnamefont {Hunt}},
  \bibinfo {author} {\bibfnamefont {H.}~\bibnamefont {Keller}},\ and\ \bibinfo
  {author} {\bibfnamefont {K.~A.}\ \bibnamefont {M{\"{u}}ller}},\ }\bibfield
  {title} {\bibinfo {title} {{Evidence for polaronic supercarriers in the
  copper oxide superconductors La$_{2–x}$Sr$_x$CuO$_4$}},\ }\href
  {https://doi.org/10.1038/385236a0} {\bibfield  {journal} {\bibinfo  {journal}
  {Nature}\ }\textbf {\bibinfo {volume} {385}},\ \bibinfo {pages} {236}
  (\bibinfo {year} {1997})}\BibitemShut {NoStop}%
\bibitem [{\citenamefont {Gerber}\ \emph {et~al.}(2017)\citenamefont {Gerber},
  \citenamefont {Yang}, \citenamefont {Zhu}, \citenamefont {Soifer},
  \citenamefont {Sobota}, \citenamefont {Rebec}, \citenamefont {Lee},
  \citenamefont {Jia}, \citenamefont {Moritz}, \citenamefont {Jia},
  \citenamefont {Gauthier}, \citenamefont {Li}, \citenamefont {Leuenberger},
  \citenamefont {Zhang}, \citenamefont {Chaix}, \citenamefont {Li},
  \citenamefont {Jang}, \citenamefont {Lee}, \citenamefont {Yi}, \citenamefont
  {Dakovski}, \citenamefont {Song}, \citenamefont {Glownia}, \citenamefont
  {Nelson}, \citenamefont {Kim}, \citenamefont {Chuang}, \citenamefont
  {Hussain}, \citenamefont {Moore}, \citenamefont {Devereaux}, \citenamefont
  {Lee}, \citenamefont {Kirchmann},\ and\ \citenamefont {Shen}}]{Gerber2017}%
  \BibitemOpen
  \bibfield  {author} {\bibinfo {author} {\bibfnamefont {S.}~\bibnamefont
  {Gerber}}, \bibinfo {author} {\bibfnamefont {S.-L.}\ \bibnamefont {Yang}},
  \bibinfo {author} {\bibfnamefont {D.}~\bibnamefont {Zhu}}, \bibinfo {author}
  {\bibfnamefont {H.}~\bibnamefont {Soifer}}, \bibinfo {author} {\bibfnamefont
  {J.~A.}\ \bibnamefont {Sobota}}, \bibinfo {author} {\bibfnamefont
  {S.}~\bibnamefont {Rebec}}, \bibinfo {author} {\bibfnamefont {J.~J.}\
  \bibnamefont {Lee}}, \bibinfo {author} {\bibfnamefont {T.}~\bibnamefont
  {Jia}}, \bibinfo {author} {\bibfnamefont {B.}~\bibnamefont {Moritz}},
  \bibinfo {author} {\bibfnamefont {C.}~\bibnamefont {Jia}}, \bibinfo {author}
  {\bibfnamefont {A.}~\bibnamefont {Gauthier}}, \bibinfo {author}
  {\bibfnamefont {Y.}~\bibnamefont {Li}}, \bibinfo {author} {\bibfnamefont
  {D.}~\bibnamefont {Leuenberger}}, \bibinfo {author} {\bibfnamefont
  {Y.}~\bibnamefont {Zhang}}, \bibinfo {author} {\bibfnamefont
  {L.}~\bibnamefont {Chaix}}, \bibinfo {author} {\bibfnamefont
  {W.}~\bibnamefont {Li}}, \bibinfo {author} {\bibfnamefont {H.}~\bibnamefont
  {Jang}}, \bibinfo {author} {\bibfnamefont {J.-S.}\ \bibnamefont {Lee}},
  \bibinfo {author} {\bibfnamefont {M.}~\bibnamefont {Yi}}, \bibinfo {author}
  {\bibfnamefont {G.~L.}\ \bibnamefont {Dakovski}}, \bibinfo {author}
  {\bibfnamefont {S.}~\bibnamefont {Song}}, \bibinfo {author} {\bibfnamefont
  {J.~M.}\ \bibnamefont {Glownia}}, \bibinfo {author} {\bibfnamefont
  {S.}~\bibnamefont {Nelson}}, \bibinfo {author} {\bibfnamefont {K.~W.}\
  \bibnamefont {Kim}}, \bibinfo {author} {\bibfnamefont {Y.-D.}\ \bibnamefont
  {Chuang}}, \bibinfo {author} {\bibfnamefont {Z.}~\bibnamefont {Hussain}},
  \bibinfo {author} {\bibfnamefont {R.~G.}\ \bibnamefont {Moore}}, \bibinfo
  {author} {\bibfnamefont {T.~P.}\ \bibnamefont {Devereaux}}, \bibinfo {author}
  {\bibfnamefont {W.-S.}\ \bibnamefont {Lee}}, \bibinfo {author} {\bibfnamefont
  {P.~S.}\ \bibnamefont {Kirchmann}},\ and\ \bibinfo {author} {\bibfnamefont
  {Z.-X.}\ \bibnamefont {Shen}},\ }\bibfield  {title} {\bibinfo {title}
  {{Femtosecond electron-phonon lock-in by photoemission and x-ray
  free-electron laser}},\ }\href {https://doi.org/10.1126/science.aak9946}
  {\bibfield  {journal} {\bibinfo  {journal} {Science}\ }\textbf {\bibinfo
  {volume} {357}},\ \bibinfo {pages} {71} (\bibinfo {year} {2017})}\BibitemShut
  {NoStop}%
\bibitem [{\citenamefont {Ding}\ \emph {et~al.}(2022)\citenamefont {Ding},
  \citenamefont {Wang}, \citenamefont {Wei}, \citenamefont {Gao}, \citenamefont
  {Cui},\ and\ \citenamefont {Zhang}}]{Ding2022}%
  \BibitemOpen
  \bibfield  {author} {\bibinfo {author} {\bibfnamefont {W.}~\bibnamefont
  {Ding}}, \bibinfo {author} {\bibfnamefont {Y.}~\bibnamefont {Wang}}, \bibinfo
  {author} {\bibfnamefont {T.}~\bibnamefont {Wei}}, \bibinfo {author}
  {\bibfnamefont {J.}~\bibnamefont {Gao}}, \bibinfo {author} {\bibfnamefont
  {P.}~\bibnamefont {Cui}},\ and\ \bibinfo {author} {\bibfnamefont
  {Z.}~\bibnamefont {Zhang}},\ }\bibfield  {title} {\bibinfo {title}
  {{Correlation-enhanced electron-phonon coupling for accurate evaluation of
  the superconducting transition temperature in bulk FeSe}},\ }\href
  {https://doi.org/10.1007/s11433-022-1888-2} {\bibfield  {journal} {\bibinfo
  {journal} {Science China Physics, Mechanics and Astronomy}\ }\textbf
  {\bibinfo {volume} {65}},\ \bibinfo {pages} {267412} (\bibinfo {year}
  {2022})}\BibitemShut {NoStop}%
\bibitem [{\citenamefont {B{\"{o}}hmer}\ \emph {et~al.}(2013)\citenamefont
  {B{\"{o}}hmer}, \citenamefont {Hardy}, \citenamefont {Eilers}, \citenamefont
  {Ernst}, \citenamefont {Adelmann}, \citenamefont {Schweiss}, \citenamefont
  {Wolf},\ and\ \citenamefont {Meingast}}]{Bohmer2013}%
  \BibitemOpen
  \bibfield  {author} {\bibinfo {author} {\bibfnamefont {A.~E.}\ \bibnamefont
  {B{\"{o}}hmer}}, \bibinfo {author} {\bibfnamefont {F.}~\bibnamefont {Hardy}},
  \bibinfo {author} {\bibfnamefont {F.}~\bibnamefont {Eilers}}, \bibinfo
  {author} {\bibfnamefont {D.}~\bibnamefont {Ernst}}, \bibinfo {author}
  {\bibfnamefont {P.}~\bibnamefont {Adelmann}}, \bibinfo {author}
  {\bibfnamefont {P.}~\bibnamefont {Schweiss}}, \bibinfo {author}
  {\bibfnamefont {T.}~\bibnamefont {Wolf}},\ and\ \bibinfo {author}
  {\bibfnamefont {C.}~\bibnamefont {Meingast}},\ }\bibfield  {title} {\bibinfo
  {title} {{Lack of coupling between superconductivity and orthorhombic
  distortion in stoichiometric single-crystalline FeSe}},\ }\href
  {https://doi.org/10.1103/PhysRevB.87.180505} {\bibfield  {journal} {\bibinfo
  {journal} {Phys. Rev. B}\ }\textbf {\bibinfo {volume} {87}},\ \bibinfo
  {pages} {180505} (\bibinfo {year} {2013})}\BibitemShut {NoStop}%
\bibitem [{\citenamefont {B\"ohmer}\ \emph
  {et~al.}(2016{\natexlab{b}})\citenamefont {B\"ohmer}, \citenamefont
  {Taufour}, \citenamefont {Straszheim}, \citenamefont {Wolf},\ and\
  \citenamefont {Canfield}}]{Bohmer2016g}%
  \BibitemOpen
  \bibfield  {author} {\bibinfo {author} {\bibfnamefont {A.~E.}\ \bibnamefont
  {B\"ohmer}}, \bibinfo {author} {\bibfnamefont {V.}~\bibnamefont {Taufour}},
  \bibinfo {author} {\bibfnamefont {W.~E.}\ \bibnamefont {Straszheim}},
  \bibinfo {author} {\bibfnamefont {T.}~\bibnamefont {Wolf}},\ and\ \bibinfo
  {author} {\bibfnamefont {P.~C.}\ \bibnamefont {Canfield}},\ }\bibfield
  {title} {\bibinfo {title} {{Variation of transition temperatures and residual
  resistivity ratio in vapor-grown FeSe}},\ }\href
  {https://doi.org/10.1103/PhysRevB.94.024526} {\bibfield  {journal} {\bibinfo
  {journal} {Phys. Rev. B}\ }\textbf {\bibinfo {volume} {94}},\ \bibinfo
  {pages} {024526} (\bibinfo {year} {2016}{\natexlab{b}})}\BibitemShut
  {NoStop}%
\bibitem [{\citenamefont {Carrington}(2011)}]{Carrington2011}%
  \BibitemOpen
  \bibfield  {author} {\bibinfo {author} {\bibfnamefont {A.}~\bibnamefont
  {Carrington}},\ }\bibfield  {title} {\bibinfo {title} {{Quantum oscillation
  studies of the Fermi surface of iron-pnictide superconductors}},\ }\href
  {https://doi.org/10.1088/0034-4885/74/12/124507} {\bibfield  {journal}
  {\bibinfo  {journal} {Reports Prog. Phys.}\ }\textbf {\bibinfo {volume}
  {74}},\ \bibinfo {pages} {124507} (\bibinfo {year} {2011})}\BibitemShut
  {NoStop}%
\bibitem [{\citenamefont {Rourke}\ \emph {et~al.}(2010)\citenamefont {Rourke},
  \citenamefont {Bangura}, \citenamefont {Benseman}, \citenamefont {Matusiak},
  \citenamefont {Cooper}, \citenamefont {Carrington},\ and\ \citenamefont
  {Hussey}}]{Rourke2010}%
  \BibitemOpen
  \bibfield  {author} {\bibinfo {author} {\bibfnamefont {P.~M.~C.}\
  \bibnamefont {Rourke}}, \bibinfo {author} {\bibfnamefont {A.~F.}\
  \bibnamefont {Bangura}}, \bibinfo {author} {\bibfnamefont {T.~M.}\
  \bibnamefont {Benseman}}, \bibinfo {author} {\bibfnamefont {M.}~\bibnamefont
  {Matusiak}}, \bibinfo {author} {\bibfnamefont {J.~R.}\ \bibnamefont
  {Cooper}}, \bibinfo {author} {\bibfnamefont {A.}~\bibnamefont {Carrington}},\
  and\ \bibinfo {author} {\bibfnamefont {N.~E.}\ \bibnamefont {Hussey}},\
  }\bibfield  {title} {\bibinfo {title} {{A detailed de Haas–van Alphen
  effect study of the overdoped cuprate Tl$_2$Ba$_2$CuO$_{6+\delta}$}},\ }\href
  {https://doi.org/10.1088/1367-2630/12/10/105009} {\bibfield  {journal}
  {\bibinfo  {journal} {New Journal of Physics}\ }\textbf {\bibinfo {volume}
  {12}},\ \bibinfo {pages} {105009} (\bibinfo {year} {2010})}\BibitemShut
  {NoStop}%
\bibitem [{\citenamefont {Bates}(2022)}]{Bates2020}%
  \BibitemOpen
  \bibfield  {author} {\bibinfo {author} {\bibfnamefont {J.}~\bibnamefont
  {Bates}},\ }\bibfield  {title} {\bibinfo {title} {{Fermi surface topology of
  iron-based superconductors}},\ }\href@noop {} {\bibfield  {journal} {\bibinfo
   {journal} {MPhys project {\rm (University of Oxford)}}\ } (\bibinfo {year}
  {2022})}\BibitemShut {NoStop}%
\bibitem [{\citenamefont {Tsujii}\ \emph {et~al.}(2022)\citenamefont {Tsujii},
  \citenamefont {Ishida}, \citenamefont {Ishida}, \citenamefont {Mizukami},
  \citenamefont {Iyo}, \citenamefont {Eisaki},\ and\ \citenamefont
  {Shibauchi}}]{Tsujii2022}%
  \BibitemOpen
  \bibfield  {author} {\bibinfo {author} {\bibfnamefont {M.}~\bibnamefont
  {Tsujii}}, \bibinfo {author} {\bibfnamefont {K.}~\bibnamefont {Ishida}},
  \bibinfo {author} {\bibfnamefont {S.}~\bibnamefont {Ishida}}, \bibinfo
  {author} {\bibfnamefont {Y.}~\bibnamefont {Mizukami}}, \bibinfo {author}
  {\bibfnamefont {A.}~\bibnamefont {Iyo}}, \bibinfo {author} {\bibfnamefont
  {H.}~\bibnamefont {Eisaki}},\ and\ \bibinfo {author} {\bibfnamefont
  {T.}~\bibnamefont {Shibauchi}},\ }\bibfield  {title} {\bibinfo {title}
  {{Charge Transport in Ba$_{1−x}$Rb$_{x}$Fe$_{2}$As$_{2}$ Single
  Crystals}},\ }\href {https://doi.org/10.7566/JPSJ.91.104706} {\bibfield
  {journal} {\bibinfo  {journal} {Journal of the Physical Society of Japan}\
  }\textbf {\bibinfo {volume} {91}},\ \bibinfo {pages} {104706} (\bibinfo
  {year} {2022})}\BibitemShut {NoStop}%
\bibitem [{\citenamefont {E.~Hussey}(2005)}]{Hussey2005}%
  \BibitemOpen
  \bibfield  {author} {\bibinfo {author} {\bibfnamefont {N.}~\bibnamefont
  {E.~Hussey}},\ }\bibfield  {title} {\bibinfo {title} {{Non-generality of the
  Kadowaki–Woods Ratio in Correlated Oxides}},\ }\href
  {https://doi.org/10.1143/JPSJ.74.1107} {\bibfield  {journal} {\bibinfo
  {journal} {Journal of the Physical Society of Japan}\ }\textbf {\bibinfo
  {volume} {74}},\ \bibinfo {pages} {1107} (\bibinfo {year}
  {2005})}\BibitemShut {NoStop}%
\bibitem [{\citenamefont {Watson}\ \emph
  {et~al.}(2015{\natexlab{c}})\citenamefont {Watson}, \citenamefont {Kim},
  \citenamefont {Haghighirad}, \citenamefont {Blake}, \citenamefont {Davies},
  \citenamefont {Hoesch}, \citenamefont {Wolf},\ and\ \citenamefont
  {Coldea}}]{Watson2015c}%
  \BibitemOpen
  \bibfield  {author} {\bibinfo {author} {\bibfnamefont {M.~D.}\ \bibnamefont
  {Watson}}, \bibinfo {author} {\bibfnamefont {T.~K.}\ \bibnamefont {Kim}},
  \bibinfo {author} {\bibfnamefont {A.~A.}\ \bibnamefont {Haghighirad}},
  \bibinfo {author} {\bibfnamefont {S.~F.}\ \bibnamefont {Blake}}, \bibinfo
  {author} {\bibfnamefont {N.~R.}\ \bibnamefont {Davies}}, \bibinfo {author}
  {\bibfnamefont {M.}~\bibnamefont {Hoesch}}, \bibinfo {author} {\bibfnamefont
  {T.}~\bibnamefont {Wolf}},\ and\ \bibinfo {author} {\bibfnamefont {A.~I.}\
  \bibnamefont {Coldea}},\ }\bibfield  {title} {\bibinfo {title} {{Suppression
  of orbital ordering by chemical pressure in FeSe$_{1-x}$S$_x$}},\ }\href
  {https://doi.org/10.1103/PhysRevB.92.121108} {\bibfield  {journal} {\bibinfo
  {journal} {Phys. Rev. B}\ }\textbf {\bibinfo {volume} {92}},\ \bibinfo
  {pages} {121108} (\bibinfo {year} {2015}{\natexlab{c}})}\BibitemShut
  {NoStop}%
\end{thebibliography}%

\newcommand{\blue}{\textcolor{blue}}
\newcommand{\bdm}[1]{\mbox{\boldmath $#1$}}

\renewcommand{\thefigure}{S\arabic{figure}} 
\renewcommand{\thetable}{S\arabic{table}} 

\newlength{\figwidth}
\figwidth=0.48\textwidth

\setcounter{figure}{0}

\newcommand{\fig}[3]
{
	\begin{figure}[!tb]
		\vspace*{-0.1cm}
		\[
		\includegraphics[width=\figwidth]{#1}
		\]
		\vskip -0.2cm
		\caption{\label{#2}
			\small#3
		}
\end{figure}}

\vspace{40cm}

{\bf Supplemental Materials}
\begin{table*}[h!]
	\caption{{\bf The parameters used in the Fermi surface simulation}. The parameters
are extracted by comparing the experimental data to the expansion $k_F(\psi,\kappa) = k_{0,0} + k_{0,4} \cos 4\psi + k_{1,0} \cos \kappa$.
The isotropic term is $  k_{0,0}$, $k_{04}$ is a four-fold symmetric term that affects the in-plane anisotropy
and $ k_{1,0}$ describes the interplane distortion of the Fermi surface.
   In FeS, the electron pockets are larger but significantly warped  with $k_{10}=-0.063$
   for the outer electron pockets and $k_{10}= -0.02$ for the outer hole cylindrical pocket \cite{Bates2020}. }
\begin{tabular}{lllllllllll}
\hline
\multicolumn{1}{l}{} & \multicolumn{6}{l}{Electron}                                                                                                                                                                                                           & \multicolumn{4}{l}{Hole}
\\ \hline
                       & \multicolumn{3}{l}{Inner}                                                                                          & \multicolumn{3}{l}{Outer}                                                                                          & \multicolumn{2}{l}{Inner}
                       & \multicolumn{2}{l}{Outer}                                                                                          \\ \hline
\textit{p} (kbar)      & $k_{\rm 00}$ & $k_{\rm 04}$ & $k_{\rm 10}$ & $k_{\rm 00}$ & $k_{\rm 04}$ & $k_{\rm 10}$ & $k_{\rm 00}$ & $k_{\rm 10}$ & $k_{\rm 00}$ & $k_{\rm 10}$\\ \hline
0 & 0.0575 & 0.005 & -0.0025 & 0.13 & -0.04 & -0.035 & 0.041 & -0.041 & 0.1365 & -0.0335 \\
4.6 & 0.0625 & 0.005 & -0.004 & 0.14 & -0.04 & -0.035 & 0.043 & -0.043 & 0.148 & -0.0305 \\
11 & 0.0675 & 0.005 & -0.007 & 0.1535 & -0.04 & -0.0355 & 0.0495 & -0.0495 & 0.1605 & -0.029 \\
17 & 0.0725 & 0.005 & -0.011 & 0.163 & -0.04 & -0.037 & 0.0515 & -0.0515 & 0.171 & -0.02675 \\
22 & 0.078 & 0.005 & -0.015 & 0.171475 & -0.04 & -0.0365 & 0.0566 & -0.0566 & 0.17975 & -0.025 \\ \hline
\end{tabular}
	\label{Table:FS_SimulationParameters}
\end{table*}

\begin{table*}[htpb]
\small
\caption{{\bf The extracted Fermi surface parameters from quantum oscillations studies}.
The effective masses are extracted using the Lifshitz-Kosevich formula \cite{Shoenberg1984}, as shown in Fig.~\ref{FigSM_FFT_LK}.
The Fermi energy is estimated using $E_{\rm F}= \hbar k^2/(2 m^*)$ and the Osanger relationship $F=\hbar (\pi k^2)/(2\pi e)$
 for different applied pressures.}
\begin{tabular}{lllllllllllllllll}
\hline
\hline
$p$ & \multicolumn{3}{l}{0 kbar} & \multicolumn{3}{l}{4.6 kbar} & \multicolumn{3}{l}{11 kbar} & \multicolumn{3}{l}{17 kbar} & \multicolumn{3}{l}{22 kbar} \\
\hline
Orbit     & $F$ (T)  & $m^{*}$ (m$_{\rm e}$)    & $E_{\rm F}$ (meV) & $F$ & $m^{*}$ & $E_{\rm F}$ & $F$ & $m^{*}$ & $E_{\rm F}$ & $F$ & $m^{*}$ & $E_{\rm F}$ & $F$ & $m^{*}$ & $E_{\rm F}$ \\ \hline
$\alpha_1$ &   &  &   & 100(10)  &   &      									& 103(12)   &        &        & 110(8)   &        &         & 120  &         &        \\
$\alpha_2$ &  100(10) &  &   & 130(10)  &    &									& 180(15)   &        &        & 230(15)  &  		&          & 260  &         &       \\
$\chi$       & 237(6) & 2.3(2)   & 11.7(10)     & 276(5)    & 2.1(2)   & 15(2) 	& 347(10)   & 2.0(2) & 20(2)  & 408(4)   & 2.0(2) &  24(2)  & 453  & 1.9(2)  &  28(2)\\
$\epsilon$    &    &  &    &     &        &      								& 447(3)    & 6.0(10)& 8.6(14)& 511(6)   & 4.5(10)&  13(3)  & 588  & 6.5(13) &  11(2)\\
$\beta$    &  359(4)   &   &  & 483(6)   & 3.7(2)   & 15(1)  					& 610(8)    & 4.05(20) & 18(1)  & 732(10)  & 4.2(2) &  20(1)& 835  & 4.5(3)  &  22(2)\\
2$\chi$    &   470(15)    &  &     & 572(6)     & 3.8(3)   &  18(1)   			& 719(10)   & 3.6(4) &  23(2) & 828(12)  & 3.4(2) &  28(2)  & 930  & 3.4(5)  &  32(5)\\
$\gamma$     & 873(24)    & 5.8(15)  &  18(5)   &  1068   &    &      			& 1158(18)  & 4.1(5) &  33(4) & 1285(11) &        &         & 1376 &         &        \\
$\delta$     & 971(11)   & 3.9(1)  & 29(4) & 1091(7)  & 3.8(3)  & 33(3) 		& 1236(10)  & 3.6(2) &  40(3) & 1343(5)  & 3.5(2) &  44(3)  & 1431 & 3.8(3)  &  44(3) \\ \hline \hline
\end{tabular}
	\label{Table:QOexperimental}
\end{table*}

\clearpage

\begin{figure*}[htbp]
	\centering
	\includegraphics[trim={0cm 0cm 0cm 0cm}, width=0.8\linewidth,clip=true]{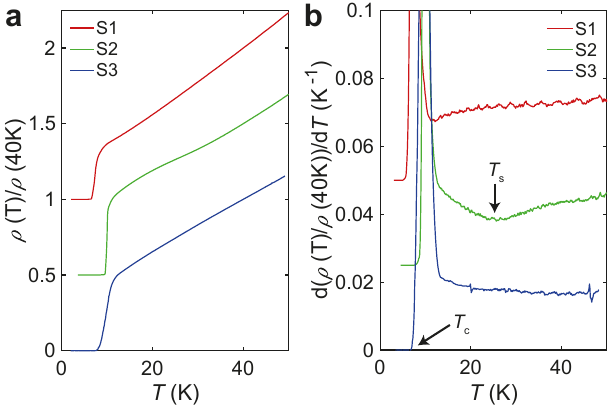}
	\caption{{\bf The low temperature resistivity for different crystals of FeSe$_{0.82}$S$_{0.18}$ from the same batch.}
	(a) The temperature dependence of the normalized resistivity at 40\,K and (b) the corresponding first derivative of the
resistivity  for the different single crystals investigated for this study.
The arrows indicate the position of the nematic transition $T_{\rm s}$ for sample S2, and the superconducting transition $T_{\rm c}$ for sample S3.
	}
	\label{FeSeSx18_RvT_AllSamples} %
\end{figure*}

\begin{figure*}[htbp]
	\centering
	\includegraphics[trim={0cm 0cm 0cm 0cm}, width=\linewidth,clip=true]{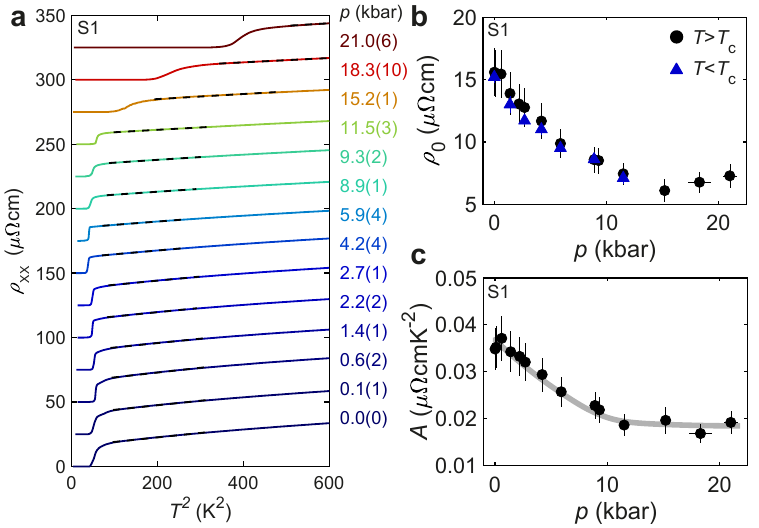}
	\caption{{\bf Low temperature Fermi liquid behaviour of FeSe$_{0.82}$S$_{0.18}$ for sample S1}. (a)
The temperature dependence of resistivity at low temperature for sample S1. Curves are shifted vertically
for clarity and the dashed lines are fits to a Fermi liquid behaviour, $\rho_0$+$A T^2$, over a limited temperature
range ( $\Delta T \sim 8$\,K)  above the onset of superconducting transition ($T_{\rm on}+2$\,K).
(b) Residual resistivity, $\rho_0$, extracted from the fits in (a) for $T > T_{\rm c}$.
For $T < T_{\rm c}$, the low temperature resistivity is extrapolated from quadratic behaviour in magnetic field (see Fig.~\ref{FigureSM_raw_RxxvB_S1})
before it is fitted to a Fermi liquid dependence.
 (c) The $A$ coefficient of the $T^2$ resistivity dependence
 as a function of pressure for sample S1.}
	\label{Fig_x18_Avp_T2_Fits}
\end{figure*}

\begin{figure*}[htbp]
	\centering
	\includegraphics[trim={0cm 0cm 0cm 0cm}, width=1\linewidth,clip=true]{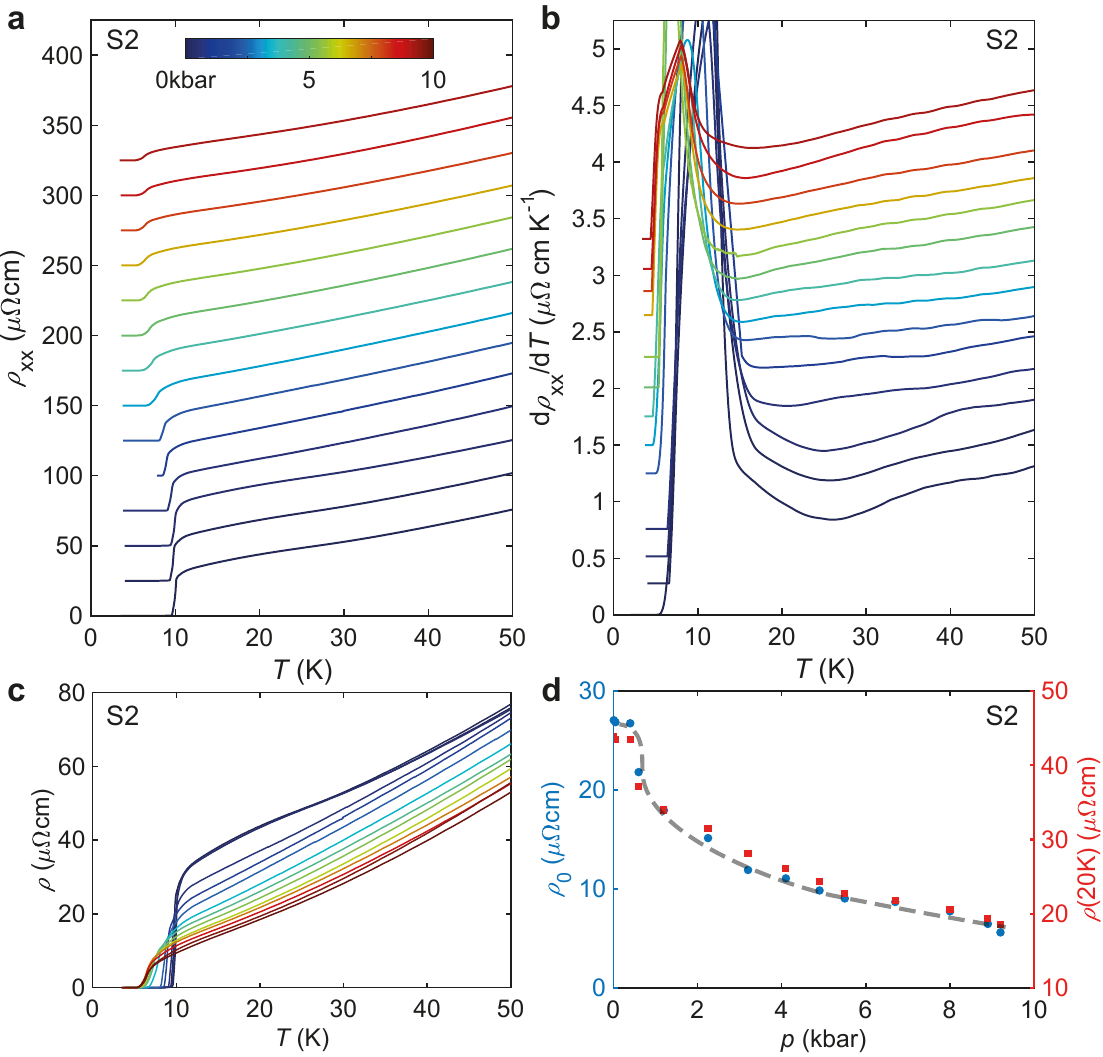}
	\caption{{\bf  Transport properties under pressure of FeSe$_{0.82}$S$_{0.18}$ for sample S2.}
(a) Temperature dependence of the resistivity up to 9\,kbar and (b) the corresponding derivative used for determining the transition temperatures.
Curves are shifted vertically for clarity. (c) The same as (a) without a vertical shift.
 (d) The resistivity extrapolated to zero temperature from a fit to the normal state resistivity above the onset of superconductivity, $\rho_0$ (blue circles, left vertical axis), and the value
 of resistivity at 20 K, (red squares, right vertical axis), as a function of applied pressure. The dashed line is a guide to the eye.
 The corresponding $A$ coefficient extracted from the $T^2$ dependence from panel (a)  is plotted in Fig.~\ref{FigSM_Acoefficient_all}.
  }
\label{FigSM_S2Transport}
\end{figure*}

\begin{figure*}[htbp]
	\centering
	\includegraphics[trim={0cm 0cm 0cm 0cm}, width=1\linewidth,clip=true]{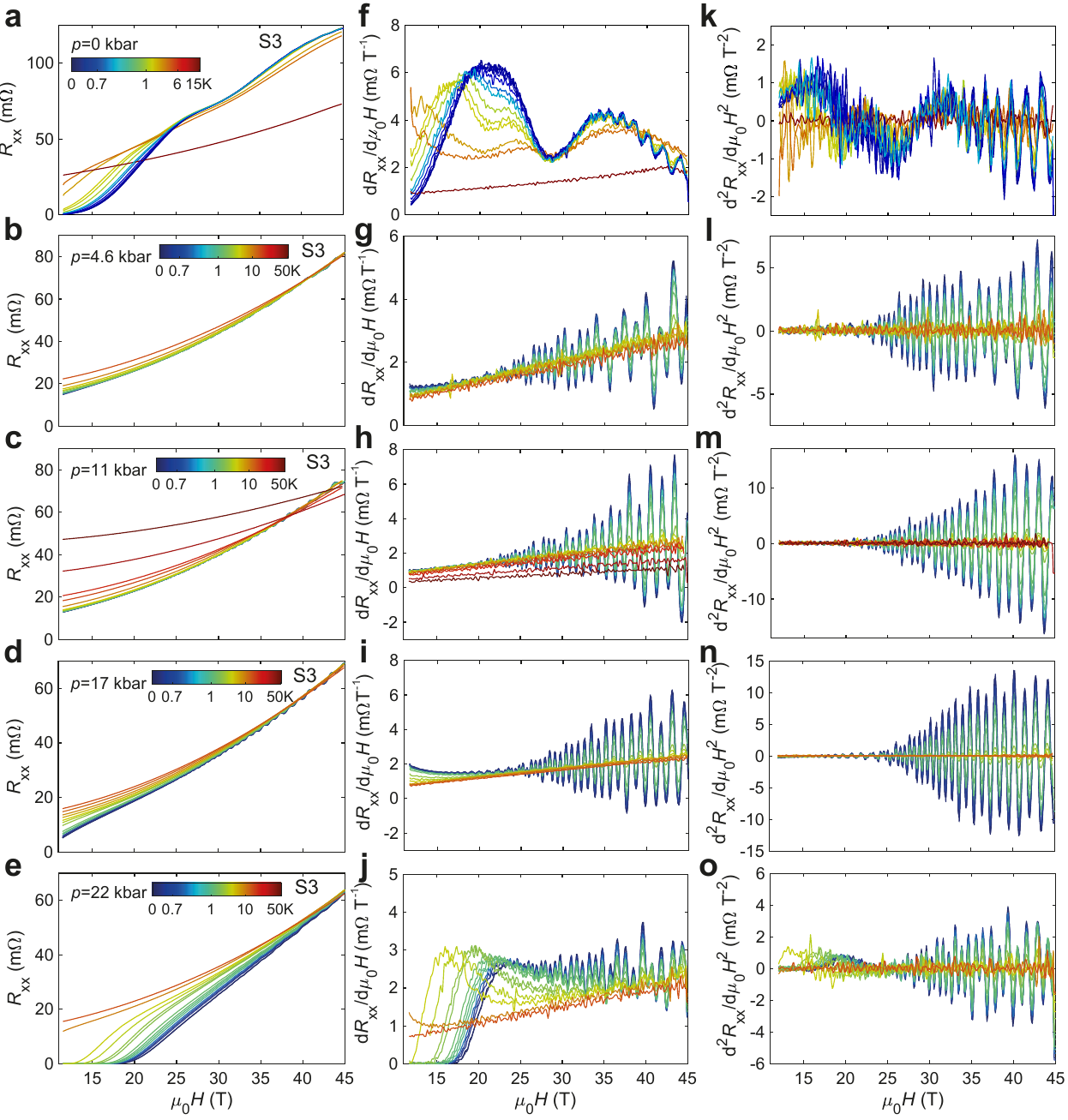}
	\caption{{\bf High-magnetic field longitudinal magnetoresistance of FeSe$_{0.82}$S$_{0.18}$ for sample S3.}
	The field dependence of the (a - e) magnetoresistance, (f - j) its first derivative
in field, and (k - o) its second derivative in magnetic field. Each row corresponds to a different pressure, and the
colour scale is used to define the temperature for each run applied across each row.
	}
	\label{FigureSM_FeSeSx18_S3_Derivatives}
\end{figure*}

\begin{figure*}[htbp]
	\centering
	\includegraphics[trim={0cm 0cm 0cm 0cm}, width=1\linewidth,clip=true]{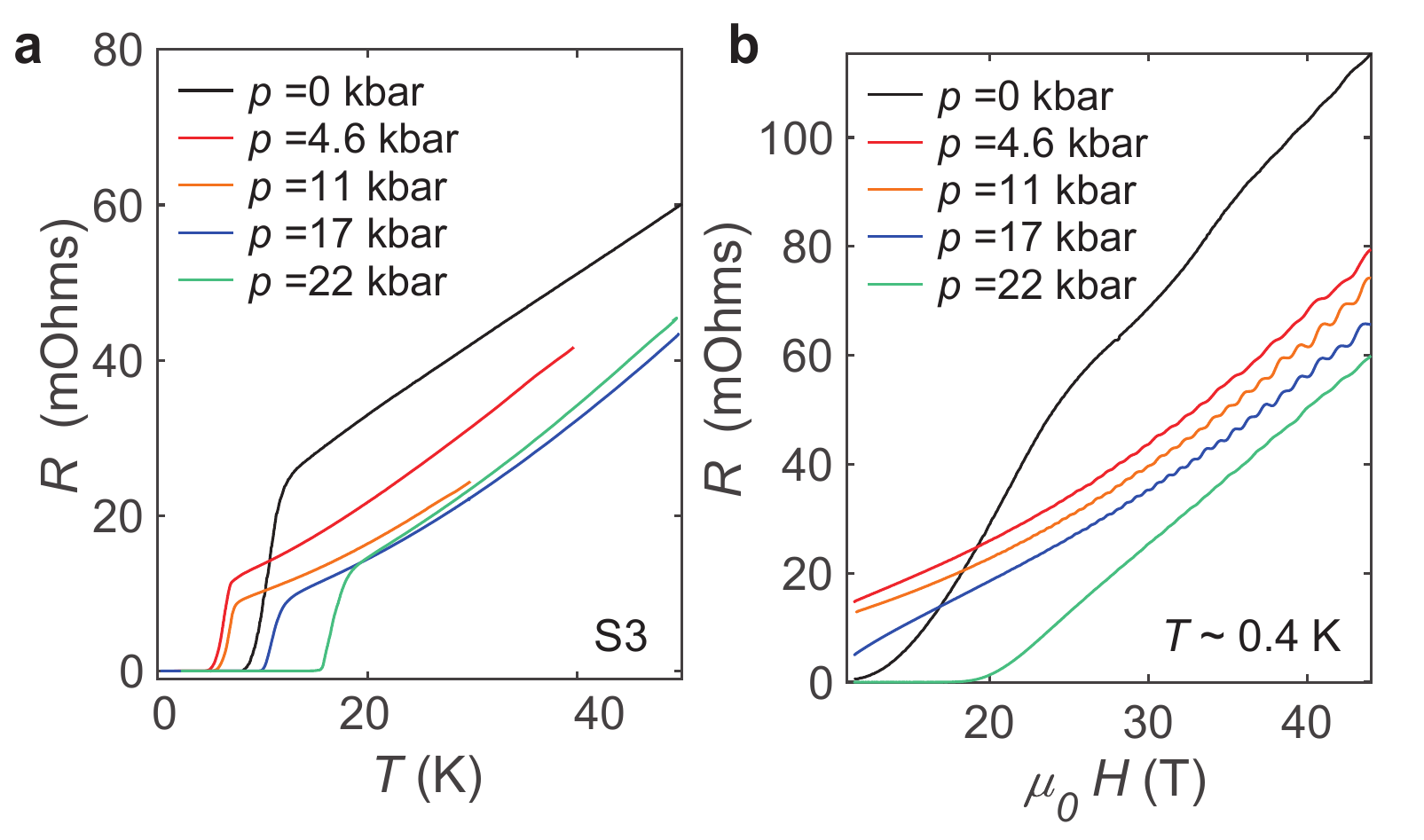}
	\caption{{\bf The transport properties of FeSe$_{0.82}$S$_{0.18}$ for sample S3.}
(a) The low temperature resistance in zero-magnetic field and (b) the field-dependence magnetoresistance at 0.4~K
for sample S3 measured at different applied pressures.
	}
	\label{SMFig_RvsB_T0p4K_S3}
\end{figure*}

\begin{figure*}[htbp]
	\centering
	\includegraphics[trim={0cm 0cm 0cm 0cm}, width=0.7\linewidth,clip=true]{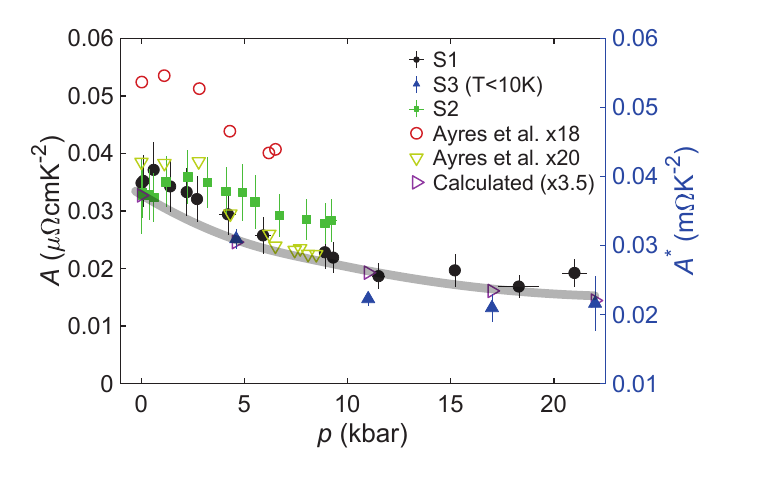}
	\caption{{\bf  The pressure dependence of the  $A$ coefficient.}
The pressure dependence of the  $A$ coefficient extracted from fitting to a low
temperature Fermi liquid behaviour for different crystals
and compared with $x$ = 0.18 and 0.20 from Ref.~\cite{Ayres2022}.
The $A$ values for samples S1 and S2 are
estimated in zero field from a temperature range above the
onset of superconductivity (Fig.~\ref{Fig_x18_Avp_T2_Fits} in the SM).
The $A^*$ values for sample S3 values are extracted from resistance below 10\,K and
are plotted on the right axis (Fig.~\ref{FigureSM_raw_S3}). Calculated $A $ values use
$A \sim (\sum_i k_{\rm F_i}^3/m^{*2}_i)^{-1}$ \cite{Tsujii2022,Hussey2005}
 and the $k_{\rm 00}$ values from Table.~\ref{Table:FS_SimulationParameters}.
 The solid line is a guide to the eye for the calculated $A$ values.
  }
\label{FigSM_Acoefficient_all}
\end{figure*}

\begin{figure*}[htbp]
	\centering
	\includegraphics[trim={0cm 0cm 0cm 0cm}, width=1\linewidth,clip=true]{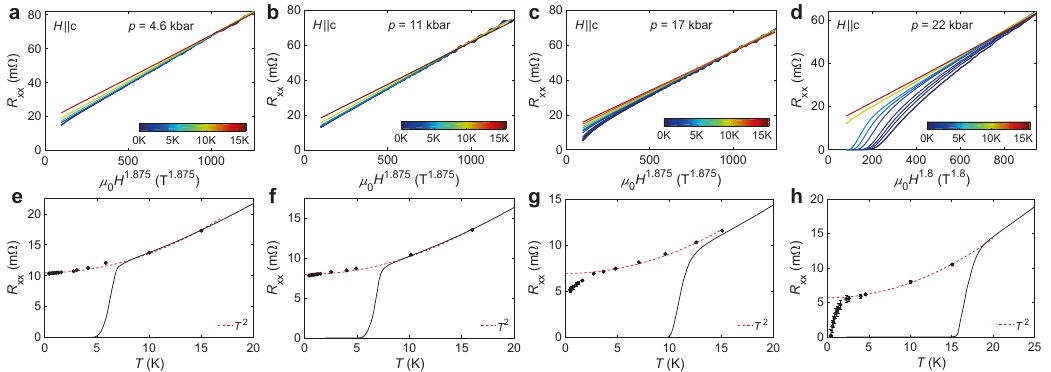}
	\caption{{\bf Magnetoresistance in magnetic field for FeSe$_{0.82}$S$_{0.18}$ for sample S3.}
(a)-(d) The longitudinal resistivity, $\rho_{\rm xx}$, as a function of $H^{1.875}$ field dependence for
different applied pressures. (e)-(h) The zero field resistivity at low temperatures for
each pressure (solid curve). Solid circles correspond to extrapolated zero
field values at low temperatures from (a)-(d). Red dashed lines are fits at low temperature resistivity
to a Fermi liquid $T^2$ dependence to extract the coefficient, $A^*$, shown in Fig.~\ref{FigSM_Acoefficient_all}.
	}
	\label{FigureSM_raw_S3}
\end{figure*}

\begin{figure*}[htbp]
	\centering
	\includegraphics[trim={0cm 0cm 0cm 0cm}, width=0.8\linewidth,clip=true]{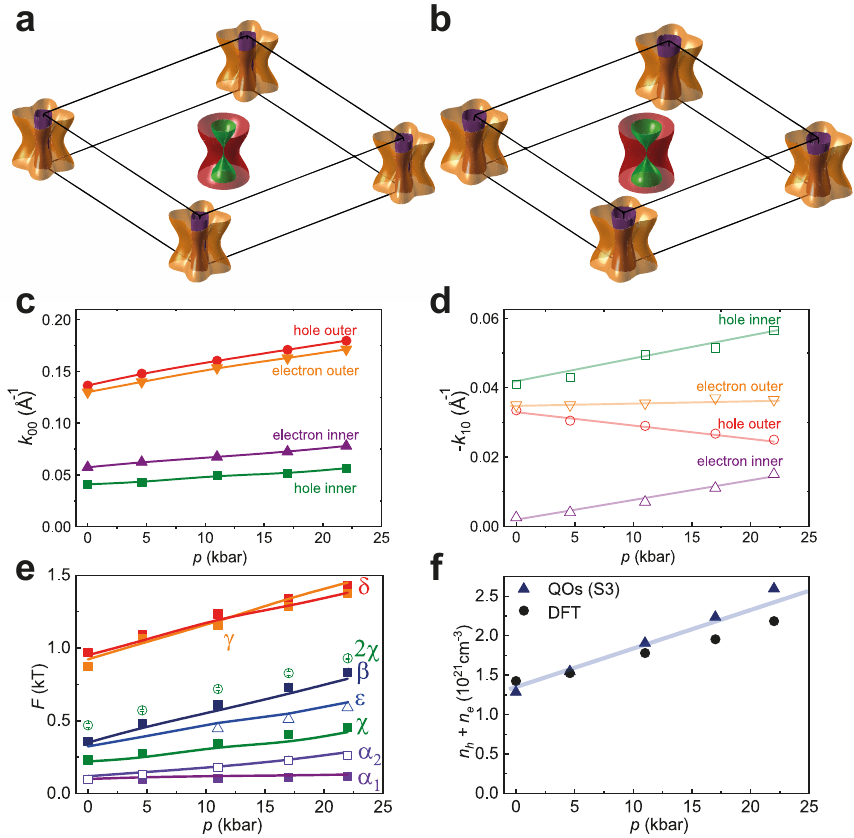}
	\caption{{\bf Modelling of the evolution of the Fermi surface of FeSe$_{0.82}$S$_{0.18}$ with pressure.}
Fermi surface topography using the expansion $k_F(\psi,\kappa) = k_{0,0} + k_{0,4} \cos 4\psi + k_{1,0} \cos \kappa$
and matching the experimental parameters corresponding at (a) 0 kbar and (b) 22 kbar.
The model expands the Fermi cylinders using symmetric parameters and match the observed frequencies and
Fermi surface maps seen in  ARPES \cite{Watson2015c}, where the hole pocket suffers
a Lifshitz transition \cite{Coldea2019}.
The variation of the $k_{\rm 00}$ in (c) and $k_{\rm 10}$  parameters in (d) with applied pressure,
  as listed  in Table~\ref{Table:FS_SimulationParameters}.
   Solid lines are guide to the eye.
(e) The pressure dependence of the quantum oscillations frequencies from experiments (symbols) and the model
(solid lines). (f) The estimation of the carrier density, $n$, using the Fermi surface model (solid triangles),
and compared to the
estimated values from shifted DFT calculations (solid circles).
	}
	\label{FigSM_FermiSurfaceModel}
\end{figure*}

\begin{figure*}[htbp]
	\centering
	\includegraphics[trim={0cm 0cm 0cm 0cm}, width=1\linewidth,clip=true]{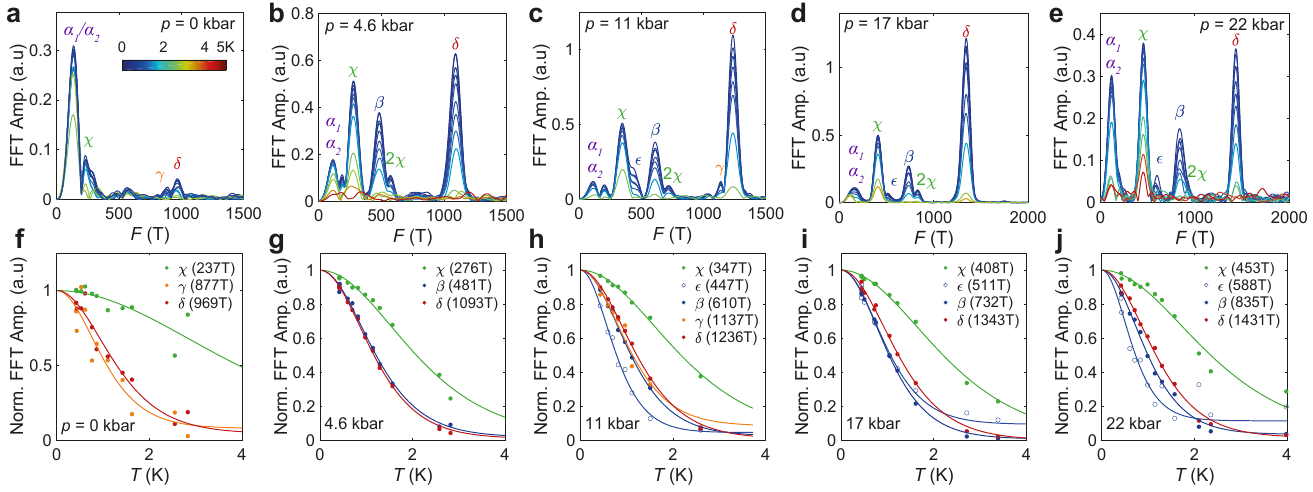}
	\caption{{\bf The determination of the quasiparticle effective masses.}
(a-e) The temperature dependence of the  Fast fourier frequency spectrum  dependence for different pressures for sample S3
using a field window $\Delta B$=22 to 44.5\,T and a Hanning window after a background of polynomial order 7 was removed.
 (f-j) The temperature dependence of the amplitude of the main peaks. The solid lines are fits to the Lifshitz-Kosevich formula which extracts the orbitally averaged effective mass of a particular orbit.
 and each plot is normalised to the base temperature for easier comparison.
 The main features correspond to hole pockets, $\chi, \beta$ and $\delta$ and the
 first harmonic of $\chi$ is clearly observed in most pressures, however, the $\epsilon$ and  $\gamma$ pocket with heavier mass  is much weaker.
	}
	\label{FigSM_FFT_LK}
\end{figure*}

\begin{figure*}[htbp]
  \centering
    \includegraphics[trim={0cm 0cm 0cm 0cm},clip,width=0.5\textwidth]{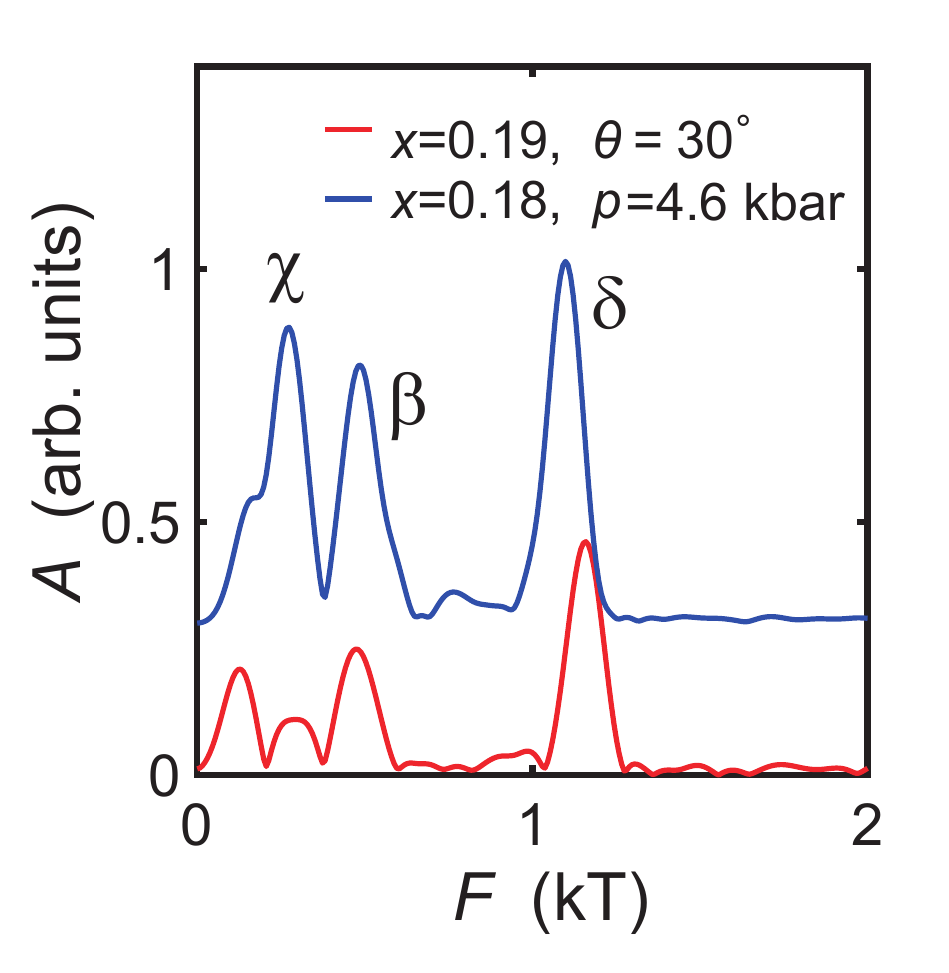}
  \caption{ {\bf Comparison between the FFT spectra outside the nematic phase at $T \sim 0.4$\,K.}
  The FFT spectra  using the field window 25-45\,T
  for FeSe$_{1-x}$S$_x$  for $x=0.18$ at $p=4.6$\,kbar  (top curve)
  and $x=0.19$, $\theta=30^\circ$ (bottom curve) from Ref.~\cite{Coldea2019}.
    }
  \label{fig:FFT_comparison}
\end{figure*}

\begin{figure*}[htbp]
	\centering
		\includegraphics[trim={0cm 0cm 0cm 0cm}, width=1\linewidth,clip=true]{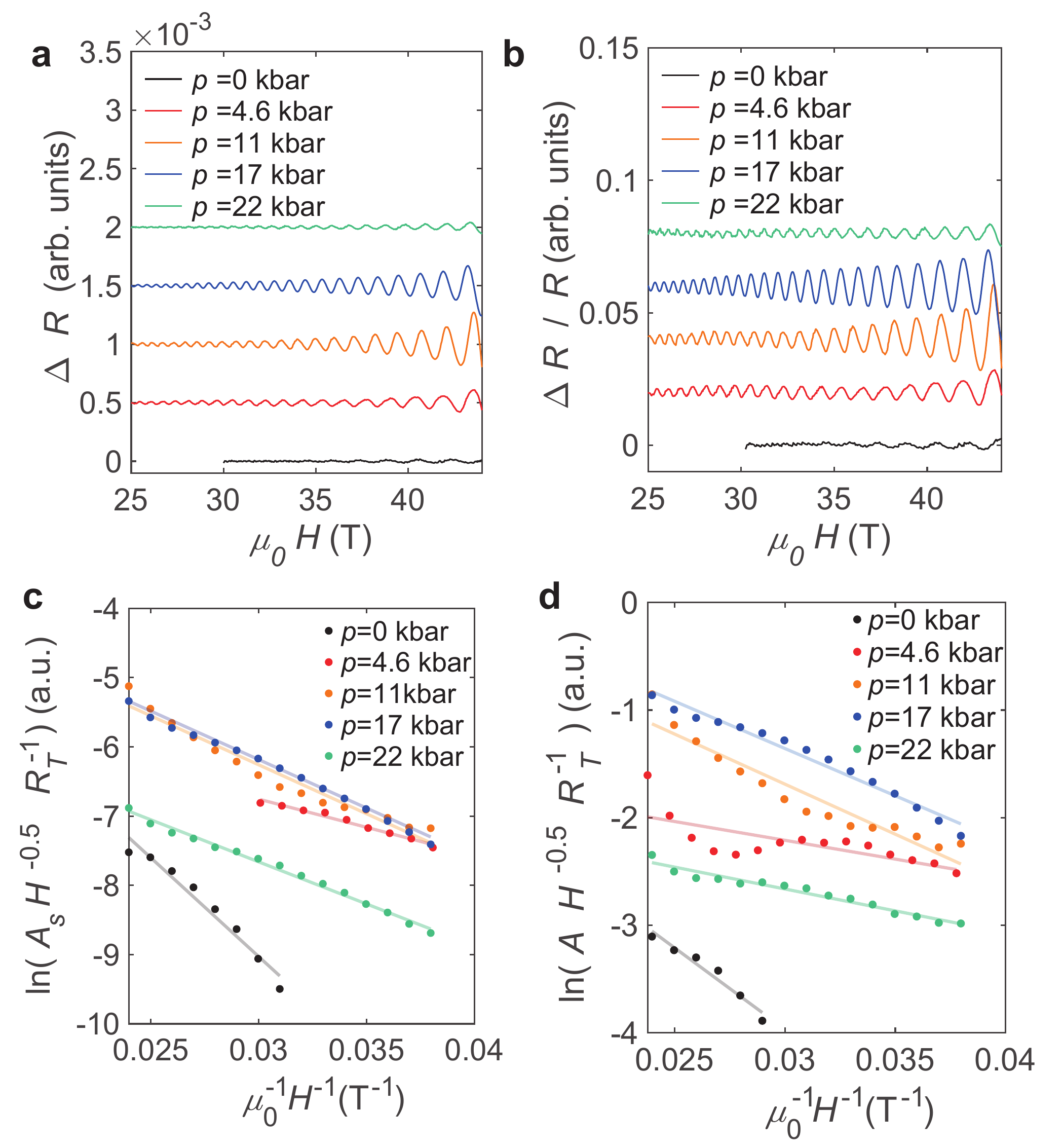}
	\caption{{\bf Dingle analysis of $\delta$ orbit.}
	Oscillatory part of the signal with a bandpass filter around the frequency of the largest orbit, $\delta$, as found from FFT spectra,
obtained by subtracting a high order polynomial from the raw data, $\Delta R $ in (a) or additionally dividing the raw background $\Delta R /R$.
The Dingle plots to extract the corresponding scattering times from (a) and (b) for the $\delta$ frequency at $\sim 0.4$~K.
Significant deviation from linear dependence is observed for 4.6~kbar, due to potential interference with $\gamma$ frequency.
}
	\label{FigSM_DeltaDingle}
\end{figure*}

\begin{figure*}[htbp]
	\centering
	\includegraphics[trim={0cm 0cm 0cm 0cm}, width=1\linewidth,clip=true]{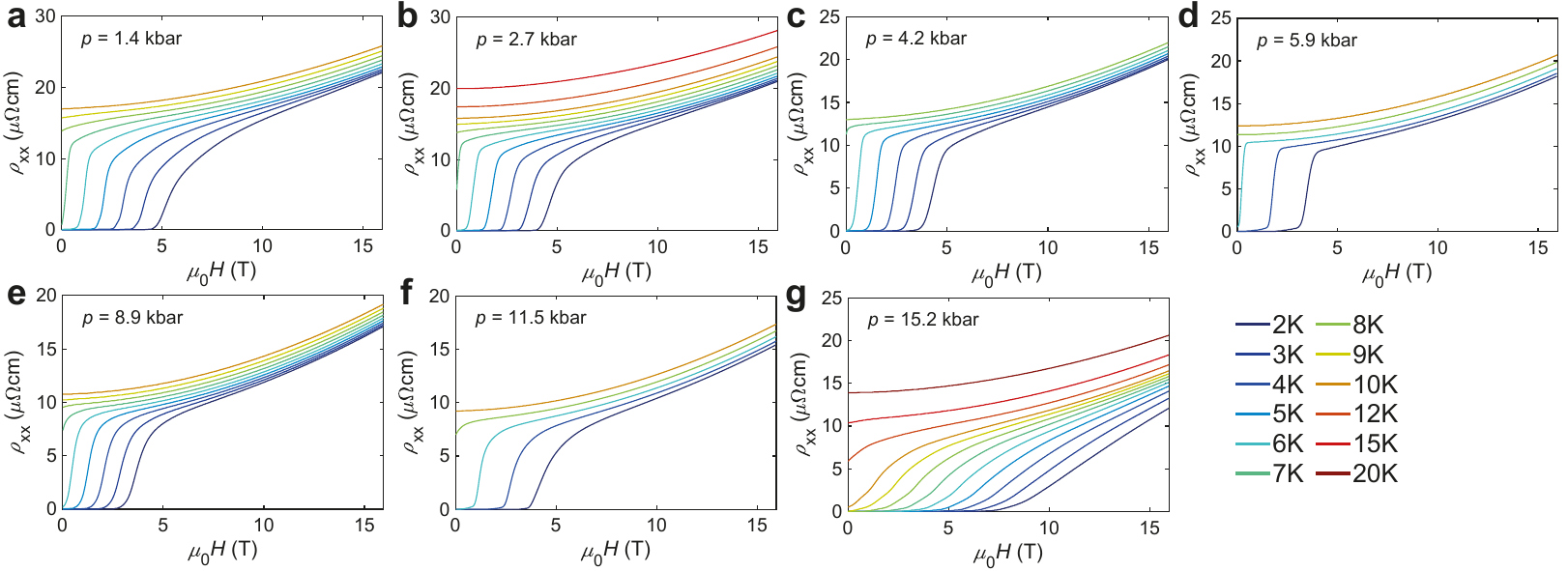}
	\caption{{\bf Low-magnetic field longitudinal resistivity for FeSe$_{0.82}$S$_{0.18}$ for sample S1.}
 (a)-(g) Longitudinal resistivity field dependence measured at fixed temperatures for different pressures.
The transitions in magnetic field become broader  at lowest temperature above 15~kbar
 in the region of the phase diagram where $T_{\rm c}$($p$) increases fastest (1.2\,K/kbar).
	}
	\label{FigureSM_raw_RxxvB_S1}
\end{figure*}

\begin{figure*}[htbp]
  \centering
    \includegraphics[trim={0cm 0cm 0cm 0cm},clip,width=0.6\textwidth]{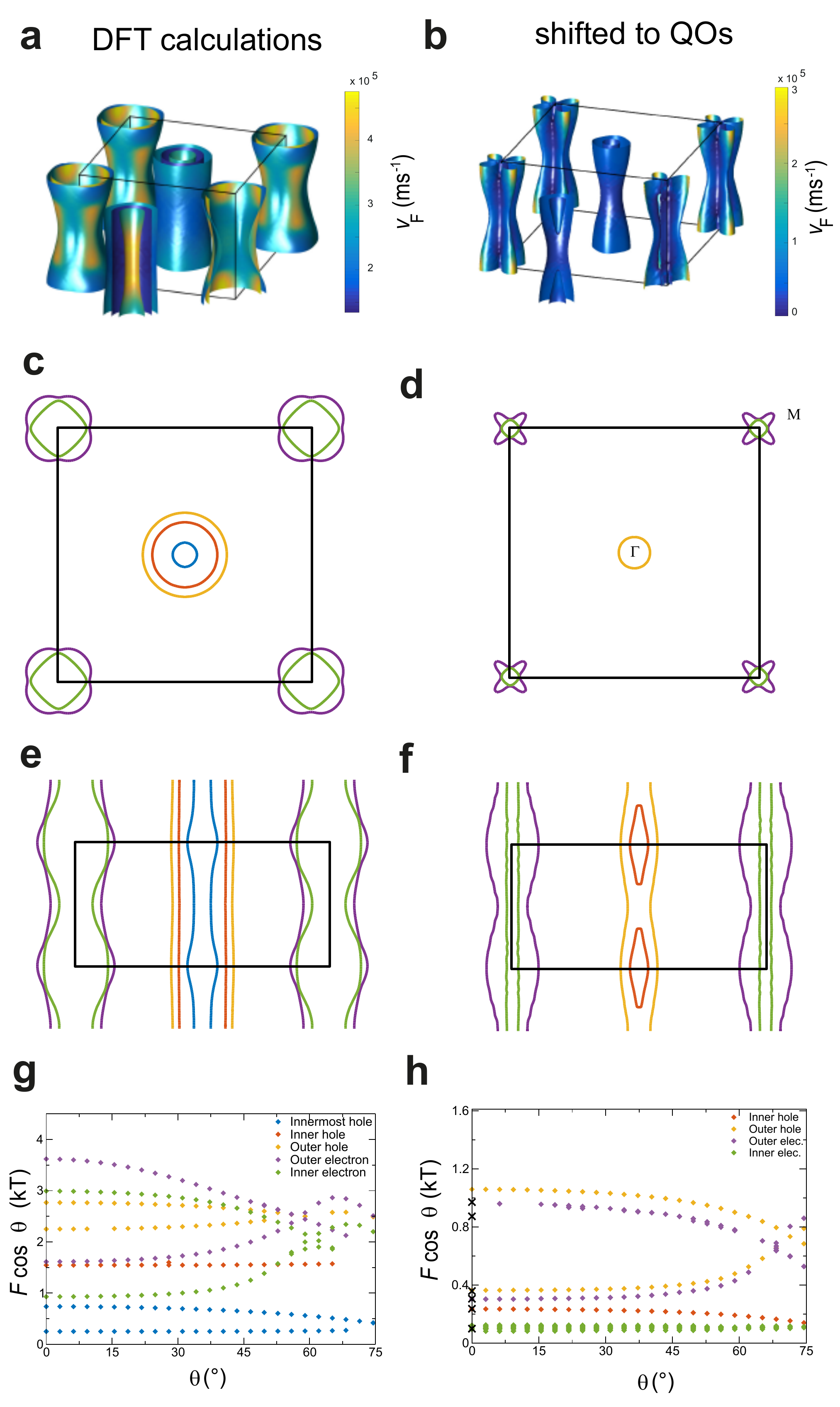}
   \caption{
\textbf{Fermi surface of FeSe$_{0.82}$S$_{0.18}$ from DFT calculations
  and the resulting one to match experimental data at ambient
  pressure.} Density functional theory (DFT) calculations were
performed using Wien2k with the PBE functional (GGA approximation) for
the non-magnetic case and including the spin-orbit coupling.
We used the following lattice parameters
$a = b = 3.754$~\AA, $c=5.442$~\AA, and $z$=0.266343,
where $z$ is the position of the
chalcogen atoms above Fe plane.
(a) The three-dimensional representation of the Fermi surface from DFT
calculations shows three cylindrical hole pockets in the centre of
the Brillouin zone and two electron pockets at the corners.
 (b-c) The Fermi surface after applying shifts and renormalisation to match to the
experimental data from quantum oscillations (QOs) in (b).
(c-d) The Fermi surface cuts perpendicular to the c-axis at the centre of the Brillouin zone
($k_z=0$) corresponding to the Fermi surfaces in (a-b). (e-f) The
corresponding Fermi surface cuts along the $(110)$ plane and (g-h) the
frequency rotation plots corresponding to the extremal areas on the
Fermi surface from $B||c$ ($\theta = 0^\circ$) towards
$B||a$ ($\theta = 90^\circ$). The  shifted Fermi surface contains two
electron and two hole pockets having mainly cylindrical shape
except for the inner hole pocket which is a 3D pocket
centred at $k_z = 0.5$. The electron pockets are cylindrical with different
degree of in-plane and out-of-plane warping, with the outer electron
pocket having a flower shape with four-fold symmetry and the inner
electron pocket being a small and almost 2D cylinder.
}
  \label{fig:QOShiftedFS}
\end{figure*}

\begin{figure*}[htbp]
  \centering
    \includegraphics[trim={0cm 0cm 0cm 0cm},clip,width=0.8\textwidth]{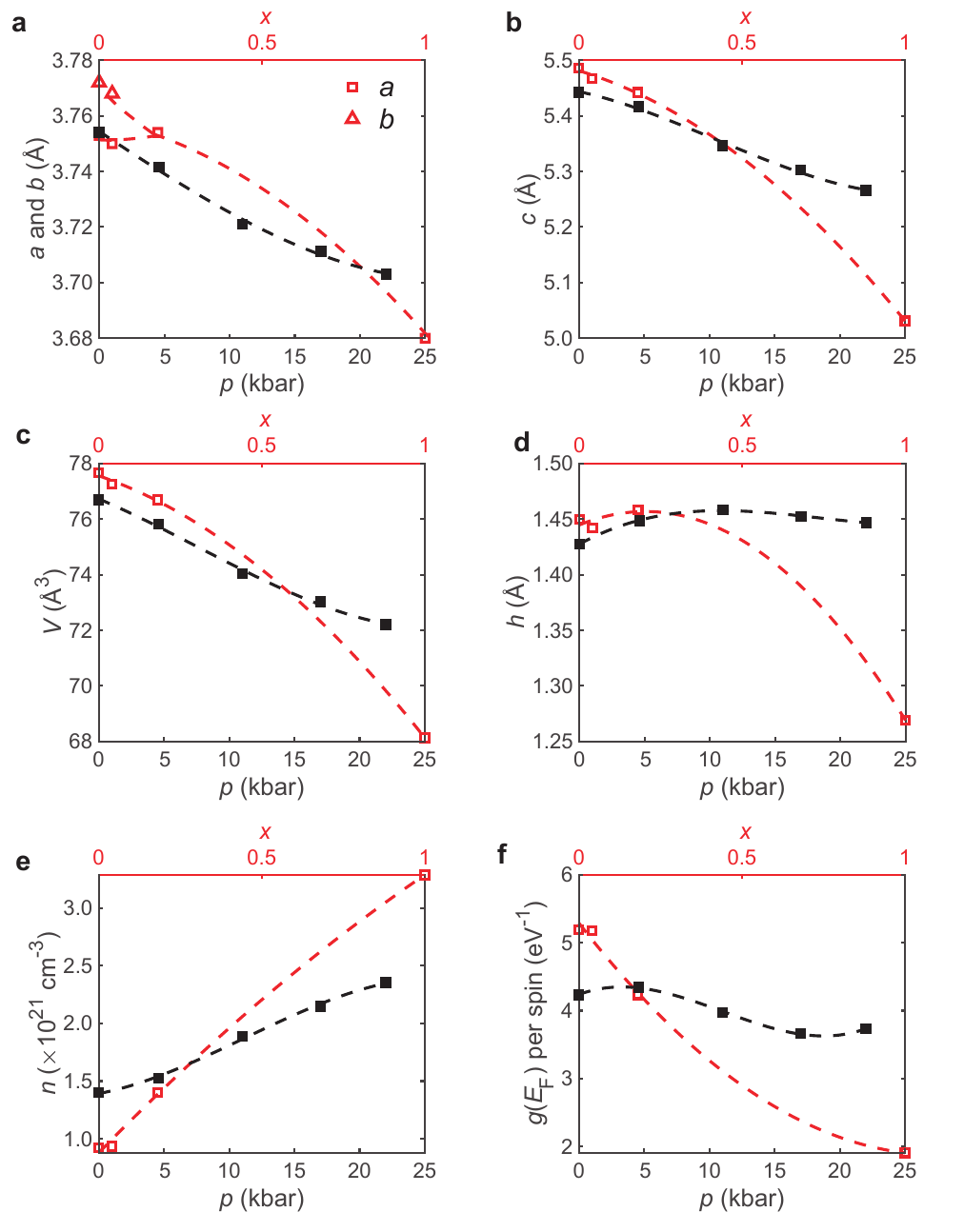}
  \caption{
\textbf{The evolution of the structural parameters as a function of
  hydrostatic pressure, $p$, and isovalent substitution, $x$ in
  FeSe$_{1−x}$S$_x$.} (a-d) Lattice constants $a$ (or $b$) (in panel
a) and $c$ (panel b) as well as the volume of the unit cell $V$ (c)
and the chalcogen height, $h=z \times c$ (d),
as a function of S content, $x$ at ambient pressure (open red symbols,
top axis), and as a function of applied pressure, $p$ for $x=0.18$
(black squares, bottom axis) estimated using the same pressure trends as those
for powder samples of $x=0.2$
from Ref.~\cite{Tomita2015}. (e-f) The trends of the estimated total carrier density
(e) and the density of states at the Fermi level (f) as a function of
$x$ and $p$ after shifting and renormalising the calculated electronic
band structure to match experimental quantum oscillations data. 
DFT calculations assume that the chalcogen is populated by
 Se ions, and modify the lattice parameters and $z$ values,
 except for FeS, where S is the chalcogen.
    }
  \label{fig:structural_details}
\end{figure*}

\end{document}